\documentclass[longauth]{aa}
\usepackage[utf8]{inputenc}
\usepackage[varg]{txfonts}
\usepackage[breaklinks, colorlinks, citecolor=blue, linkcolor=blue]{hyperref}
\usepackage[usenames, dvipsnames]{color}
\usepackage{amsmath}
\usepackage{graphicx}
\usepackage[english]{babel}
\usepackage{siunitx}
\usepackage{float}
\usepackage{url}
\usepackage{longtable}
\usepackage{fancyhdr}
\usepackage{natbib}
\usepackage[normalem]{ulem}
\usepackage[version=4]{mhchem}

\usepackage{gensymb}

\makeatletter

\def\instrefs#1{{\def\scsep{\def\scsep{,}}\@for\w:=#1\do{\scsep\ref{inst:\w}}}}
\renewcommand{\inst}[1]{\unskip$^{\instrefs{#1}}$}

\let\orgautoref\autoref
\renewcommand{\autoref}
        {\def\equationautorefname{Eq.}%
         \def\figureautorefname{Fig.}%
         \def\sectionautorefname{Sect.}%
         \def\subsectionautorefname{Sect.}%
         \def\subsubsectionautorefname{Sect.}%
         \orgautoref}

\renewcommand*\aa@pageof{, page \thepage{} of \pageref*{LastPage}} 

\makeatother
\begin{document}

\title{The HD~260655 system: \\ Two rocky worlds transiting a bright M~dwarf at 10\,pc
			}
\titlerunning{Multi-planetary system orbiting HD~260655}

\author{R.~Luque\inst{iaa,uchicago} 
\and B.\,J.~Fulton\inst{caltech}
\and M.~Kunimoto\inst{mit2} 
\and P.\,J.~Amado\inst{iaa}
\and P.~Gorrini\inst{iag}
\and S.~Dreizler\inst{iag}
\and C.~Hellier\inst{keele}
\and G.\,W.~Henry\inst{tsu}
\and K.~Molaverdikhani\inst{lmu,origins,mpia,lsw}
\and G.~Morello\inst{iac,ull}
\and L.~Peña-Moñino\inst{iaa}
\and M.~Pérez-Torres\inst{iaa}
\and F.\,J.~Pozuelos\inst{liege1,liege2}
\and Y.~Shan\inst{oslo}
\and G.~Anglada-Escudé\inst{ice} 
\and V.\,J.\,S.~B\'ejar\inst{iac,ull} 
\and G.~Bergond\inst{caha} 
\and A.\,W.~Boyle\inst{nexsci} 
\and J.\,A.~Caballero\inst{cabesac}
\and D.~Charbonneau\inst{cfa}
\and D.\,R.~Ciardi\inst{nexsci}
\and S.~Dufoer\inst{oudebleken}
\and N.~Espinoza\inst{stsci}
\and M.~Everett\inst{noirlab}
\and D.~Fischer\inst{yale}
\and A.\,P.~Hatzes\inst{tls}
\and Th.~Henning\inst{mpia}
\and K.~Hesse\inst{mit2}
\and A.\,W.~Howard\inst{caltech}
\and S.\,B.~Howell\inst{Ames}
\and H.~Isaacson\inst{berkeley}
\and S.\,V.~Jeffers\inst{mps}
\and J.\,M.~Jenkins\inst{Ames} 
\and S.\,R.~Kane\inst{riverside} 
\and J.~Kemmer\inst{lsw} 
\and S.~Khalafinejad\inst{lsw}
\and R.\,C.~Kidwell~Jr.\inst{stsci}
\and D.~Kossakowski\inst{mpia}
\and D.\,W.~Latham\inst{cfa}
\and J.~Lillo-Box\inst{cabesac}
\and J.\,J.~Lissauer\inst{Ames}
\and D.~Montes\inst{ucm}
\and J.~Orell-Miquel\inst{iac,ull}
\and E.~Pall\'e\inst{iac,ull}
\and D.~Pollacco\inst{warwick}
\and A.~Quirrenbach\inst{lsw}
\and S.~Reffert\inst{lsw}
\and A.~Reiners\inst{iag}
\and I.~Ribas\inst{ice,ieec}
\and G.\,R.~Ricker\inst{mit2} 
\and L.\,A.~Rogers\inst{uchicago}
\and J.~Sanz-Forcada\inst{cabesac}
\and M.~Schlecker\inst{arizona}
\and A.~Schweitzer\inst{hs}
\and S.~Seager\inst{mit2,mit1,mit3}
\and A.~Shporer\inst{mit2}
\and K.\,G.~Stassun\inst{vanderbilt} 
\and S.~Stock\inst{lsw}
\and L.~Tal-Or\inst{ariel,iag}
\and E.\,B.~Ting\inst{Ames} 
\and T.~Trifonov\inst{mpia,sofia} 
\and S.~Vanaverbeke\inst{astrolab}
\and R.~Vanderspek\inst{mit2}
\and J.~Villase\~nor\inst{mit2}
\and J.\,N.~Winn\inst{princeton}
\and J.\,G.~Winters\inst{cfa}
\and M.\,R.~Zapatero Osorio\inst{cab}
}

\institute{
\label{inst:iaa}Instituto de Astrof\'isica de Andaluc\'ia (IAA-CSIC), Glorieta de la Astronom\'ia s/n, 18008 Granada, Spain
\and
\label{inst:uchicago}Department of Astronomy \& Astrophysics, University of Chicago, Chicago, IL 60637, USA; \email{rluque@uchicago.edu}
\and
\label{inst:caltech}Division of Geological and Planetary Sciences, California Institute of Technology, 1200 East California Blvd, Pasadena, CA, 91125, USA
\and
\label{inst:mit2}Department of Physics and Kavli Institute for Astrophysics and Space Research, Massachusetts Institute of Technology, Cambridge, MA 02139, USA
\and
\label{inst:iag}Institut f\"ur Astrophysik, Georg-August-Universit\"at, Friedrich-Hund-Platz 1, 37077 G\"ottingen, Germany
\and
\label{inst:keele}Astrophysics Group, Keele University, Keele, ST5 6BG, U.K.
\and
\label{inst:tsu}Center of Excellence in Information Systems, Tennessee State University, Nashville, TN  37209-1561,  USA
\and 
\label{inst:lmu}Universitäts-Sternwarte, Ludwig-Maximilians-Universität München, Scheinerstrasse 1, D-81679 München, Germany
\and
\label{inst:origins}Exzellenzcluster Origins, Boltzmannstraße 2, 85748 Garching, Germany
\and 
\label{inst:mpia}Max-Planck-Institut f\"ur Astronomie, K\"onigstuhl 17, 69117 Heidelberg, Germany
\and
\label{inst:lsw}Landessternwarte, Zentrum für Astronomie der Universität Heidelberg, Königstuhl 12, 69117 Heidelberg, Germany
\and 
\label{inst:iac}Instituto de Astrof\'isica de Canarias (IAC), 38205 La Laguna, Tenerife, Spain
\and 
\label{inst:ull}Departamento de Astrof\'isica, Universidad de La Laguna (ULL), 38206, La Laguna, Tenerife, Spain
\and
\label{inst:liege1}Space Sciences, Technologies and Astrophysics Research (STAR) Institute, Universit\' de Li\`ege, All\'ee du 6 Ao\^ut 19C, B-4000 Li\`ege, Belgium
\and
\label{inst:liege2}Astrobiology Research Unit, Universit\'e de Li\`ege, All\'ee du 6 Ao\^ut 19C, B-4000 Li\`ege, Belgium
\and
\label{inst:oslo}Center for Earth Evolution and Dynamics, Department of Geosciences, University of Oslo, Sem Saelands vei 2b 0315 Oslo, Norway
\and
\label{inst:ice}Institut de Ciències de l’Espai (CSIC), Campus UAB, c/ de Can Magrans s/n, E-08193 Bellaterra, Barcelona, Spain
\and
\label{inst:caha}Centro Astronómico Hispano Alemán, Observatorio de Calar Alto, Sierra de los Filabres, 04550 Gérgal, Almería, Spain
\and
\label{inst:nexsci}NASA Exoplanet Science Institute, Caltech/IPAC, Mail Code 100-22, 1200 E. California Blvd., Pasadena, CA 91125, USA
\and 
\label{inst:cabesac}Centro de Astrobiolog\'ia (INTA-CSIC), ESAC Campus, Camino Bajo del Castillo s/n, 28692 Villanueva de la Ca\~nada, Madrid, Spain
\and
\label{inst:cfa}Center for Astrophysics | Harvard \& Smithsonian, 60 Garden St, Cambridge, MA 02138, USA
\and
\label{inst:oudebleken}Vereniging Voor Sterrenkunde, Oostmeers 122 C, 8000 Brugge, Belgium
\and
\label{inst:stsci}Space Telescope Science Institute, 3700 San Martin Drive, Baltimore, MD, 21218, USA
\and
\label{inst:noirlab}NSF’s Optical Infrared Astronomy Research Laboratory, 950 North Cherry Avenue Tucson, AZ 85719, USA
\and
\label{inst:yale}Department of Astronomy, Yale University, 52 Hillhouse Avenue, New Haven, CT 06511, USA
\and 
\label{inst:tls}Th\"uringer Landessternwarte Tautenburg, Sternwarte 5, 07778 Tautenburg, Germany
\and
\label{inst:Ames}Space Science \& Astrobiology Division, NASA Ames Research Center, Moffett Field, CA 94035, USA
\and
\label{inst:berkeley}Department of Astronomy, The University of California, Berkeley, CA 94720, USA
\and
\label{inst:mps}Max Planck Institute for Solar System Research, Justus-von-Liebig-Weg 3, D-37077 Göttingen, Germany
\and
\label{inst:riverside}Department of Earth and Planetary Sciences, University of California, Riverside, CA 92521, USA
\and
\label{inst:ucm}Departamento de F{\'i}sica de la Tierra y Astrof{\'i}sica \& IPARCOS-UCM (Instituto de F\'{i}sica de Part\'{i}culas y del Cosmos de la UCM), Facultad de Ciencias F{\'i}sicas, Universidad Complutense de Madrid, E-28040 Madrid, Spain
\and
\label{inst:warwick}Department of Physics, University of Warwick, Gibbet Hill Road, Coventry CV4 7AL, UK
\and
\label{inst:ieec}Institut d’Estudis Espacials de Catalunya, E-08034 Barcelona, Spain
\and
\label{inst:arizona}Department of Astronomy/Steward Observatory, The University of Arizona, 933 North Cherry Avenue, Tucson, AZ 85721, USA
\and
\label{inst:hs}Hamburger Sternwarte, Universität Hamburg, Gojenbergsweg 112, 21029 Hamburg, Germany
\and
\label{inst:mit1}Department of Earth, Atmospheric and Planetary Sciences, Massachusetts Institute of Technology, Cambridge, MA 02139, USA
\and
\label{inst:mit3}Department of Aeronautics and Astronautics, MIT, 77 Massachusetts Avenue, Cambridge, MA 02139, USA
\and
\label{inst:vanderbilt}Department of Physics and Astronomy, Vanderbilt University, Nashville, TN 37235, USA
\and
\label{inst:ariel}Department of Physics, Ariel University, Ariel 40700, Israel
\and
\label{inst:sofia}Department of Astronomy, Sofia University ``St Kliment Ohridski'', 5 James Bourchier Blvd, BG-1164 Sofia, Bulgaria
\and
\label{inst:astrolab}AstroLAB IRIS, Provinciaal Domein “De Palingbeek”, Verbrandemolenstraat 5, 8902 Zillebeke, Ieper, Belgium
\and
\label{inst:princeton}Department of Astrophysical Sciences, Princeton University, 4 Ivy Lane, Princeton, NJ 08544, USA 
\and
\label{inst:cab}Centro de Astrobiología (CSIC-INTA). Carretera de Ajalvir km 4. E-28850 Torrejón de Ardoz, Madrid. Spain.
}
\date{Received 22 April 2022 / Accepted 13 June 2022}

\abstract{We report the discovery of a multi-planetary system transiting the M0\,V dwarf HD~260655 (GJ~239, TOI-4599). The system consists of at least two transiting planets, namely HD~260655~b, with a period of 2.77\,d, a radius of $R_{\rm b}=1.240\pm0.023\,R_\oplus$, a mass of $M_{\rm b}=2.14\pm0.34\,M_\oplus$, and a bulk density of $\rho_{\rm b}=6.2\pm1.0\,\mathrm{g\,cm^{-3}}$, and HD~260655~c, with a period of 5.71\,d, a radius of $R_{\rm c}=1.533^{+0.051}_{-0.046}\,R_\oplus$, a mass of $M_{\rm c}=3.09\pm0.48\,M_\oplus$, and a bulk density of $\rho_{\rm c}=4.7^{+0.9}_{-0.8}\,\mathrm{g\,cm^{-3}}$. The planets were detected in transit by the \textit{TESS} mission and confirmed independently with archival and new precise radial velocities obtained with the HIRES and CARMENES instruments since 1998 and 2016, respectively. At a distance of 10\,pc, HD~260655 becomes the fourth closest known multi-transiting planet system after HD~219134, LTT~1445~A, and AU~Mic. Due to the apparent brightness of the host star ($J=6.7\,\mathrm{mag}$), both planets are among the most suitable rocky worlds known today for atmospheric studies with the \textit{JWST}, both in transmission and emission. }


\maketitle

\section{Introduction} \label{sec:intro}

Space missions devoted to exoplanet research via the transit technique (\textit{CoRoT}, \citealt{COROT}; \textit{Kepler/K2}, \citealt{Kepler,Howell2014}; \textit{TESS}, \citealt{TESS}) are providing a wealth of discoveries and precisely measured parameters for planets with radii between 1 and $4\,R_\oplus$. These planets, with no counterpart in the solar system, have been found to be very abundant around early M~dwarfs, more so than around solar-type stars \citep{Howard2012,Mulders2015,Sabotta2021}. M~dwarfs are interesting as planetary hosts because of their relative sizes and masses with respect to their planets, which make these systems more easily detectable by transit and radial velocity (RV) techniques.

Measuring precise planetary bulk densities by combining both techniques is especially important for these planets, since a wide diversity of compositions and atmospheric evolution paths are possible a priori \citep[e.g.,][]{RogersSeager2010,Zeng2019}. In this way it is possible to constrain the mass fraction of the gas and water of the planet, which informs formation and evolution processes \citep[e.g.,][]{Luque2021,Delrez2021,Wilson2022}. Multiplanetary systems offer a unique opportunity for characterization via comparative planetology, as they have formed within the same protoplanetary disk.

In what follows, we present the work that led to the discovery, confirmation and characterization of a multi-planetary system with at least two transiting planets around the bright, early-type M~dwarf HD~260655. The paper is organized as follows. Section~\ref{sec:tess} presents the space-based photometry from \textit{TESS} and discusses its processing and analysis. Section~\ref{sec:obs} provides similar information for all the ground-based observations, which include seeing-limited photometry, high-angular resolution imaging and high-spectral resolution spectroscopy. Section~\ref{sec:star} provides all available information on the host star, including the determination of its rotational period. Section~\ref{sec:fit} shows our analysis of all data. In Sect.~\ref{sec:discussion}, we discuss several aspects of our results, including the detection limits of our data sets, the internal composition and formation history of the planets, the detectability of their atmospheres, and the possibility of interaction between the star and the planets detectable at radio wavelengths. Section~\ref{sec:conclusions} summarizes our results and presents our conclusions.

\section{\textit{TESS} photometry} \label{sec:tess}

\begin{figure*}[ht!]
    \centering
    \includegraphics[width=0.33\hsize]{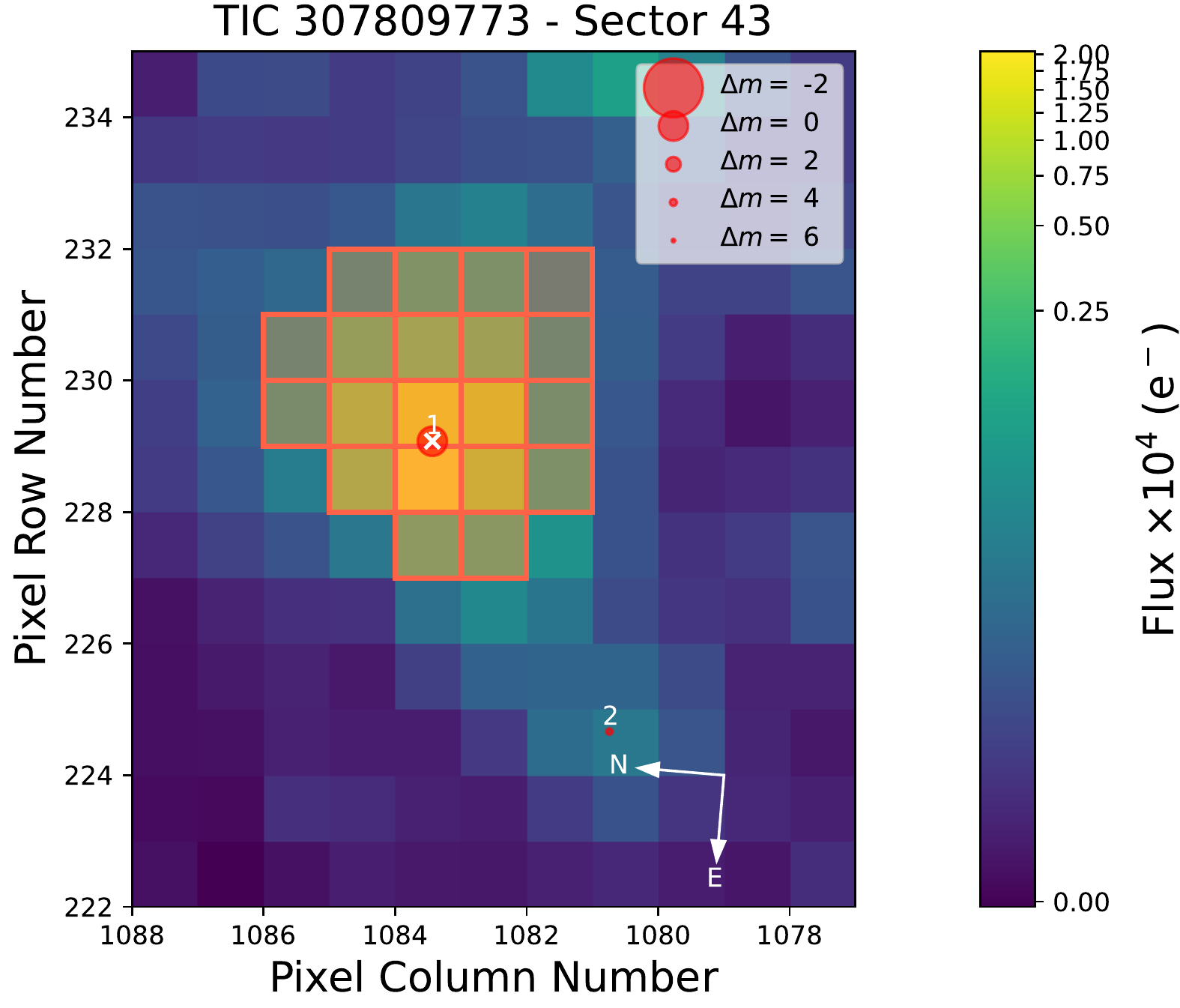}
    \includegraphics[width=0.33\hsize]{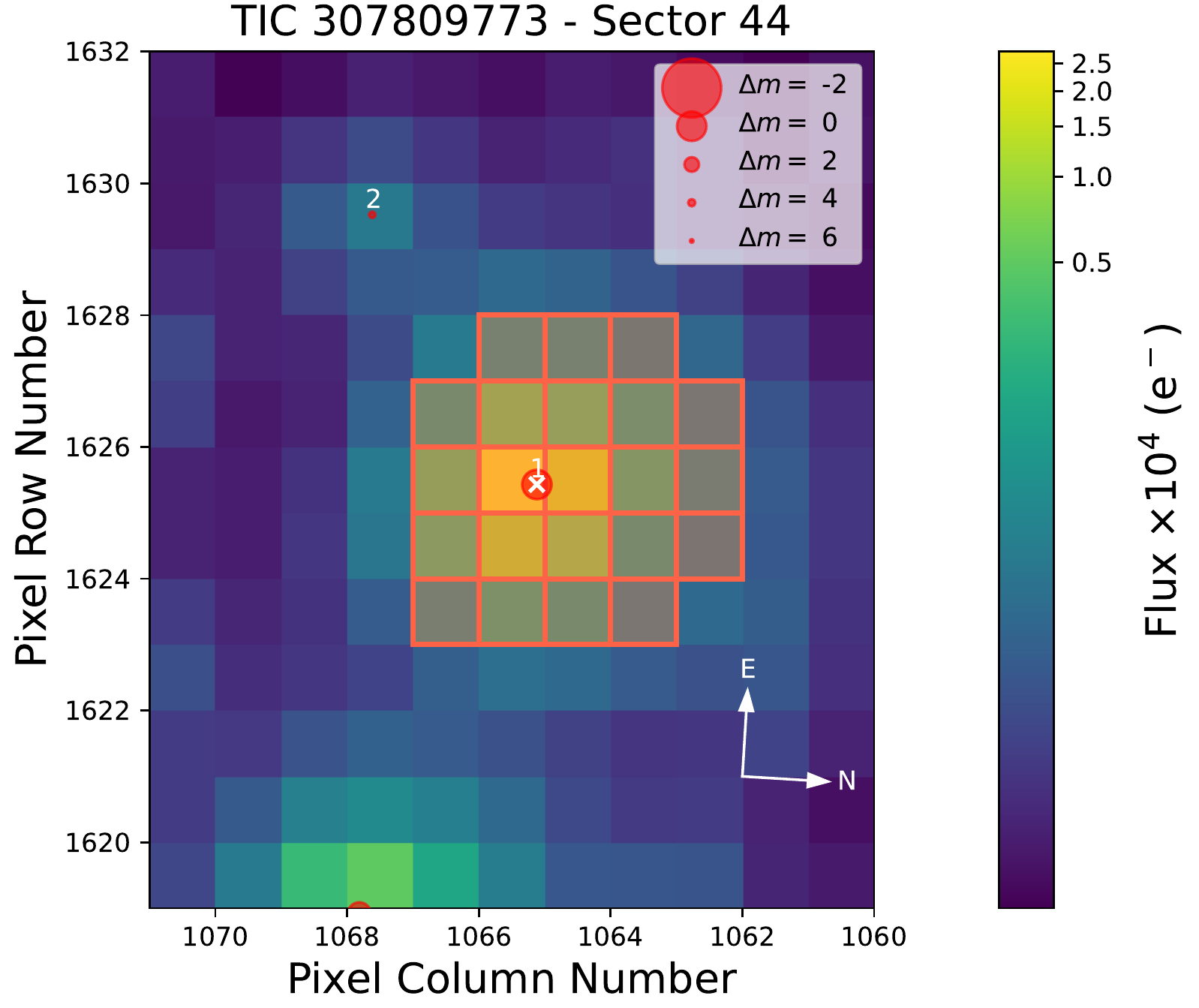}
    \includegraphics[width=0.33\hsize]{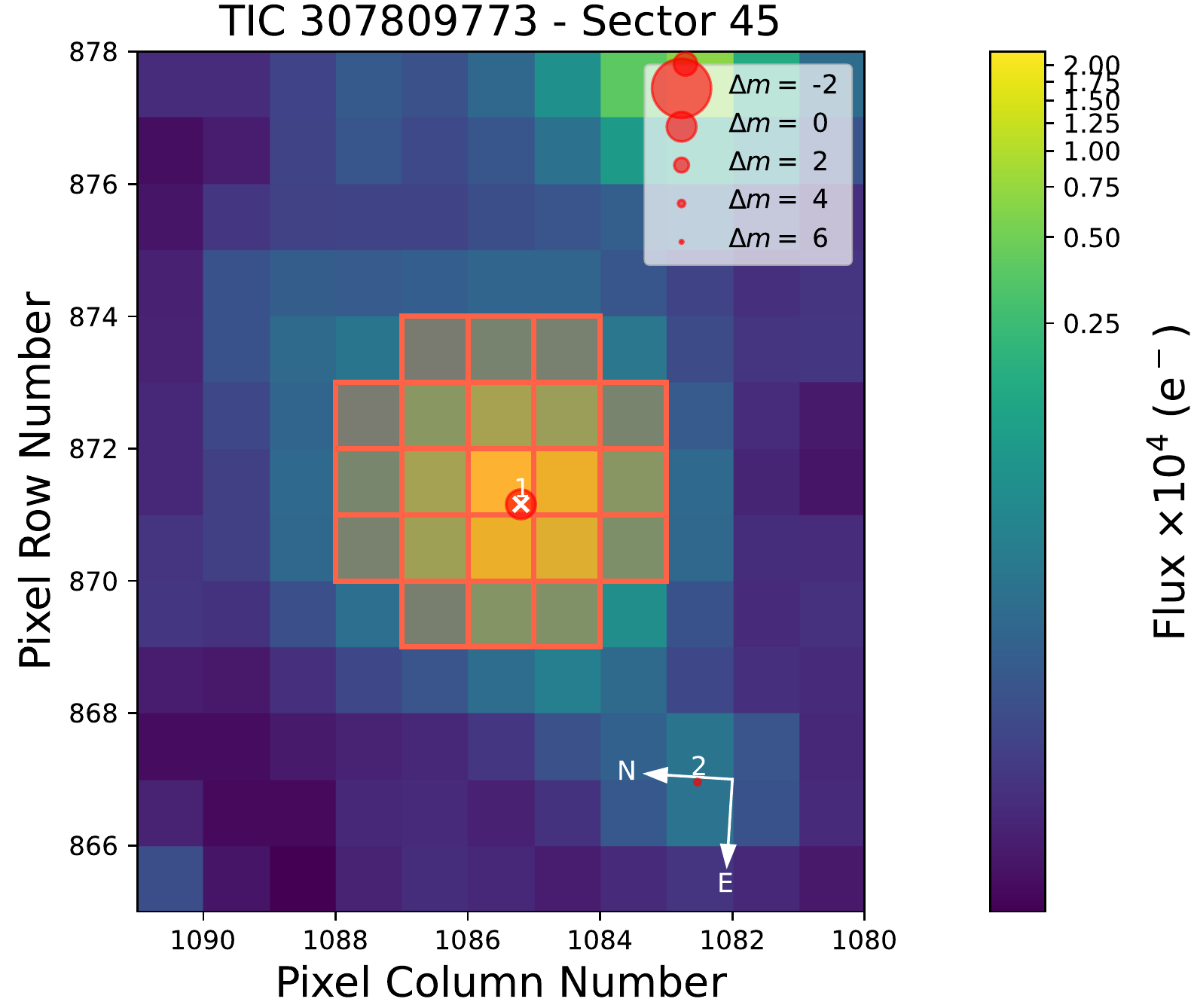}
    \caption{\textit{TESS} target pixel file image of HD~260655 in Sectors~43 (left), 44 (center), and 45 (right) created with \texttt{tpfplotter} \citep{tpfplotter}. The electron counts are color-coded. The bordered pixels are used in the simple aperture photometry. The size of the red circles indicates the \textit{TESS} magnitudes of all nearby stars and host{} (label \#1 with the "$\times$"). The \textit{TESS} pixel scale is approximately 21\arcsec. }
    \label{fig:tpfgaia}
\end{figure*}

\begin{figure*}[ht!]
    \centering
    \includegraphics[width=\hsize]{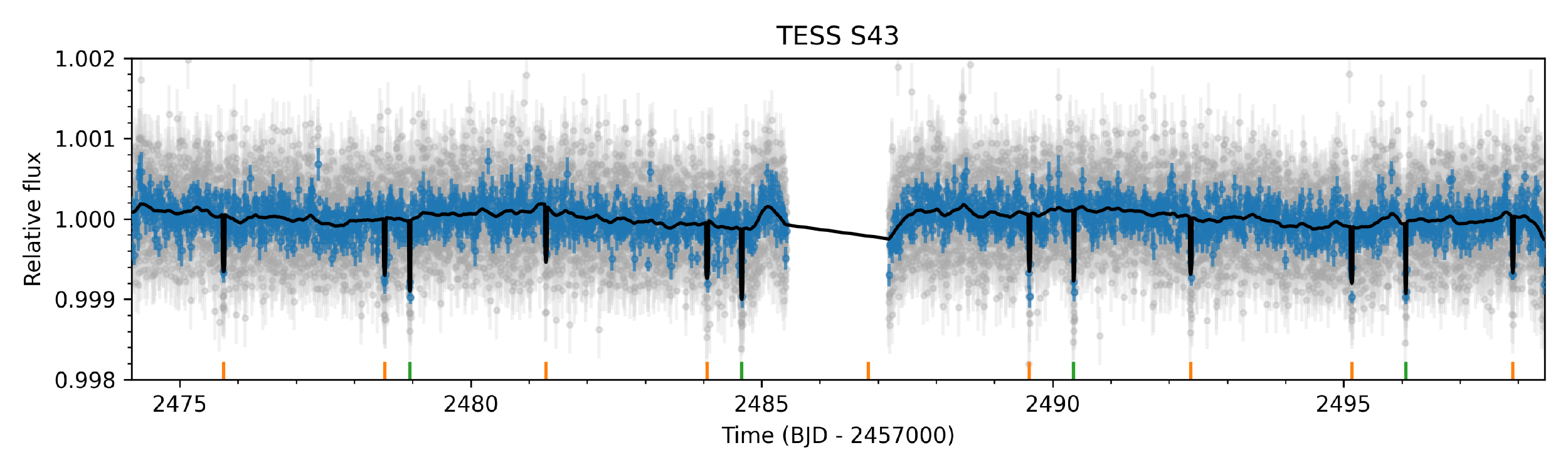}
    \includegraphics[width=\hsize]{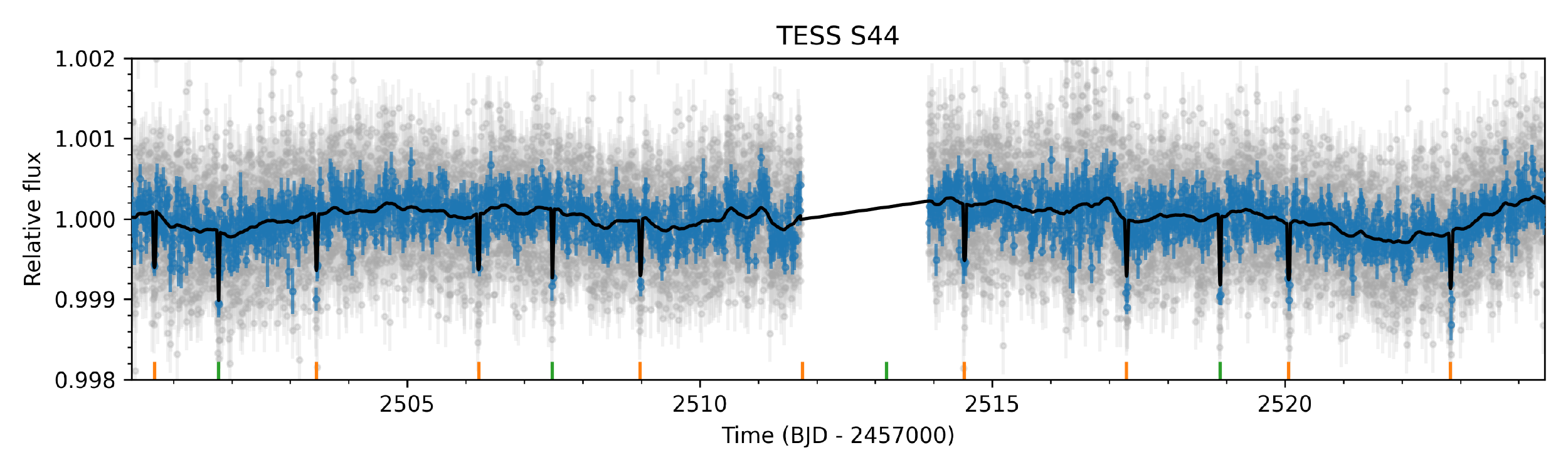}
    \includegraphics[width=\hsize]{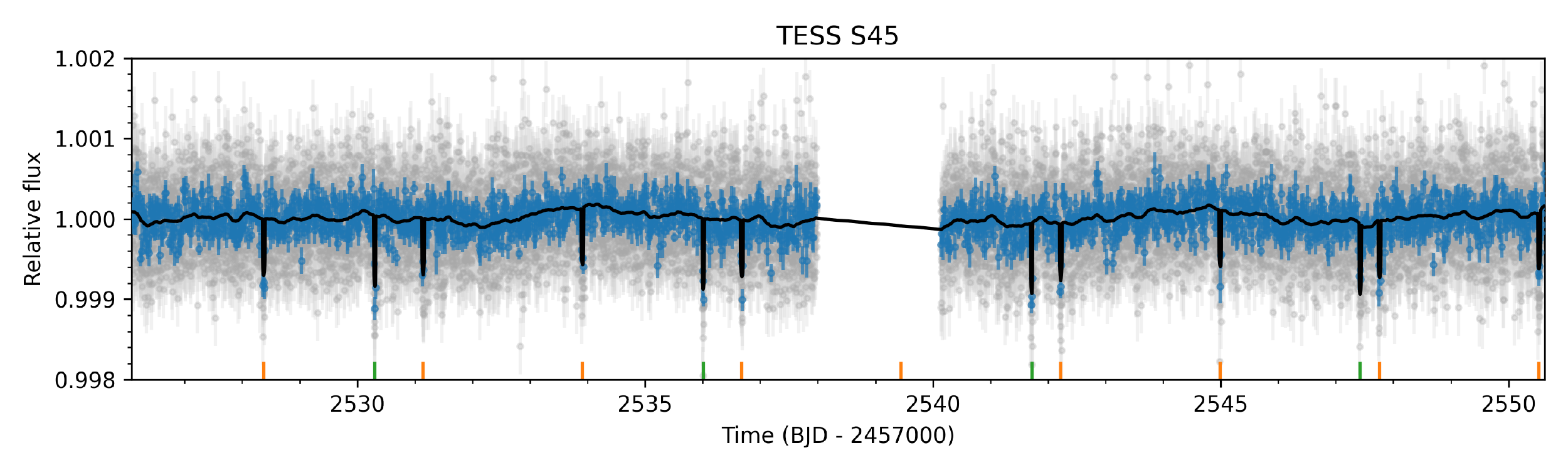}
    \caption{PDC-corrected photometry of HD~260655 from {\it TESS}. Grey points are the original 2-min cadence \textit{TESS} data, while blue shows 15-min binned photometric data. Transits of the planet candidates TOI-4599.01 and TOI-4599.02 are marked in orange and green, respectively. The black line shows the best-fit model to the data (see Sect.~\ref{sec:fit} for details)}
    \label{fig:tess_lc}
\end{figure*}

The Transiting Exoplanet Survey Satellite \citep[\textit{TESS;}][]{Ricker2015} was launched in April 2018 with a goal of discovering transiting exoplanets around nearby, bright stars. The four cameras that constitute \textit{TESS} observe $24\degree\times96\degree$ strips of the sky known as sectors for $\sim$27.4 days each. Observations of its entire field of view (FOV) are recorded as full-frame images (FFIs), with data sampled every 30 minutes (in the \textit{TESS} Prime Mission, spanning Sectors 1 -- 26) or 10 minutes (in the current Extended Mission, from Sector 27 to Sector 55). 

\textit{TESS} observed HD~260655 (TIC 307809773) during Cycle 4 in Sectors 43 (Camera 4, CCD 3), 44 (Camera 3, CCD 2), and 45 (Camera 1, CCD 1), covering 77 days (September 16 to December 2, 2021) almost continuously. The 10-minute FFIs were calibrated using the \textit{TESS} Image CAlibrator \citep[TICA;][]{TICA}, and further processed by the Quick-Look Pipeline \citep[QLP;][]{QLP1,QLP2} at MIT. The star was also observed at two-minute cadence as part of the approved Guest Investigator programs G04211 (PI: Marocco) and G04191 (PI: Burt) in all three sectors. The data were reduced by the Science Processing Operations Center \citep[SPOC;][]{SPOC} at the NASA Ames Research Center. Both SPOC and QLP independently extracted light curves, performed a transit search, and produced vetting reports for this target, which were then delivered to the \textit{TESS} Objects of Interest (TOIs) vetting team for further inspection.


Two transiting planet candidates around HD 260655 were alerted from Sector 43 as TOIs in the \textit{TESS} data alerts public website\footnote{\url{https://tev.mit.edu/data/}} on November 1, 2021. The candidates were announced in the \textit{TESS} Alerts Data Validation Report \citep[DVR;][]{Twicken2018PASP..130f4502T, Li2019PASP..131b4506L} with the following properties: TOI-4599.01 was alerted as a planet candidate with a period of 2.77\,d and a transit depth of $675\pm37$\, parts-per-million (ppm), corresponding to a planet radius of $1.44\pm0.09\,R_\oplus$, while TOI-4599.02 was announced as a planet candidate with a period of 5.70\,d and a transit depth of $931.6\pm58.8$\,ppm, corresponding to a planet radius of $1.74\pm0.14\,R_\oplus$. The two transiting signals were detected by both SPOC and QLP, passing all tests in the DVR, such as even-odd transits comparison, eclipsing binary discrimination tests, tests against effects associated with scattered light, and tests against background eclipsing binaries. The signals were re-detected and passed all tests in Sectors 44 and 45. The multi-sector search of Sectors 42--46 by SPOC with a noise-compensating, adaptive matched filter \citep{Jenkins2020} recovered both planetary signatures and passed all the diagnostic tests, including the difference image centroid analysis test, which located the transit sources on target. 

For our analyses, we use the light curves produced by SPOC, which were uploaded to the Mikulski Archive for Space Telescopes (MAST)\footnote{\url{https://mast.stsci.edu}}. SPOC computed simple aperture photometry (SAP) and systematics-corrected photometry (PDC; Pre-Search Data Conditioning) for this target. Figure~\ref{fig:tpfgaia} shows the \textit{TESS} pixels included in the computation of the SAP and PDC-corrected data. For the remainder of this work, we use the PDC-corrected \textit{TESS} photometry, which has been processed with an adaptation of the Kepler Presearch Data Conditioning algorithm \citep[PDC,][]{Smith2012PASP..124.1000S, Stumpe2012PASP..124..985S, Stumpe2014PASP..126..100S}. Data points with quality flag equal zero were considered reliable. The light curves for each Sector are shown in Fig.~\ref{fig:tess_lc}.

\section{Ground-based observations} \label{sec:obs}




\subsection{Seeing-limited photometric monitoring} \label{subsec:phot_data}

We compiled ground-based, long-timeline photometric series of HD~260655 with the goals of determining its rotation period and of discriminating between signals induced by rotation from those induced by the presence of planets. The telescope location, instrument configurations, and photometric bands of each survey are the following. A summary of the main properties of the different data sets is presented in Table~\ref{tab:phot_data}.

\paragraph{ASAS.}
The All-Sky Automated Survey \citep[ASAS;][]{Pojmanski2002AcA....52..397P} is a Polish project devoted to constant photometric monitoring of all stars brighter than $V ~\sim 14\,\mathrm{mag}$. It consists of two observing stations at Las Campanas Observatory, Chile, and Haleakala Observatory, Hawai'i. \citet{DiezAlonso2019A&A...621A.126D} analyzed $V$ band data of HD~260655 taken in Las Campanas Observatory between December 2002 and November 2009. The authors were not able to derive a reliable rotational period, but they indicate that the star shows a significant signal (false alarm probability around 2\%) with a period between 750 and 1100\,d.

\paragraph{SuperWASP.}
HD~260655 was observed by SuperWASP, the transit-search camera array located at the Observatorio Roque de los Muchachos on La Palma. Each of the 8 cameras in the array consist of a 200-mm, f/1.8 Canon lens backed by a $\text{2k}\times\text{2k}$ CCD, observing with a broad, white-light filter, and rastering fields with a typical 15-min cadence \citep{2006PASP..118.1407P}. Observations in the 2009-10 season spanned 132 days, accumulating 1566 photometric data points; observations in the 2010-11 season then spanned 123 days, accumulating a further 3024 data points.  

\paragraph{T8 APT.}
HD~260655 has been monitored since 2009 with the Tennessee State University T8 0.8-m Automatic Photoelectric Telescope (APT) located at Fairborn Observatory in southern Arizona \citep{APT}. The telescope is equipped with a two-channel precision photometer that uses a dichroic mirror and two EMI 9124QB bi-alkali photomultiplier tubes to measure the Str\"omgren $b$ and $y$ passbands simultaneously.  The observations cover 9 seasons, from 2009 until 2018, with 620 total observations and 50--100 measurements per season. The differential magnitudes of HD~260655 are determined with respect to the two comparison stars HD~46781 (F5\,V, $J=5.82\,\mathrm{mag}$) and HD~45506 (G5\,V, $J=4.67\,\mathrm{mag}$). The final magnitudes are the average of the differential magnitudes between each pair of stars and are also averaged in the Str\"omgren $b$ and $y$ filters. This averaging of the two comparison stars and the two passbands improves the precision of the data, as outlined in \citet{APT} and \citet{FekelHenry2005AJ....129.1669F}.

\paragraph{ASAS-SN.}
We also used public data from the All-Sky Automated Survey for Supernovae \citep[ASAS-SN;][]{Kochanek2017PASP..129j4502K} in the $g’$ and $V$ bands collected between January 2012 and November 2021. As in \citet{GJ486}, we retrieved the calculated real-time magnitudes using aperture photometry centered on the expected equatorial coordinates of HD~260655 at the middle of every observing season. Thanks to this retrieval, we took into account the large total proper motion of the star, of about 840\,mas\,yr$^{-1}$. The ASAS-SN $V$- and $g’$-band magnitudes are zero-point calibrated with the American Association of Variable Star Observers Photometric All Sky Survey APASS catalogue \citep{Henden2012JAVSO..40..430H}.

\paragraph{e-EYE.}
We collected new observations between November 2021 and March 2022 using the 40-cm ODK Corrected-Dall-Kirkham telescope hosted at the e-EYE\footnote{\url{https://www.e-eye.es/en/hosting/}} observatory in Fregenal de la Sierra, Spain. The telescope is equipped with a QHY 16803 monochrome CCD camera and $B$, $V$, and $R$ Astrodon filters. Differential photometry is carried out with the \texttt{LesvePhotometry} software.

\begin{table}
\caption{Summary of the main properties of the different ground-based photometric data sets. }
\label{tab:phot_data}
\centering
\begin{tabular}{lccc}
\hline\hline
\noalign{\smallskip}
 Survey     & Time span (d)  & $\sigma$ (mag)    & $N_{\mathrm{obs}}$ \\
\noalign{\smallskip}
\hline
\noalign{\smallskip}
 ASAS           & 2544          & 0.040             & 565 \\ 
 SuperWASP      & 486           & 0.025             & 4590 \\
 T8 APT         & 3126          & 0.006             & 620 \\
 ASAS-SN ($g'$) & 833           & 0.036             & 1191 \\ 
 e-EYE ($B$)    & 120           & 0.015             & 72\\ 
 e-EYE ($V$)    & 120           & 0.009             & 72\\ 
 e-EYE ($R$)    & 120           & 0.010             & 71\\ 
\noalign{\smallskip}
\hline     
\end{tabular}
\end{table}

\subsection{High angular resolution imaging} \label{subsec:hri_data}

HD~260655 has been the subject of previous searches for close companions. \citet{Balega2007AstBu..62..339B} cataloged the star as a ``possible non-single system'' after speckle interferometry observations with the BTA 6-m telescope of the Special Astrophysical Observatory of the Russian Academy of Sciences in October 2004. The star was also observed in 2002 with the Advanced Camera for Surveys installed on the Hubble Space Telescope \citep{ACS}, but the observations were carried out in order to subtract the stellar PSF from the observations of the disk around GG~Tau, not to look for companions \citep{Krist2005AJ....130.2778K}. However, \citet{Ward-Duong2015MNRAS.449.2618W} included HD~260655 in their M-dwarfs in Multiples survey of late-K to mid-M dwarfs within 15\,pc. Observations with NAOS-CONICA on the Very Large Telescope \citep{NACO} were able to rule out the presence of close companions 6\,mag fainter at separations 1\arcsec to 5\arcsec from the host. 

As part of our standard process for validating transiting exoplanets to assess the possible contamination of bound or unbound companions and its impact on the derived planetary radii \citep{ciardi2015}, we observed HD~260655 with a combination of high-resolution instruments including the near-infrared adaptive optics (AO) imager PHARO at Palomar Observatory and the speckle camera NESSI on WIYN. \textit{Gaia} EDR3 is also used to provide additional constraints on the presence of undetected stellar companions as well as wide companions. Our new observations constitute the deepest search for companions of this star to date.

\begin{figure}
    \centering
    \includegraphics[width=\hsize]{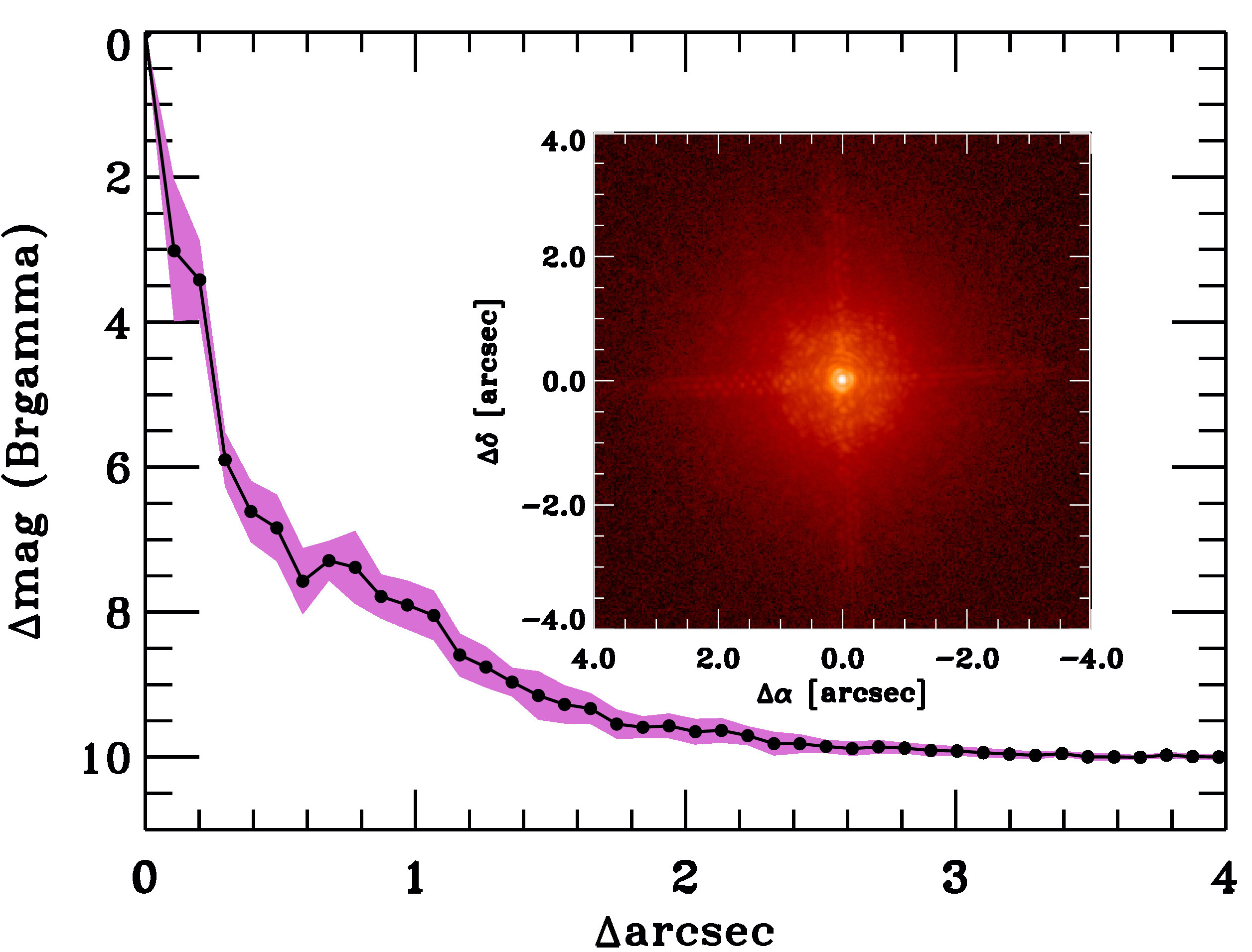} \\
    \includegraphics[width=\hsize]{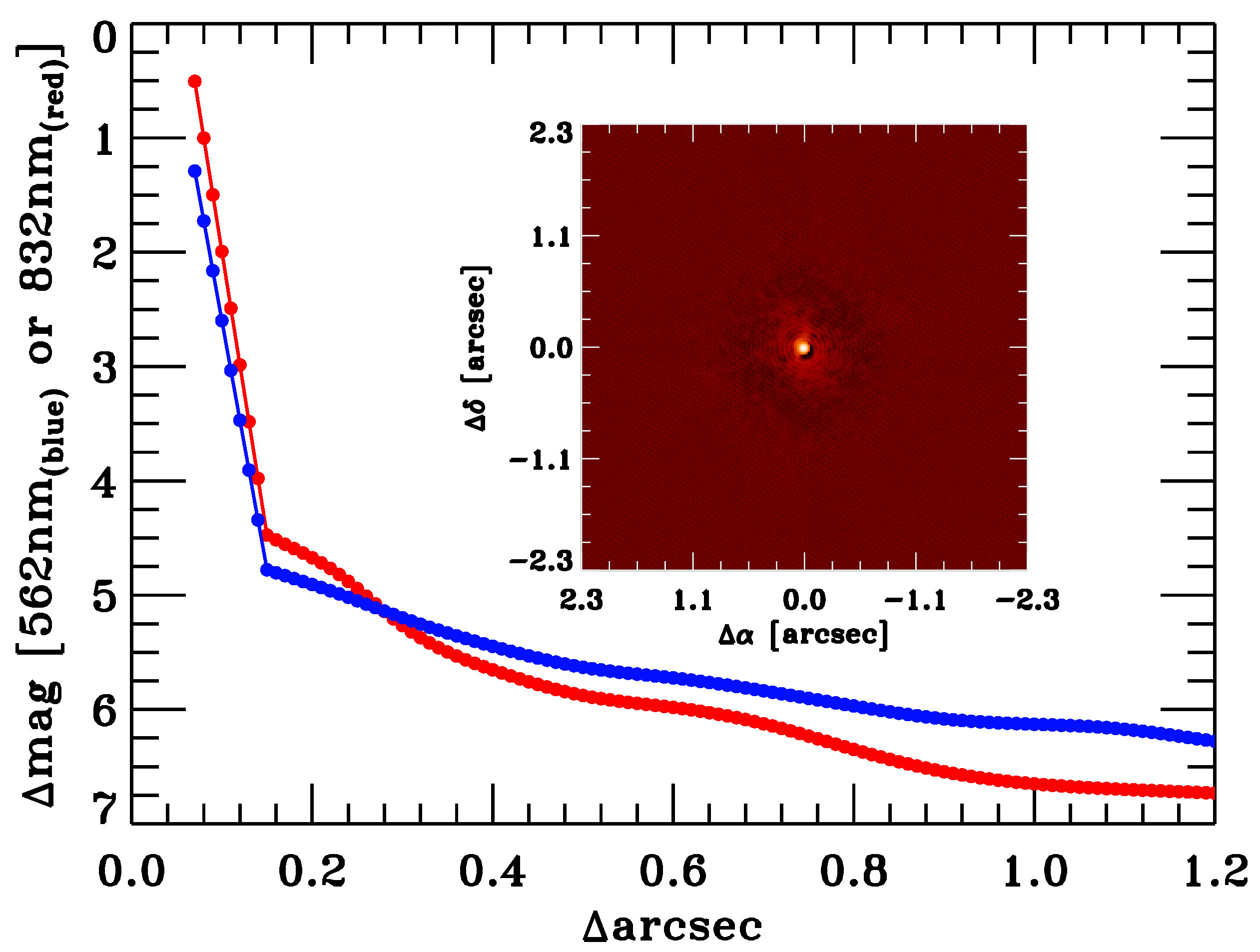}
    \caption{{\textit Top}: Palomar NIR AO imaging and sensitivity curves for HD~260655 taken in the Br$\gamma$ filter. The images were taken in good seeing conditions, and we reach a contrast of 7 magnitudes fainter than the host star at 0\farcs5. The inset shows an image of the central portion of the data, centered on the star. \textit{Bottom}: NESSI sensitivity curves and reconstructed image of HD~260655 taken with filters centered at 562\,nm (blue) and 832\,nm (red). We reach a contrast of almost 5 magnitudes fainter than the host star at separations 0\farcs15 to 1\farcs2 from HD~260655. The inset shows an image of the central portion of the data, centered on the star.  }
    \label{fig:hri}
\end{figure}

\subsubsection{PHARO}
The Palomar Observatory observations of HD~260655 were carried out with the PHARO instrument \citep{hayward2001} behind the natural guide star AO system P3K \citep{dekany2013} on November 11, 2021, in a standard 5-point quincunx dither pattern with steps of 5\arcsec\ in the narrow-band Br$\gamma$ filter $(\lambda_o = \SI{2.1686}{\micro\metre}; \Delta\lambda = \SI{0.0326}{\micro\metre}$). Each dither position was observed three times, offset in position from each other by 0.5\arcsec\ for a total of 15 frames; with an integration time of 1.4\,s per frame, for total on-source times of 21\,s. PHARO has a pixel scale of $0.025\arcsec$ per pixel for a total FOV of $\sim25\arcsec$.
    
The science frames were flat-fielded and sky-subtracted. The flat fields were generated from a median average of dark subtracted flats taken on-sky. The flats were normalized such that the median value of the flats is unity. The sky frames were generated from the median of the 15 dithered science frames; each science image was then sky-subtracted and flat-fielded. The reduced science frames were combined into a single image using an intra-pixel interpolation that conserves flux, shifts the individual dithered frames by the appropriate fractional pixels, and median-coadds the frames. The final resolutions of the combined dithers were determined from the full-width half-maximum of the point spread functions: 0.097\arcsec. The sensitivity curve of the final combined AO image was determined by injecting simulated sources azimuthally around the primary target every $20^\circ$ at separations of integer multiples of the central source's FWHM \citep{furlan2017}. The brightness of each injected source was scaled until standard aperture photometry detected it with $5\sigma$ significance. The resulting brightness of the injected sources relative to HD~260655 set the contrast limits at that injection location. The final $5\sigma$ limit at each separation was determined from the average of all of the determined limits at that separation and the uncertainty on the limit was set by the quadratic mean dispersion of the azimuthal slices at a given radial distance. The final sensitivity curve for the Palomar data is shown in Fig.~\ref{fig:hri} (top panel); no additional stellar companions 7\,mag fainter than the host star down to 0\farcs5 were detected.

\subsubsection{NESSI}
High-resolution speckle data were obtained using the NN-EXPLORE Exoplanet and Stellar Speckle Imager \citep[NESSI;][]{Scott(2018)}, a dual-channel imager that allows simultaneous observations in two narrow-band filters using EMCCDs with plate scales of 0\farcs0182. We used the filters centered at 562\,nm ($\Delta\lambda = 44$\,nm) and 832\,nm ($\Delta\lambda = 40$\,nm). Observations were obtained at the WIYN 3.5-m telescope on January 23, 2019, under the long-term program 2017-0006 (PI: S.~Howell) as a part of an ongoing M-dwarf companion survey. The data were taken in sets of $1000\times40$\,ms exposures. Nearby point sources were observed either before or after each science target observation and were used in the image reconstruction. More details on the data reduction techniques and uncertainty calculations are provided in \citet{Horch(2011a)} and \citet{Howell(2011)}.

The correlation of speckle patterns breaks down for objects separated by more than the roughly one-arcsecond FOV over which isoplanicity can be assumed for the atmosphere. Thus, we show contrast curves in both filters for only the inner 1\farcs2 of our NESSI data in Fig.~\ref{fig:hri}, along with an inset of the reconstructed image. No companions were detected in either filter within the detection limits of NESSI, which reached a contrast of almost 5\,mag fainter than the host star at separations 0\farcs15 to 1\farcs2 from HD~260655.

\subsubsection{\textit{Gaia} EDR3}
In addition to the high-resolution imaging, we have used \textit{Gaia} to identify any wide stellar companions that may be bound members of the system. Typically, these stars are already in the \textit{TESS} Input Catalog \citep{Stassun2018AJ....156..102S} and their flux dilution to the transit has already been accounted for in the transit fits and associated derived parameters. Based upon similar parallaxes and proper motions \citep{mugrauer2020,mugrauer2021}, there are no widely separated companions to HD~260655 within 1\,deg ($\sim 36000\,\mathrm{au}$).
    
The \textit{Gaia} EDR3 astrometry provides additional information on the possibility of inner companions that may have gone undetected by either \textit{Gaia} or the high angular resolution imaging. The \textit{Gaia} Renormalized Unit Weight Error (RUWE) is a metric, similar to a reduced chi-square, where values that are $\lesssim 1.4$  indicate that the \textit{Gaia} astrometric solution is consistent with the star being single whereas RUWE values $\gtrsim 1.4$ may indicate an astrometric excess noise, possibily caused the presence of an unseen companion \citep[e.g., ][]{ziegler2020}. HD~260655 has a \textit{Gaia} EDR3 RUWE value of 1.12 indicating that there are no indications in the astrometric fits for binarity or for a spurious source.

\subsection{Precise radial velocities} \label{subsec:rv_data}

\subsubsection{HIRES}

HD~260655 is part of the search for exoplanets around bright dwarf stars carried out by the high-resolution spectrograph HIRES \citep{HIRES} for more than two decades \citep[e.g.,][]{2000ApJ...536..902V,2008PASP..120..531C}. The instrument, installed on the 10-m Keck-I telescope on Mauna Kea, Hawai'i, is a general purpose slit spectrograph that allows precise radial velocity measurements by placing an iodine absorption cell in front of the slit, reaching a precision down to $\sim 1\,\mathrm{m\,s^{-1}}$ \citep{Butler2006ApJ...646..505B}. We also collected an iodine-free observation of the star to use as a template in the RV extraction pipeline.

The raw spectra are extracted and broken up into approximately $700$, 2\,\AA-wide, chunks. We forward model the observed spectrum in each chunk using a lab-measured spectrum of the HIRES iodine cell convolved with a sum-of-Gaussians model for the instrumental point-spread function multiplied by the observed stellar template. An RV is measured for each chunk and the final RV and uncertainty are derived using a weighted mean over all chunks. After all RVs are extracted for the star, an additional night-to-night offset correction is applied by \citet{Tal-Or19}.

HD~260655 was observed 92 
times between 26 January 1998 and 18 January 2014. 
The RVs show a mean internal uncertainty of $1.4\,\mathrm{m\,s^{-1}}$ 
and a standard deviation of $4.1\,\mathrm{m\,s^{-1}}$. 

\subsubsection{CARMENES}

HD~260655 (Karmn J06371+175) was one of the 324 initial stars monitored in the CARMENES Guaranteed Time Observation program to search for exoplanets around M dwarfs started in January 2016 \citep{Reiners17}. The CARMENES instrument is a dual-channel high-resolution spectrograph installed at the 3.5\,m telescope at the Calar Alto Observatory in Spain that covers the spectral range $0.52$--$\SI{0.96}{\micro\metre}$ in the visible (VIS) and $0.96$--$\SI{1.71}{\micro\metre}$ in the near-infrared (NIR) \citep{CARMENES20}. The overall performance of CARMENES, its data reduction, and wavelength calibration were all described in \citet{Trifonov18} and \citet{Kaminski18}.

HD~260655 was observed 88 times between 8 January 2016 and 23 February 2022. Relative RV values, chromatic index (CRX), differential line width (dLW), and spectral index values were obtained using {\tt serval} \citep{SERVAL}. 
The RV measurements were corrected for barycentric motion, secular acceleration, and nightly zero points. The RVs show a mean internal uncertainty of $1.7\,\mathrm{m\,s^{-1}}$ and a standard deviation of $3.5\,\mathrm{m\,s^{-1}}$.

\section{Stellar properties} \label{sec:star}

HD~260655 is a high proper motion M0.0\,V star in the Gemini constellation located at a distance of about 10\,pc \citep{GaiaEDR3}. It is among the brightest early-type M dwarfs in the sky, with an apparent magnitude in the $J$ band of $6.7$\,mag. We took the photospheric stellar parameters of HD~260655, namely the effective temperature $T_{\rm eff}$, surface gravity $\log g$, and metallicity $\mathrm{[Fe/H]}$, from \citet{Marfil2021A&A...656A.162M}. They used a co-added high-S/N template obtained with \texttt{serval} of the CARMENES VIS and NIR spectra corrected for telluric features as in \citet{Passegger18,Passegger19}. The tabulated values, especially those of $T_{\rm eff}$, match the ones previously determined by \citet{Gaidos2014MNRAS.443.2561G,Mann2015ApJ...804...64M,Houdebine2019AJ....158...56H}, and \citet{Schweitzer2019}. The apparently small uncertainty comes from the methodology and does not take into account the ``synthetic gap'' \citep{Passegger2022A&A...658A.194P}. Realistic $T_{\rm eff}$ uncertainties in the literature point to about 50--60\,K. However, most works agree on the slightly sub-solar metallicity of the star. The luminosity was computed from integrating broadband photometry together with {\it Gaia} EDR3 parallaxes as in \citet{Cifuentes2020}. Finally, the stellar radius was obtained from the Stefan-Boltzmann law, and the mass from the empirical mass-radius relationship presented in \citet{Schweitzer2019}. Furthermore, we computed Galactocentric space velocities $UVW$ as in \cite{2001MNRAS.328...45M} and \cite{Cortes2016UCM-PhD}, which place HD~260655 in the thick disk-thin disk transition \citep{Marfil2021A&A...656A.162M}. A summary of all stellar properties, including multi-band photometry compiled by \citet{Cifuentes2020}, can be found in Table~\ref{tab:star}.


\begin{table}
\centering
\small
\caption{Stellar parameters of HD~260655.} \label{tab:star}
\begin{tabular}{lcr}
\hline\hline
\noalign{\smallskip}
Parameter                               & Value                 & Reference \\ 
\hline
\noalign{\smallskip}
\multicolumn{3}{c}{\it Name and identifiers}\\
\noalign{\smallskip}
Name                            & HD~260655                     & {\citet{1931AnHar.100...61C}}      \\
GJ                              & 239                           & {\citet{1957MiABA...8....1G}}      \\
Karmn                           & J06371+175                    & {AF15}      \\    
TOI                             & 4599                          & {\it TESS} Alerts      \\  
TIC                             & 307809773                     & {\citet{Stassun2018AJ....156..102S}}      \\  
\noalign{\smallskip}
\multicolumn{3}{c}{\it Coordinates and spectral type}\\
\noalign{\smallskip}
$\alpha$                                & 06:37:10.80       & {\it Gaia} EDR3     \\
$\delta$                                & +17:33:53.3       & {\it Gaia} EDR3     \\
Epoch                                   & 2016.0            & {\it Gaia} EDR3     \\
Spectral type           & M0.0\,V           & {\citet{Lepine2013AJ....145..102L}}  \\
\noalign{\smallskip}
\multicolumn{3}{c}{\it Magnitudes}\\
\noalign{\smallskip}
$B$ [mag]                               & $11.10\pm0.07$~        & UCAC4       \\
$g$ [mag]                               & $10.52\pm0.12$~        & UCAC4       \\
$V$ [mag]                               &  $9.77\pm0.11$        & UCAC4       \\
$r'$ [mag]                              &  $9.38\pm0.05$        & UCAC4       \\
$G$ [mag]                               &  $8.878\pm0.003$      & {\it Gaia} EDR3   \\
$i'$ [mag]                              &  $8.32\pm0.04$        & UCAC4       \\
$T$ [mag]                               &  $7.899\pm0.008$      & TIC       \\
$J$ [mag]                               &  $6.674\pm0.024$      & 2MASS       \\
$H$ [mag]                               &  $6.031\pm0.016$      & 2MASS       \\
$K_s$ [mag]                             &  $5.862\pm0.024$      & 2MASS       \\
$W1$ [mag]                              &  $5.724\pm0.124$      & AllWISE       \\
$W2$ [mag]                              &  $5.524\pm0.061$      & AllWISE       \\
$W3$ [mag]                              &  $5.596\pm0.022$      & AllWISE       \\
$W4$ [mag]                              &  $5.491\pm0.043$      & AllWISE       \\
\noalign{\smallskip}
\multicolumn{3}{c}{\it Parallax and kinematics}\\
\noalign{\smallskip}
$\pi$ [mas]                             & $100.023\pm0.021$     & {\it Gaia} EDR3             \\
$d$ [pc]                                & $9.9976\pm0.0020$     & {\it Gaia} EDR3             \\
$\mu_{\alpha}\cos\delta$ [$\mathrm{mas\,yr^{-1}}$]  & $-764.41 \pm 0.02$    & {\it Gaia} EDR3          \\
$\mu_{\delta}$ [$\mathrm{mas\,yr^{-1}}$]            & $+337.88 \pm 0.02$    & {\it Gaia} EDR3          \\
$\gamma$ [$\mathrm{km\,s^{-1}}]$           & --58.75$\pm$0.02       & \citet{Lafarga2020}    \\
$U$ [$\mathrm{km\,s^{-1}}]$             &  +45.61$\pm$0.14       & This work      \\
$V$ [$\mathrm{km\,s^{-1}}]$             &  +44.67$\pm$0.04       & This work      \\
$W$ [$\mathrm{km\,s^{-1}}]$             & --29.90$\pm$0.01       & This work      \\
\noalign{\smallskip}
\multicolumn{3}{c}{\it Photospheric parameters}\\
\noalign{\smallskip}
$T_{\mathrm{eff}}$ [K]                      & $3803 \pm 10$         & {\citet{Marfil2021A&A...656A.162M}}   \\
$\log g$                                    & $5.20 \pm 0.07$       & {\citet{Marfil2021A&A...656A.162M}}   \\
{[Fe/H]}                                    & $-0.43 \pm 0.04$      & {\citet{Marfil2021A&A...656A.162M}}   \\
$v \sin i_\star$ [$\mathrm{km\,s^{-1}}$]    & $<2.0$                & \citet{Reiners17}             \\
\noalign{\smallskip}
\multicolumn{3}{c}{\it Physical parameters}\\
\noalign{\smallskip}
$M$ [$M_{\odot}$]                       & $0.439 \pm 0.011$     & This work       \\
$R$ [$R_{\odot}$]                       & $0.439 \pm 0.003$     & This work       \\
$L$ [$10^{-4}\,L_\odot$]                & $363.1 \pm 1.8$       & This work       \\
$B$ [G]                                 & $< 180$               & \citet{Reiners2022arXiv220400342R}       \\
$L_X$ [$10^{27}\,\mathrm{erg\,s^{-1}}$] & $2.0 \pm 0.7$         & \citet{vog00}       \\
$\mathrm{pEW'(H\alpha)}$ [\AA]          & $-0.067 \pm 0.021$    & \citet{Schoefer19}       \\
$\log{R_{HK}^\prime}$                         & $-4.84 \pm 0.13$      & {\citet{Perdelwitz2021A&A...652A.116P}}       \\
$P_\mathrm{rot}$ [d]                    & $37.5 \pm 0.4$        & This work       \\
Age [Gyr]                               & 2--8                  & This work       \\
\noalign{\smallskip}
\hline
\end{tabular}
\tablebib{
    AF15: \citet{2015A&A...577A.128A};
    {\it Gaia} EDR3: \citet{GaiaEDR3};
    UCAC4: \citet{UCAC4};
    TIC: \citet{TICv8p2};
    2MASS: \citet{2MASS};
    AllWISE: \citet{AllWISE}.
}
\end{table}

\subsection{Stellar activity} \label{subsec:activity}


The stellar activity indices are calculated using the high-resolution CARMENES and HIRES spectra. The overall activity level of the star from the normalized pseudo-equivalent width of the H$\alpha$ line, with a value of $\mathrm{pEW'(H\alpha)}=-0.067 \pm 0.021$\,\AA~ \citep{Schoefer19}, shows that HD~260655 is an H$\alpha$ inactive star. This result is in agreement with the reported value of the directly observed H$\alpha$ equivalent width, rather than the spectral subtraction method of \citet{Schoefer19}, $\mathrm{pEW(H\alpha)}=+0.4$\,\AA~~by \citet{Jeffers18} using CAFE spectra and, independently, by \citet{Fuhrmeister2019A&A...632A..24F} using CARMENES spectra. Similarly, \citet{Newton2017} reported an H$\alpha$ luminosity relative to the stellar bolometric luminosity of $\log{L_\mathrm{H\alpha}/L_\mathrm{bol}} = -5.2$, consistent with low levels of stellar activity.

\citet{Kiman2021} published a synthesis of various empirical relations for M~dwarfs, including relations for age as a function of H$\alpha$ equivalent width and $L_\mathrm{H\alpha}/L_\mathrm{bol}$. For an M0-type star, the equivalent width measurement is not too informative, indicating a lower limit to the age of 300\,Myr. However, the $L_\mathrm{H\alpha}/L_\mathrm{bol}$ relation is somewhat more informative, implying a lower limit of $\sim 3$\,Gyr. Moreover, we used the $L_\mathrm{H\alpha}/L_\mathrm{bol}$ and the spectral type to infer the star's likely rotation via the \citet{Newton2017} relations, which gives a predicted range of Rossby numbers of 0.8--3.0, with a most likely value of $\sim$1, giving an estimated age of $3_{-1}^{+5}$\,Gyr. 

HD~260655 was detected in X-rays (5--100\,\AA) with {\em ROSAT}/PSPC \citep{vog00}, with a $\mathrm{S/N}=2.8$ and an X-ray luminosity $L_{\rm X}=2 \cdot 10^{27}$\,erg\,s$^{-1}$. The ratio $L_{\rm X}/L_{\rm bol}=-4.8$ is indicative of a rather low activity level \citep{wri11}, consistent with a mature age of 2--5\,Gyr \citep{san11}.

The $\log{R_{HK}^\prime}$ values for HD~260655 have previously been reported by \cite{BoroSaikia2018} and \citet{Perdelwitz2021A&A...652A.116P}.  Both of these works report values of $\log{R_{HK}^\prime}$ that are consistent with HD~260655 being a low activity or inactive star. \citet{SuarezMascareno2015} and \citet{AstudilloDefru2017}  investigated the relation between  $\log(R_{HK}^\prime)$ and stellar rotation period.    For a $\log(R_{HK}^\prime)=-4.84$, both of these works estimate a rotation period 
of approximately 30 days.

\subsection{Periodicities in activity indicators} \label{subsec:period_activity}

\begin{figure}
    \centering
    \includegraphics[width=\hsize]{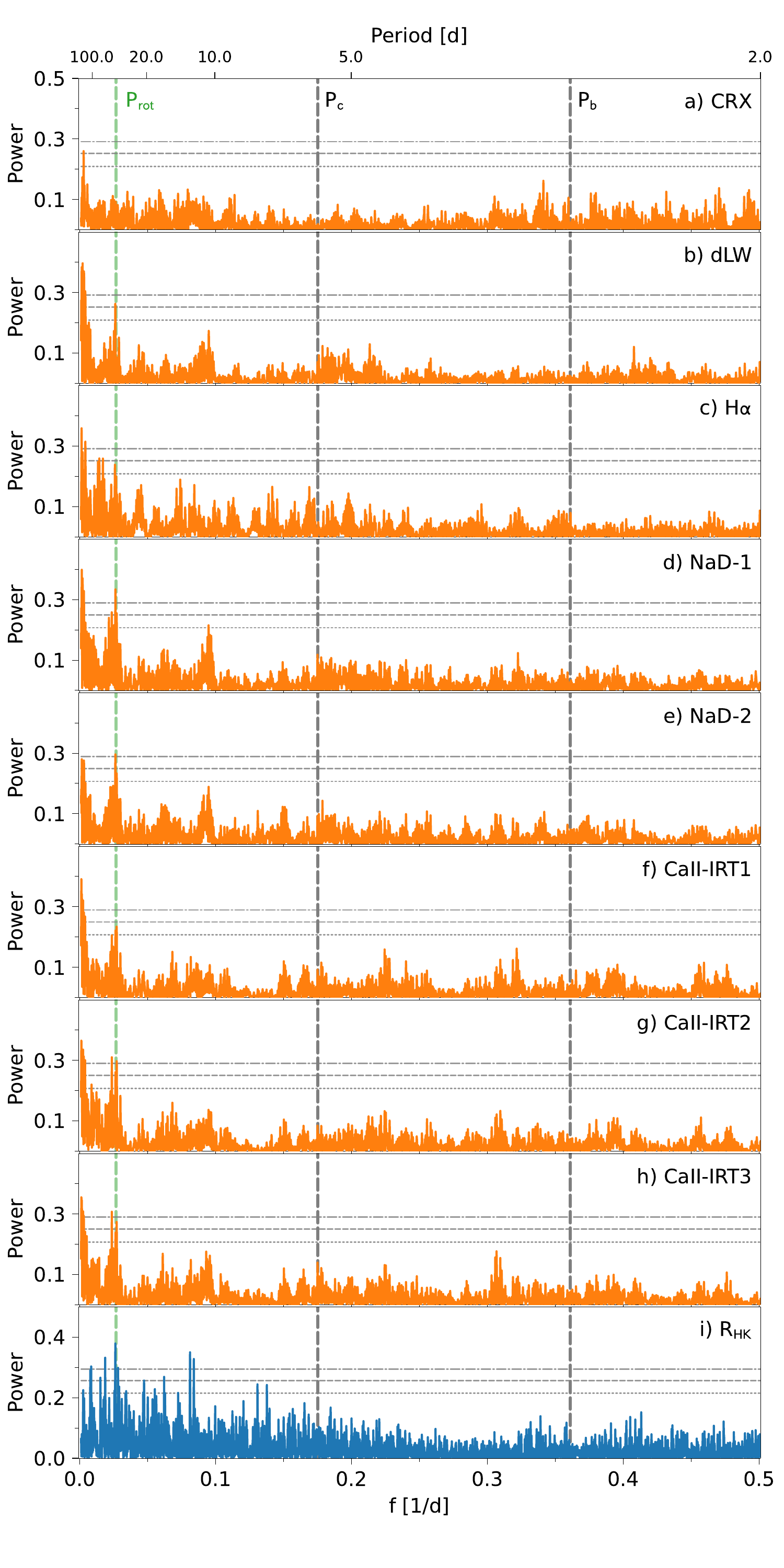}
    \caption{GLS periodograms of the spectral activity indicators from HIRES (blue) and CARMENES (orange) data. For each panel, the horizontal lines show the theoretical 10\% (short-dashed line), 1\% (long-dashed line), and 0.1\% (dot-dashed line) FAP levels. The vertical dashed lines mark the orbital frequencies of the transiting planets ($f_\mathrm{b}=0.361\,{\rm d}^{-1}$, $f_\mathrm{c}=0.175\,{\rm d}^{-1}$) and of the stellar rotation signal at $0.0266\,{\rm d}^{-1}$. \emph{Panels a--h}: Chromatic index (CRX), differential line width (dLW), H$\alpha$, Na doublet, and Ca infrared triplet lines computed with \texttt{serval} from CARMENES data. \emph{Panel i}: $R_{HK}^\prime$ index computed by \citet{Perdelwitz2021A&A...652A.116P} from HIRES data.}
    \label{fig:gls_activity}
\end{figure}

We compute a GLS periodogram \citep{GLS} for every activity indicator derived from CARMENES and HIRES spectra (Fig.~\ref{fig:gls_activity}). On the one hand, we find no signals at the periods of the transiting planet candidates. On the other hand, there is a common periodicity at $0.0266\,{\rm d}^{-1}$ ($37.6\,{\rm d}$) in the dLW, Na doublet, second and third lines of the Ca IRT, and $R_{HK}^\prime$ index. There is additional power at low frequencies that are difficult to constrain because their periods are of the same order as the time span of the observations (approximately 2000\,d). 
The detection of periodicities in the chromospheric lines and the dLW is consistent with the results of \citet{Lafarga2021} for the low-activity high-mass regime of their analysis. The detected periodicity at $37.5\pm0.4$\,d, following the empirical relations from \citet{wri11} and \citet{EngleGuinan2018}, is also consistent with the X-ray emission measured for this star and an age of $4.1 \pm 0.2$\,Gyr, respectively.

Due to the evolution of magnetic activity features over the time span of the observations, signals from individual activity indices are not necessarily periodic.  We investigate if there are any statistically significant correlations of any of the activity indices with RV by using Pearson's $r$ coefficient.  A strong correlation is defined as having a value of $r > 0.7$ or $r < -0.7$ (similar to \citealt{Jeffers2020}). We did not find any strong or moderate correlations of any of the activity indices with the measured RV values.

\subsection{Stellar rotation} \label{subsec:rotation}


The stellar rotation signal can be present in both spectroscopic and photometric data. The determination of this periodicity is important in order to disentangle it from planetary signals. We checked the \textit{TESS} photometry for variations attributable to stellar activity such as spot-induced modulations or flares, but we did not find any evidence of these signals during the spacecraft observations. Measuring stellar rotational periods longer than 13\,d using \textit{TESS} has been proven difficult due to the telescope orbit \citep[e.g.,][]{Hedges2020RNAAS...4..220H,CantoMartins2020ApJS..250...20C,Claytor2022ApJ...927..219C}. Therefore, we measured the rotation period using the ground-based long-term photometric monitoring of HD~260655 and compared it to periods derived using activity indices. For this, we performed a Gaussian Process (GP) regression model using a quasi-periodic kernel from \texttt{celerite} \citep{celerite}. 


\begin{figure}
    \centering
    \includegraphics[width=\hsize]{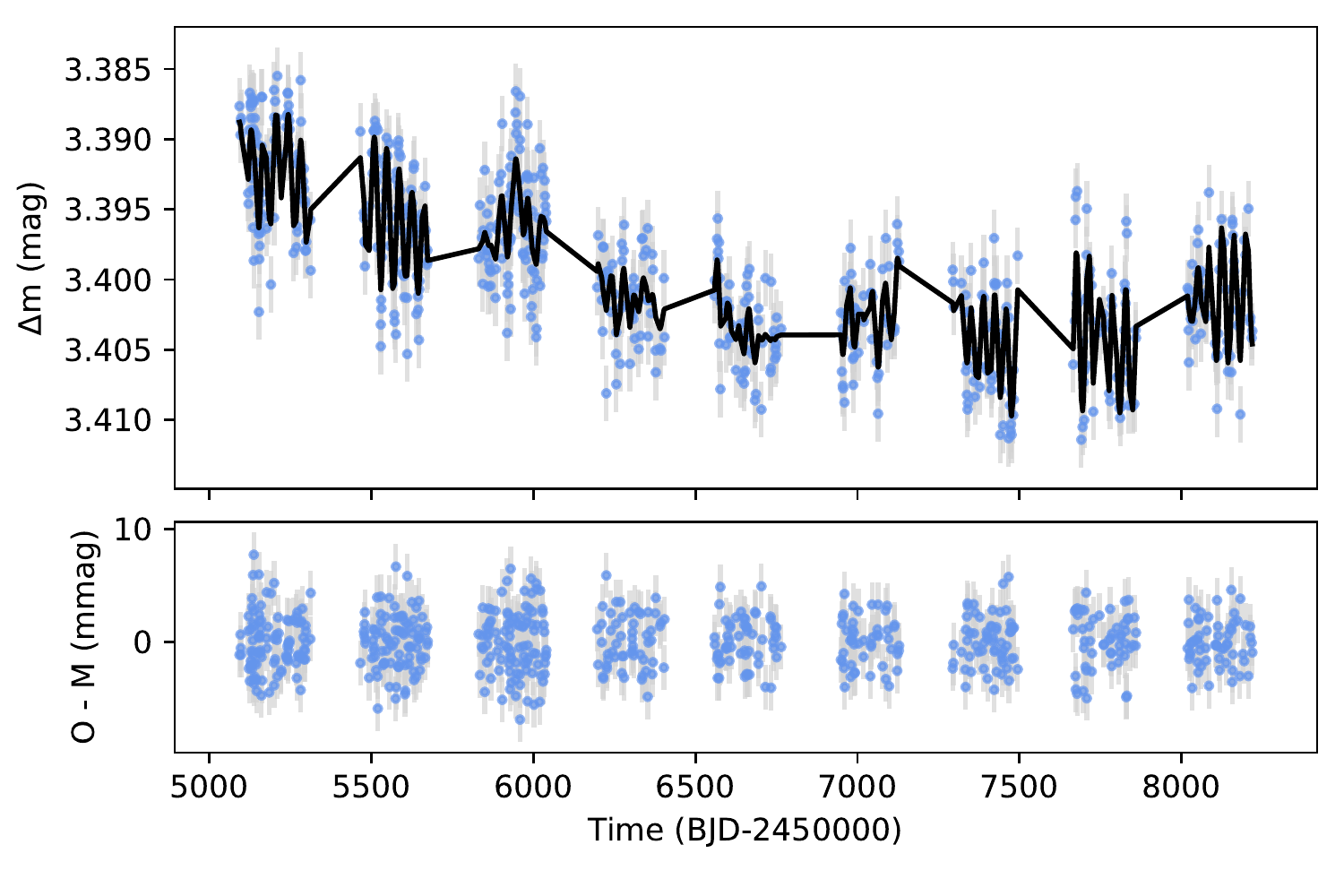}
    \caption{\textit{Top:} GP fit to the T8 APT photometric data set used to estimate the stellar rotation period of the star. The T8 APT data are shown in blue, while the model is shown in black. Our best-fit quasi-periodic GP fit reveals a rotational period of $37.5 \pm 0.4$\,d with seasonal changes in the photometric amplitude and mean magnitude. \textit{Bottom:} Residuals of the best fit model.} 
    \label{fig:stellar_rotation}
\end{figure}

\begin{table}
    \centering
    \caption{Sinusoid-modulation search of T8 APT data.} \label{tab:APT_freq}
    \begin{tabular}{cccccccc}
        \hline
        \hline
        \noalign{\smallskip}
         &      &  Mean     & $\sigma$  & Period  & Full amp.  \\ 
  Season & $N_{\rm obs}$ &  (mag)    & (mmag)    & (d)     & (mmag)       \\
        \noalign{\smallskip}
        \hline
        \noalign{\smallskip}
2009--10 &  98  &  3.3919   & 3.8      & 37.9    &  7.70    \\
2010--11 &  89  &  3.3960   & 4.2      & 38.6    &  9.17    \\
2011--12 &  98  &  3.3961   & 3.8      & 18.3    &  4.36    \\
2012--13 &  56  &  3.4013   & 3.0      & 15.2    &  4.66    \\
2013--14 &  57  &  3.4031   & 3.0      & 31.4    &  5.32    \\
2014--15 &  49  &  3.4028   & 3.2      & 37.4    &  6.10    \\
2015--16 &  60  &  3.4048   & 3.7      & 33.6    &  7.51    \\
2016--17 &  54  &  3.4037   & 4.5      & 38.3    & 10.67    \\
2017--18 &  59  &  3.4014   & 3.5      & 37.8    &  7.27    \\
        \noalign{\smallskip}
        \hline
    \end{tabular}
\end{table}

We find a robust period detection only in the T8 APT data set due to its smaller dispersion and longer baseline compared to the other surveys listed in Table \ref{tab:phot_data}. The fit to the data is shown in Fig.~\ref{fig:stellar_rotation} and a corner plot with the posterior distribution of the fitted parameters in Fig.~\ref{fig:cornerAPT} of the Appendix. We measure a rotation period of $37.5 \pm 0.4$\,d, consistent with the results from the spectral activity indicators. To confirm this result, we carried out a frequency analysis of the T8 APT data set dividing it into individual observing seasons. Table~\ref{tab:APT_freq} shows the results. Both analyses show that the periodicity of the photometric modulations remains rather constant at $\sim 38\,\mathrm{d}$ over a 9-year time span despite the changes in amplitude and mean magnitude. The measured photometric periods for the 2011-12 and 2012-13 observing seasons are approximately half the 37.5\,d period (see Table~\ref{tab:APT_freq}), indicating that the star had significant spot activity on opposite hemispheres during those two seasons.  These lightcurve changes, with a timescale of several years, are likely associated with long-term magnetic activity cycles as measured in our Sun and many other M~dwarfs \citep[e.g.,][]{Savanov2012,Robertson2013,SuarezMascareno2018,DiezAlonso2019A&A...621A.126D}. The long-term changes are in agreement with the low-frequency signals found in all the chromospheric indicators and hint toward a magnetic activity cycle of approximately 6\,yr. However, the baseline of our T8 APT observations is not sufficient to measure the magnetic cycle of HD~260655 with higher precision.

In addition, we analyzed the archival SuperWASP photometry for HD~260655, which independently confirms the rotation period measured from T8 APT data. With only 2 seasons (2009-10, 2010-11) of light curves available and relatively spotty phase coverage within each season, the data quality for this star is significantly poorer in SuperWASP than that of the T8 APT as seen in Table~\ref{tab:phot_data}. Nevertheless, Fig.~\ref{fig:swasp} shows that the top periods picked out by the GLS periodogram in each SuperWASP season are consistent with the 37.5\,d period retrieved from T8 APT. 

\section{Analysis and results} \label{sec:fit}

To model the data of HD~260655, we used \texttt{juliet} \citep{juliet}, a python library built on many publicly available tools for the modeling of transits \citep[\texttt{batman},][]{batman}, RVs \citep[\texttt{radvel},][]{radvel}, and Gaussian Processes (\texttt{george}, \citealt{Ambikasaran2015ITPAM..38..252A}; \texttt{celerite}, \citealt{celerite}). \texttt{juliet} applies nested samplers (\texttt{dynesty}, \citealt{dynesty}) to explore the parameter space of a given prior volume and to compute efficiently the Bayesian model log evidence ($\ln Z$). Thanks to this, we can compare models with different numbers of parameters accounting for the model complexity and the number of degrees of freedom with a sound statistical methodology. In our analysis, we consider that a model (M2) is greatly favored over another (M1) if $\Delta \ln Z = \ln Z_{\rm M2} - \ln Z_{M1} > 5$ \citep{2008ConPh..49...71T}. If $\Delta \ln Z < 2$, we consider that the models are statistically indistinguishable so the simpler model with less degrees of freedom would be chosen. For intermediate cases, we consider that M2 is moderately favored over M1.

\subsection{Transit photometry} \label{subsubsec:transit-fit}

First, to constrain the properties of the transiting candidates and use them for subsequent analyses, we modeled the \textit{TESS} photometry with \texttt{juliet}. We adopted a quadratic limb darkening law for \textit{TESS} parameterized by the coefficients $q_1, q_2$ introduced by \citet{Kipping13} and fitted them as free parameters with unconstrained priors. For the transiting planets, we followed the $r_1$, $r_2$ mathematical parameterization introduced by \citet{Espinoza18} to fit the planet-to-star radius ratio $p=R_p/R_*$ and the impact parameter of the orbit $b$. Additionally, rather than fitting the scaled planetary radius ($a/R_\star$) for each planet, we use the stellar density ($\rho_\star$) as a free parameter. In this way, the stellar density is shared for both planets and we reduce by one the number of free parameters in the fit \citep{Sozzetti2007}. We set a normal prior for $\rho_\star$ based on the stellar parameters from Table~\ref{tab:star}. We treated each \textit{TESS} Sector independently, adding a jitter term $\sigma$ in quadrature to the photometric uncertainties and fixing the \textit{TESS} dilution factor to $1$ for each Sector (as confirmed by our analysis in Sect.~\ref{subsec:hri_data}).

To remove the additional variability in the light curves (Fig.~\ref{fig:tess_lc}), we used the exponential GP kernel from \texttt{celerite} \citep{celerite}
\begin{equation*}
    k_{i,j} = \sigma^2_\mathrm{GP,TESS} \exp\left(- |t_i - t_j|/T_\mathrm{GP,TESS}\right),
\end{equation*}
where the characteristic timescale ($T_\mathrm{GP,TESS}$) and the amplitude of the GP modulation ($\sigma_\mathrm{GP,TESS}$) are shared hyperparameters between the three different \textit{TESS} Sectors.

For this analysis, we used wide uniform priors for the period ($P$) and mid transit time ($t_0$) of the transiting candidates based on the information reported in the \textit{TESS} DVR. We improve the uncertainties in $P$ and $t_0$ by one order of magnitude and the planet-to-star radius ratio by two orders of magnitude for both candidates with respect to the \textit{TESS} DVR. The posterior distribution of the orbital parameters from this fit are indistinguishable from the results of the final joint model, so we show only the latter results in Table~\ref{tab:posteriors} for simplicity.

\subsection{Radial velocities} \label{subsubsec:rv-fit}

\begin{figure}
    \centering
    \includegraphics[width=\hsize]{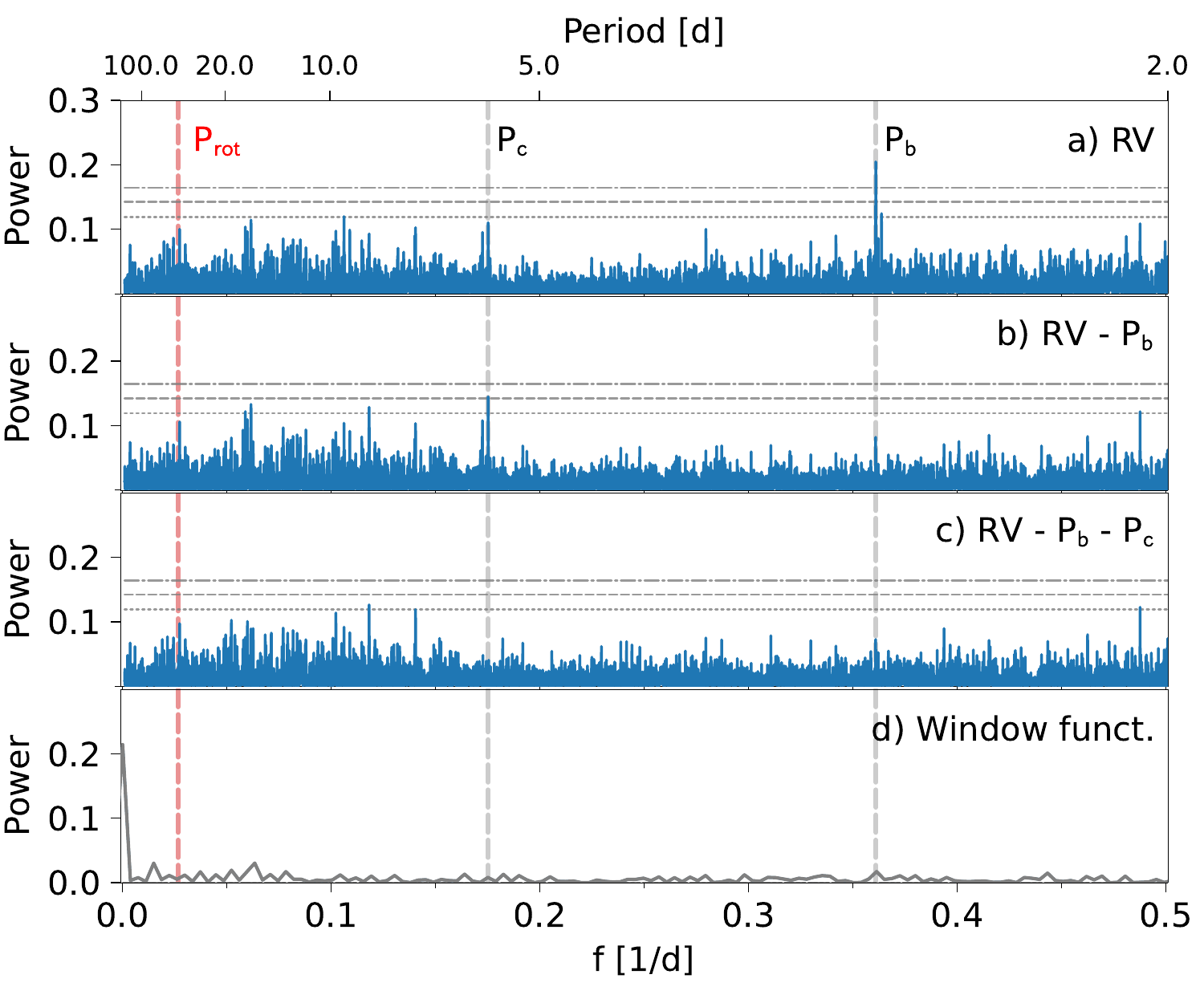}
    \caption{GLS periodograms of the CARMENES and HIRES RV data. For each panel, the horizontal and vertical lines are the same as Fig.~\ref{fig:gls_activity}. \emph{Panel a}: Combined RV data corrected for an instrumental offset. \emph{Panel b}: RV residuals following the subtraction of the signal of the transiting candidate HD~260655~b. \emph{Panel c}: RV residuals following the subtraction of the signal of both transiting candidates HD~260655~b and HD~260655~c. \emph{Panel d}: Window function.
    }
    \label{fig:gls_rvs}
\end{figure}

\begin{table}
    \centering
    \caption{Model comparison of RV-only fits with \texttt{juliet}. The GP refers to a Gaussian Process with an exponential sine-squared kernel from \texttt{george}. The model used for the final joint fit is marked in boldface.}  \label{tab:models}
    \begin{tabular}{lccc}
        \hline
        \hline
        \noalign{\smallskip}
        Model & $\Delta\ln Z$ & $K_{\rm b}$ (m\,s$^{-1}$) & $K_{\rm c}$ (m\,s$^{-1}$)  \\
        \noalign{\smallskip}
        \hline
        \noalign{\smallskip}
0p          & 0.0       & \dots & \dots    \\
0p + GP     & 4.3       & \dots & \dots    \\[0.2cm]


2p          & 15.0      & $1.85\pm0.34$ & $1.86\pm0.36$    \\
{\bf 2p + GP}     & {\bf 23.6}      & $1.69\pm0.27$ & $1.92\pm0.30$    \\[0.2cm]

3p          & 13.9      & $1.78\pm0.31$ & $1.90\pm0.29$    \\
3p + GP     & 21.5      & $1.77\pm0.24$ & $1.86\pm0.26$    \\

        \noalign{\smallskip}
        \hline
    \end{tabular}
\end{table}

We performed a detailed analysis of the RVs described in Sect.~\ref{subsec:rv_data} to constrain all the potential signals present in the data. For this analysis, we fixed the period ($P$) and the mid transit time ($t_0$) of the candidates based in our previous analysis of the \textit{TESS} data. For these two parameters, the uncertainties derived from transit models are several orders of magnitude smaller than what the RV data alone could constrain, so fixing these values does not have an impact in the RV fit while speeding up greatly the computation time. On the other hand, the eccentricity of multi-planet transiting systems is low, but not necessarily zero \citep{Xie2016,vanEylen19}. Therefore, to model the RV data we do not use circular, but Keplerian orbits with a prior on the orbital eccentricity $e$ following a beta distribution with $\alpha=1.52$ and $\beta=29$ \citep{vanEylen19}. 

First, we searched for periodic signals in the data. Figure~\ref{fig:gls_rvs}a shows a GLS periodogram of the CARMENES and HIRES RVs, corrected only for an instrumental offset to join both data sets. We find a single peak in the GLS with extremely high significance ($\mathrm{FAP} < 0.1\%$) at the orbital period of the transiting candidate HD~260655~b. The residuals after modeling this signal with a Keplerian orbit with $P$ and $t_0$ fixed from our previous analysis (Fig.~\ref{fig:gls_rvs}b) reveals a single peak with high significance ($\mathrm{FAP} < 1\%$) at the orbital period of the second transiting candidate HD~260655~c. When both signals are modeled with Keplerian orbits the residuals do not show any statistically significant periodicity (Fig.~\ref{fig:gls_rvs}c). Despite its simplicity, this analysis already shows that the RV data can confirm the presence of both transiting candidates and that they could have been detected without any prior information about their orbital period and phase.

In order to find the best fit to the RV data we tried three different sets of models: i) ``no planet'' models (0p) where the data are assumed to be consistent with a flat line or purely correlated noise modeled with a GP, ii) ``two-planet'' models (2p) that assume two planetary signals in the RV data with or without GP; and iii) ``three-planet'' models (3p) that assume the presence of three planetary signals in the RV data (all modeled as Keplerian orbits), using a wide uninformative prior on the period of the third signal absent in the transit data, with or without correlated noise. The last set of models aims to find additional signals in the RV data of non-transiting planetary origin. For our fits, we assumed an exponential sine-squared kernel of the form
\begin{equation}
    k_{i,j} = \sigma^2 \exp\left(- \alpha (t_i - t_j)^2 - \Gamma \sin^2 \left[\frac{\pi |t_i - t_j|}{P_{\rm rot}}\right]\right) \qquad .
\end{equation}
We set log-uniform priors for $\sigma$ between 0.01 and 30\,$\mathrm{m\,s^{-1}}$, $\alpha$ between $10^{-10}$ and $10^{-3}$, $\Gamma$ between 0.01 and 10, and a Gaussian prior for $P_{\rm rot}$ with a mean of 37.5\,d based on our results from the photometric determination of the stellar rotation period in Sect.~\ref{subsec:rotation}. The priors in $\alpha$ and $\Gamma$ were constrained after running a set of 0p models with wide, unconstrained hyperparameters following the approach by \citet{Stock2020a, Stock2020b} and \citet{Nava2020}.

Table~\ref{tab:models} shows the different models tested in our comparison scheme together with their Bayesian log-evidence. It is important to highlight that the derived semi-amplitude for the two transiting candidates remain the same within 1$\sigma$ for every set of models, ensuring a robust mass determination of the candidates independently of the chosen model. The set of models with the highest evidence are the 2p models. Modeling the RV data set with two Keplerian orbits whose periods and phases are set from the transit analysis implies a highly significant increase in the goodness of the fit ($\Delta \ln Z = \ln Z_{\rm 2p} - \ln Z_{\rm 0p} > 15$). Two-planet models are strongly favored over models including a GP component only (0p + GP). GP models can act as a high-pass filter that removes all the variability in the RV data, missing planetary signals that are at the level of the instrumental precision. However, we find that models accounting for the transiting candidates are favored ($\Delta \ln Z > 10$) over GP-only models, which ensures the significance of our detection and that the RV data alone would have been able to detect the transiting candidates independently without any prior information from their orbital period or phase. Considering the results from the high-resolution imaging data and the transit-only fits, we consider that the transiting planet candidates in \textit{TESS} data are bona-fide planets confirmed by independent RV measurements. 

On the other hand, 3p models are moderately disfavored compared to 2p models. The third signal in these models is assumed to be a circular orbit with a wide uniform prior in orbital period from 6 to 300\,d, to cover the range of potentially significant signals seen in Fig.~\ref{fig:gls_rvs}c. The posterior distribution in orbital period of the third signal is highly multimodal, with no preferred periods, and the RV semi-amplitude of this signal is consistent within 1$\sigma$ with zero. Therefore, our results show that there is no statistical evidence to claim additional non-transiting planets in the system with the data at hand although such systems are often detected \citep[e.g.,][]{GJ357, GJ3473, Osborn2021}. 

Finally, we find that the model with the highest evidence is a two-planet model with an additional GP to account for correlated noise. Including an informed GP kernel in the model increases significantly the goodness of the fit ($\Delta \ln Z = 8.6$) and reduces the uncertainty in the semi-amplitude of the planets. 
Therefore, considering that even small influences from stellar activity can affect the planetary parameters \citep[e.g.,][]{Stock2020a}, and that even strong activity signals may appear or not in the periodogram of the RVs \citep{Nava2020}, we consider 2p+GP as the best model to fit the RV data.

\subsection{Joint fit} \label{subsubsec:joint-fit}

\begin{figure*}
    \centering
    \includegraphics[width=0.49\hsize]{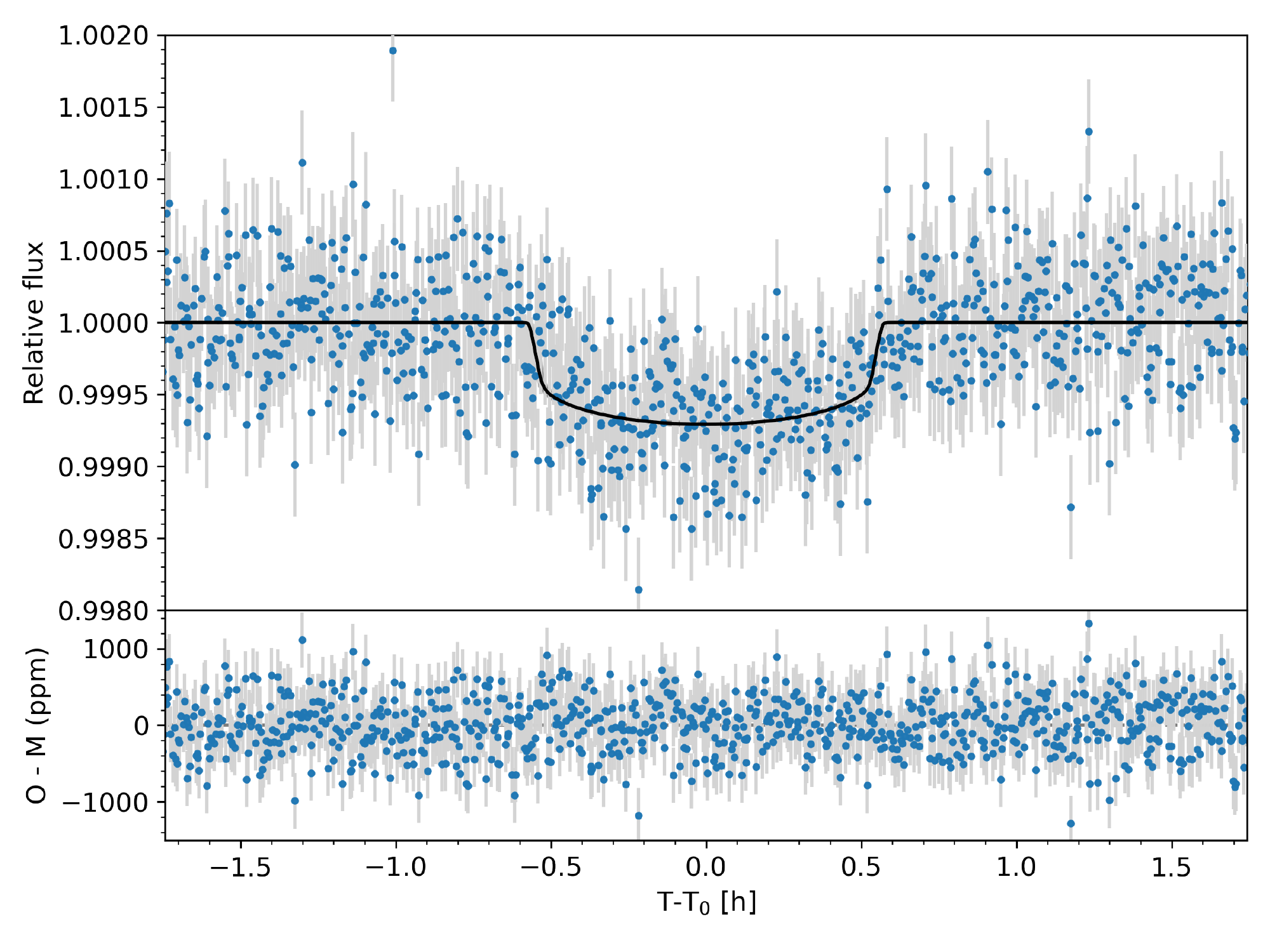}
    \includegraphics[width=0.49\hsize]{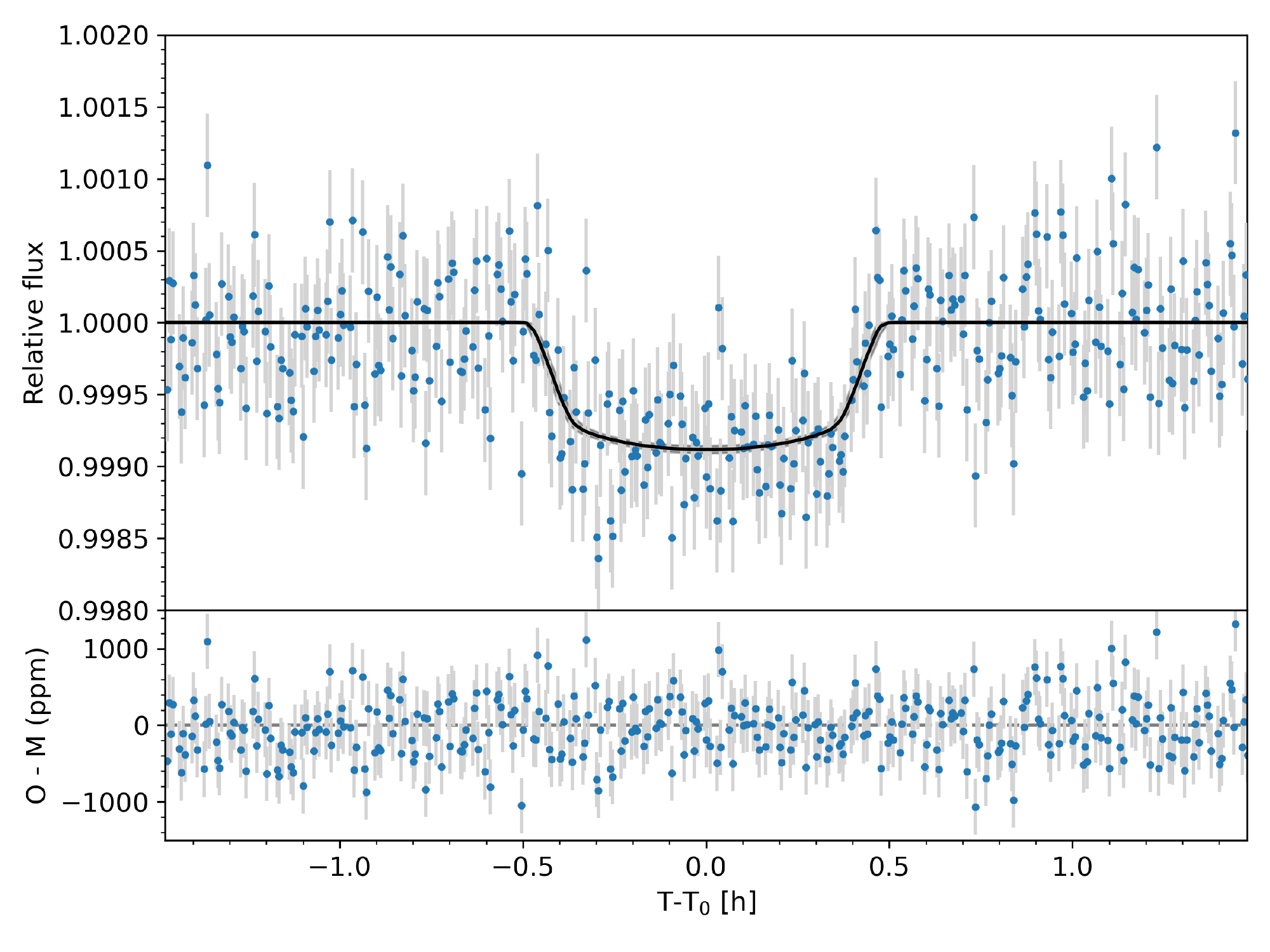}\\
    \includegraphics[width=0.49\hsize]{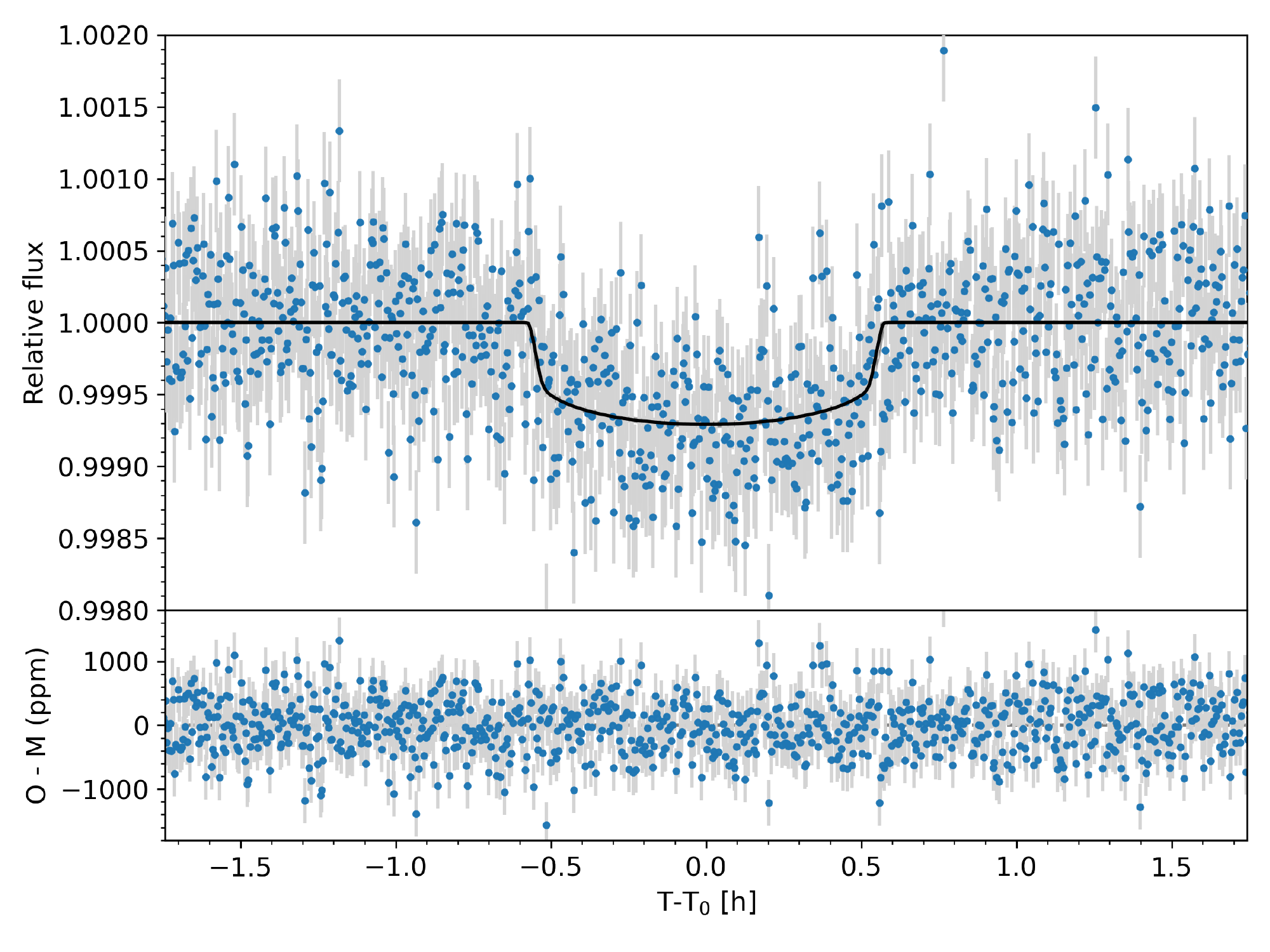}
    \includegraphics[width=0.49\hsize]{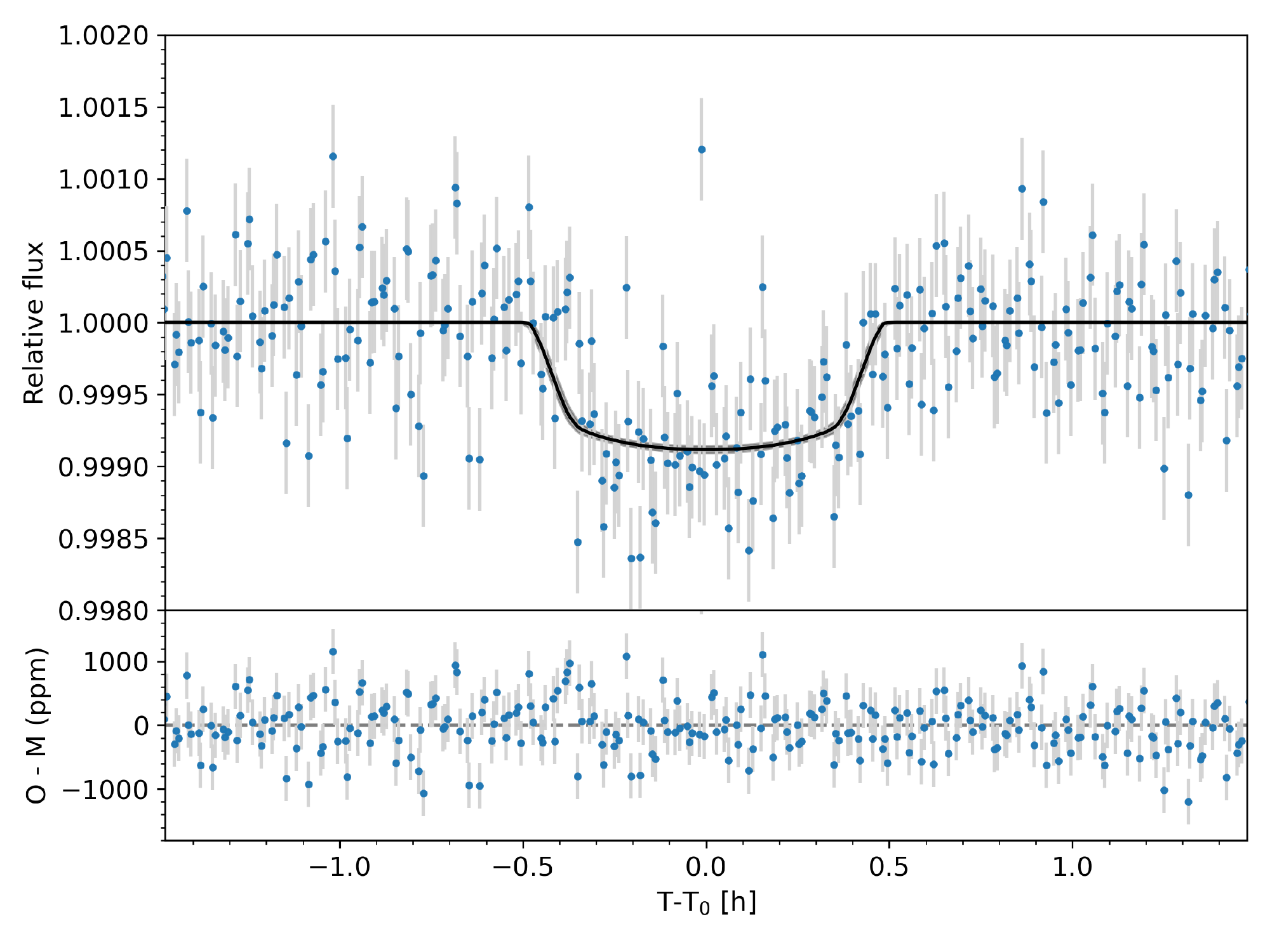}\\
    \includegraphics[width=0.49\hsize]{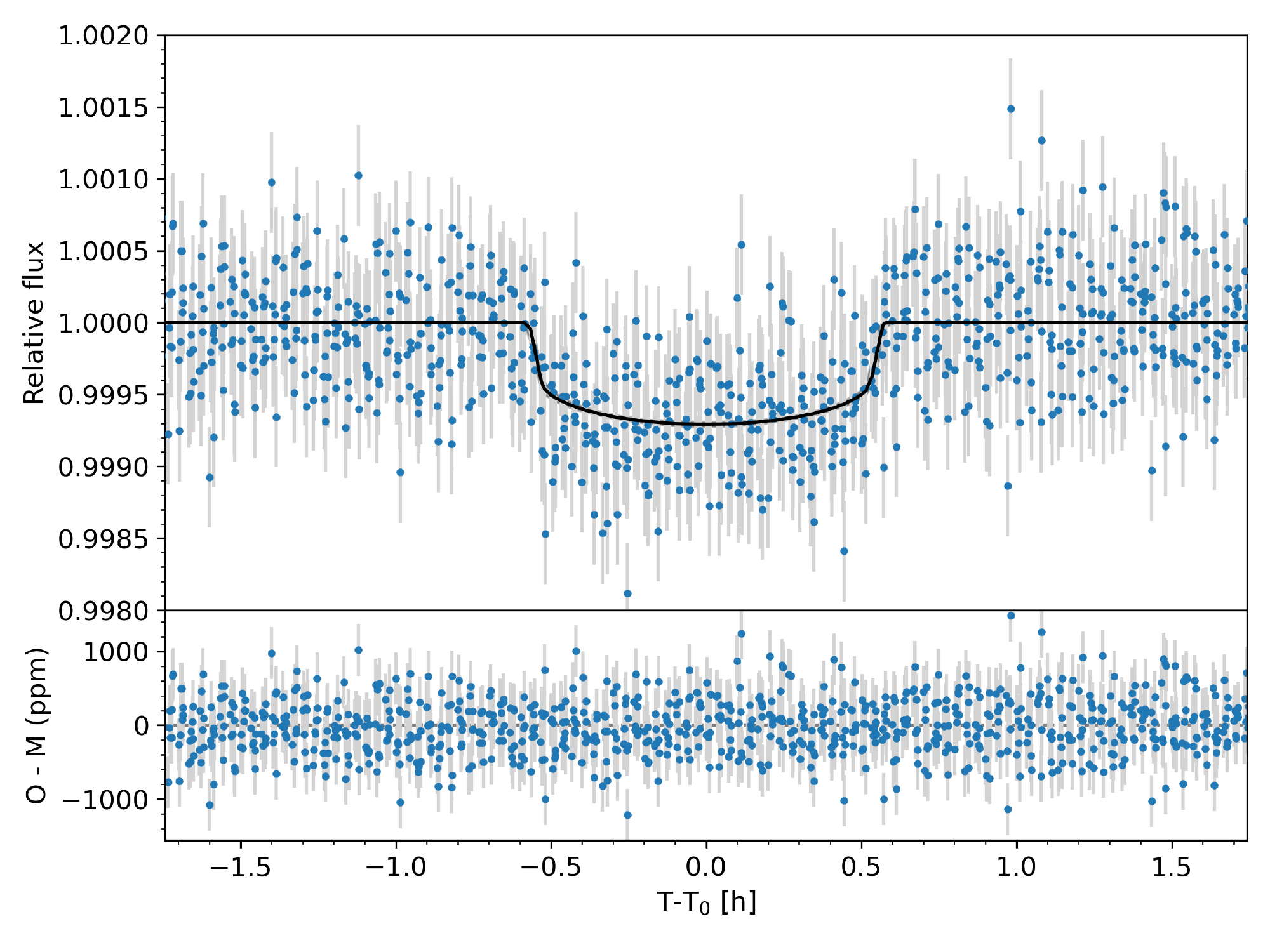}
    \includegraphics[width=0.49\hsize]{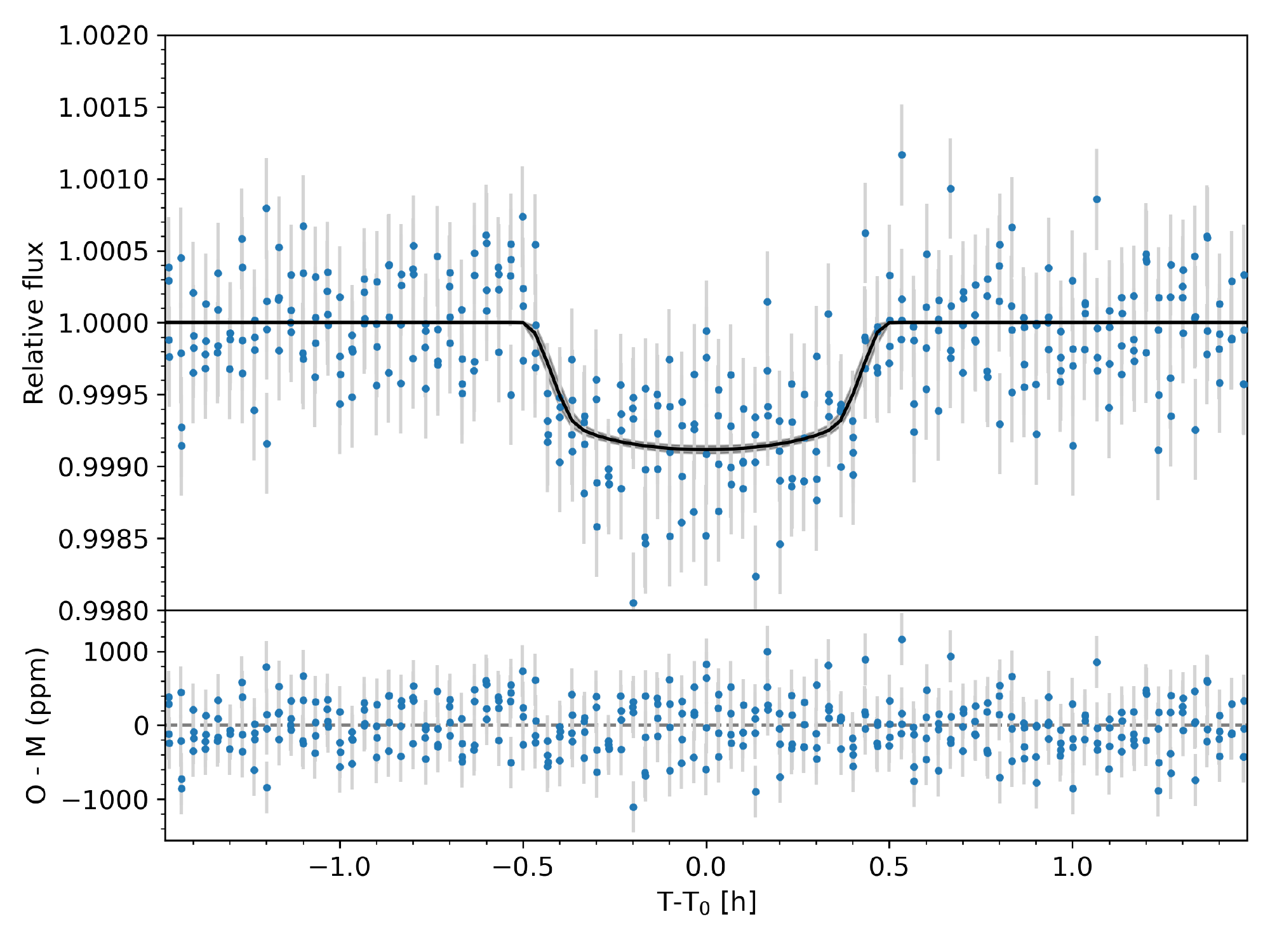}
    \caption{Best model resulting from the joint fit. {\it TESS} photometry phase-folded to the 2.77\,d period of HD~260655~b (left) and to the 5.71\,d period of HD~260655~c (right) along with best-fit transit model from the joint fit. The GP fitted to the photometry has been removed. Each row represents one Sector of {\it TESS} data (top, Sector 43; middle, Sector 44; bottom, Sector 45). } 
    \label{fig:lcfit}
\end{figure*}

\begin{figure*}
    \centering
    \includegraphics[width=\hsize]{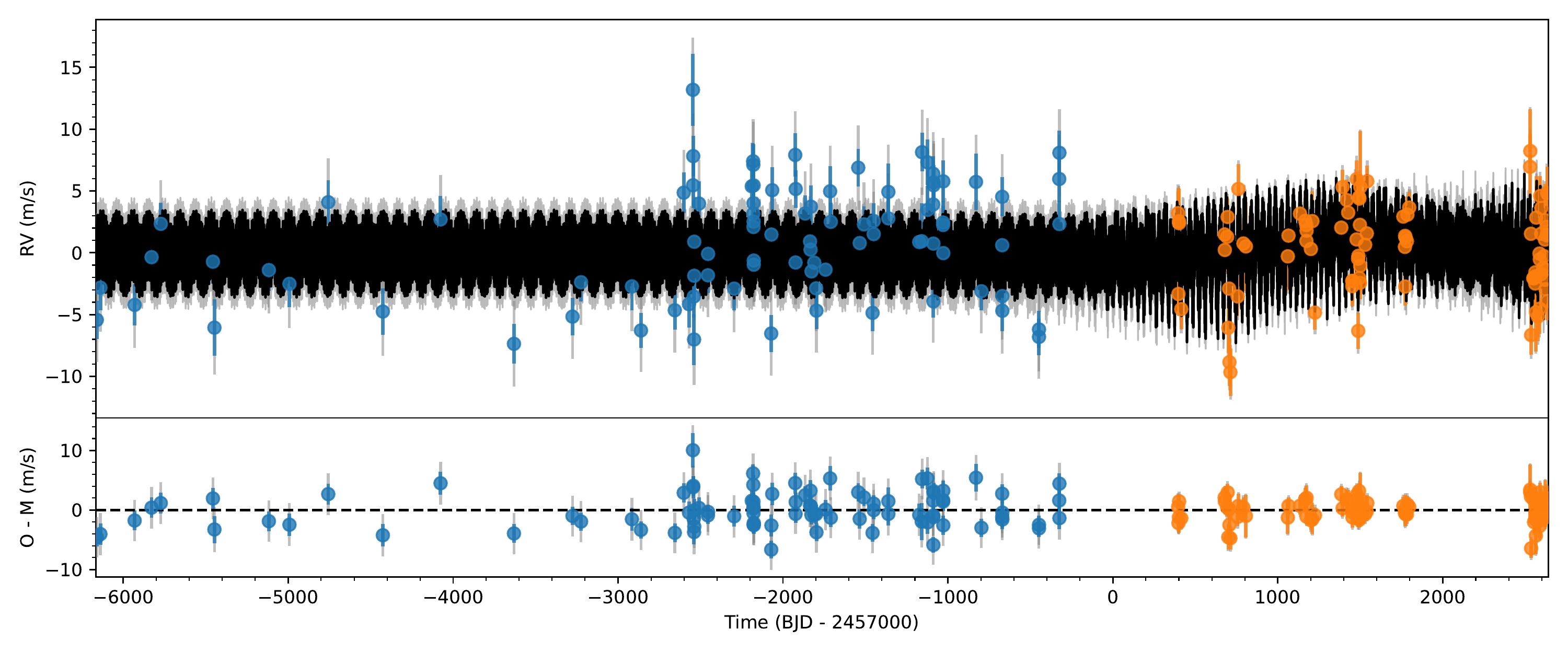}\\
    \includegraphics[width=\hsize]{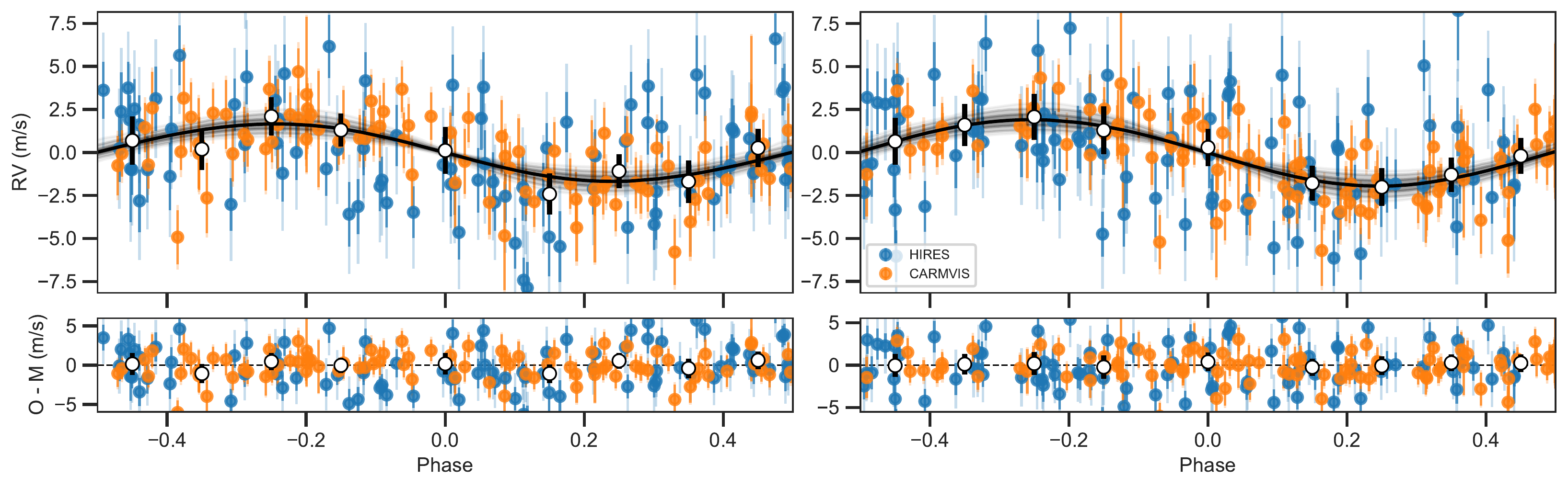}
    \caption{Results from the joint fit of the best model. Top panel: RV time series from HIRES (blue) and CARMENES (orange). Bottom panel: RVs phase-folded to the period of the two transiting planets (HD~260655~b, left; HD~260655~c, right). The error bars are broken into the measurement error and the white noise ``jitter'' with lighter color. White circles show binned data points in phase for visualization. The black line represents the best fit and the gray shaded area corresponds to the 1, 2, and 3$\sigma$ confidence intervals of the model.} 
    \label{fig:RVfit}
\end{figure*}

We carried out a global modeling of the photometric and spectroscopic data sets to jointly constrain the planetary properties of the HD~260655 system using \texttt{juliet}. We fitted two Keplerian orbits using informative --- but wide enough to derive reliable uncertainties --- priors based on the results from the previous transit- and RV-only analyses, including an exponential GP component to model the correlated noise seen in the \textit{TESS} PDC photometry and another exponential sine-squared GP component to model the imprint of the stellar rotation in the RV data. As discussed in Sect.~\ref{subsubsec:rv-fit}, there is no statistical evidence to add further planet signals. Our choice of the priors for each parameter in the joint analysis and their posterior distributions are presented in Table~\ref{tab:posteriors}. The resulting best-fit models and corresponding credibility bands are presented in Fig.~\ref{fig:lcfit} for the \textit{TESS} photometry and in Fig.~\ref{fig:RVfit} for the RVs. 

We find that the GP kernel is able to reproduce the increase of stellar activity seen in the first season of CARMENES data between 2016 and 2018, in agreement with the increase in the amplitude of the photometric variations of the T8 APT ground-based observations (Fig.~\ref{fig:stellar_rotation}). The sparse sampling of the HIRES RVs between 2009 and 2011 prevents the GP kernel to capture the variability seen in this other period of larger photometric variations in the T8 APT data. 

Table~\ref{tab:derivedparams} shows the fundamental physical parameters derived for the two planets based on the modeled transit and RV parameters from Table~\ref{tab:posteriors} and the stellar properties from Table~\ref{tab:star}. In summary, we firmly detect two transiting planets with radius and mass uncertainties below 3\% and 16\%, respectively, joining the growing population of small rocky planets orbiting low-mass, nearby stars.

\begin{table*}[t]
    \centering
    \caption{Priors, median and 68\% credibility intervals of the posterior distributions for each fit parameter of the final joint model obtained for the HD~260655 system using \texttt{juliet}.}
    \label{tab:posteriors}
    \begin{tabular}{l@{\hspace{3mm}}l@{\hspace{8mm}}c@{\hspace{3mm}}c@{\hspace{3mm}}} 
        \hline
        \hline
        \noalign{\smallskip}
        Parameter & Prior & HD~260655~b & HD~260655~c  \\
        \noalign{\smallskip}
        \hline
        \noalign{\smallskip}
        \multicolumn{4}{c}{\it Stellar parameters} \\[0.1cm]
        \noalign{\smallskip}
        $\rho_\star$ ($\mathrm{g\,cm\,^{-3}}$)  & $\mathcal{N}(7.5,0.5)$ & \multicolumn{2}{c}{$7.30^{+0.46}_{-0.43}$} \\[0.1 cm]
        \noalign{\smallskip}
        \multicolumn{4}{c}{\it Planet parameters} \\[0.1cm]
        \noalign{\smallskip}
        $P$ (d)                     & $\mathcal{U}(1,10)$       & $2.76953\pm0.00003$       & $5.70588\pm0.00007$ \\[0.1 cm]
        $t_0$\tablefootmark{(a)}    & $\mathcal{U}(9490,9500)$  & $9497.9102\pm0.0003$      & $9490.3646\pm0.0004$  \\[0.1 cm]
        $r_1$                       & $\mathcal{U}(0,1)$        & $0.776^{+0.014}_{-0.016}$ & $0.9267^{+0.0047}_{-0.0052}$ \\[0.1 cm]
        $r_2$                       & $\mathcal{U}(0,1)$        & $0.02586\pm0.00046$       & $0.0320\pm0.0010$ \\[0.1 cm]
        $e$                         & $\mathcal{B}(1.52,29.0)$  & $0.039^{+0.043}_{-0.023}$ & $0.038^{+0.036}_{-0.022}$  \\[0.1 cm]
        $\omega$ (deg)              & $\mathcal{U}(-180,180)$   & $57^{+70}_{-160}$         & $-25^{+156}_{-116}$ \\[0.1 cm]
        $K$ ($\mathrm{m\,s^{-1}}$)  & $\mathcal{U}(0,20)$       & $1.69 \pm 0.27$           & $1.92 \pm 0.30$ \\
        \noalign{\smallskip}
        \multicolumn{4}{c}{\it Photometry parameters} \\[0.1cm]
        \noalign{\smallskip}
        $\sigma_{\textnormal{TESS,S44}}$ (ppm)      & $\mathcal{J}(1,10^4)$         & \multicolumn{2}{c}{$106.9 \pm 7.3$} \\[0.1 cm] 
        $\sigma_{\textnormal{TESS,S45}}$ (ppm)      & $\mathcal{J}(1,10^4)$         & \multicolumn{2}{c}{$278.8 \pm 4.0$} \\[0.1 cm] 
        $\sigma_{\textnormal{TESS,S46}}$ (ppm)      & $\mathcal{J}(1,10^4)$         & \multicolumn{2}{c}{$172.2 \pm 4.8$} \\[0.1 cm] 
        $q_{1,\textnormal{TESS}}$                   & $\mathcal{U}(0,1)$            & \multicolumn{2}{c}{$0.31^{+0.15}_{-0.11}$} \\[0.1 cm]
        $q_{2,\textnormal{TESS}}$                   & $\mathcal{U}(0,1)$            & \multicolumn{2}{c}{$0.44^{+0.33}_{-0.28}$} \\[0.1 cm]
        \noalign{\smallskip}
        \multicolumn{4}{c}{\it RV parameters} \\[0.1cm]
        \noalign{\smallskip}
        $\gamma_{\textnormal{CARM}}$ ($\mathrm{m\,s^{-1}}$)        & $\mathcal{U}(-100,100)$   & \multicolumn{2}{c}{$-0.24^{+0.83}_{-0.95}$} \\[0.1 cm]
        $\sigma_{\textnormal{CARM}}$ ($\mathrm{m\,s^{-1}}$)     & $\mathcal{J}(0.1,100)$    & \multicolumn{2}{c}{$1.13\pm0.43$} \\[0.1 cm]
        $\gamma_{\textnormal{HIRES}}$ ($\mathrm{m\,s^{-1}}$)       & $\mathcal{U}(-100,100)$   & \multicolumn{2}{c}{$-1.12^{+0.75}_{-0.84}$} \\[0.1 cm]
        $\sigma_{\textnormal{HIRES}}$ ($\mathrm{m\,s^{-1}}$)    & $\mathcal{J}(0.1,100)$    & \multicolumn{2}{c}{$3.05\pm0.44$} \\[0.1 cm]
         \noalign{\smallskip}
        \multicolumn{4}{c}{\it GP hyperparameters} \\
        \noalign{\smallskip}
        $\sigma_\mathrm{GP,TESS}$ (ppm)       & $\mathcal{J}(10^{-10},10^{-2})$   & \multicolumn{2}{c}{$18\pm2$} \\[0.1 cm]
        $T_\mathrm{GP,TESS}$ (d)              & $\mathcal{J}(10^{-3},10^{3})$    & \multicolumn{2}{c}{$0.94^{+0.04}_{-0.07}$} \\[0.1 cm]
        
        $\sigma_\mathrm{GP,RV}$ ($\mathrm{m\,s^{-1}}$)          & $\mathcal{J}(10^{-1},30)$         & \multicolumn{2}{c}{$2.55^{+0.59}_{-0.51}$} \\[0.1 cm]
        $\alpha_\mathrm{GP,RV}$ ($10^{-6}\,\mathrm{d^{-2}}$)    & $\mathcal{J}(10^{-8},10^{-3})$    & \multicolumn{2}{c}{$3.2^{+8.9}_{-2.1}$} \\[0.1 cm]
        $\Gamma_\mathrm{GP,RV}$ (d$^{-2}$)                      & $\mathcal{J}(10^{-2},10)$         & \multicolumn{2}{c}{$6.5^{+2.3}_{-3.3}$}     \\[0.1 cm]
        $P_\mathrm{rot;GP,RV}$ (d)                              & $\mathcal{N}(37.5,0.4)$           & \multicolumn{2}{c}{$37.0^{+0.7}_{-0.3}$}   \\[0.1 cm]
        \noalign{\smallskip}
        \hline
    \end{tabular}
    \tablefoot{
        \tablefoottext{a}{Dates are BJD$-$2450000.}
        The prior labels of $\mathcal{N}$, $\mathcal{U}$, $\mathcal{B}$, and $\mathcal{J}$ represent normal, uniform, beta, and Jeffrey's distributions.

    }
\end{table*}

\begin{table}[t]
    \centering
    \caption{Derived planetary parameters obtained for the HD~260655 system using the posterior values from the joint fit in Table~\ref{tab:posteriors} and stellar parameters from Table~\ref{tab:star}.}
    \label{tab:derivedparams}
    \begin{tabular}{lrr} 
        \hline
        \hline
        \noalign{\smallskip}
        Parameter\tablefootmark{(a)} & HD~260655~b & HD~260655~c  \\
        \noalign{\smallskip}
        \hline
        \noalign{\smallskip}
        \multicolumn{3}{c}{\it Derived transit parameters} \\[0.1cm]
        \noalign{\smallskip}
        $p = R_{\rm p}/R_\star$             & $0.02586\pm0.00046$           & $0.0320\pm0.0010$ \\[0.1 cm]
        $b = (a/R_\star)\cos i_{\rm p}$     & $0.665^{+0.022}_{-0.024}$     & $0.890^{+0.007}_{-0.008}$     \\[0.1 cm]
        $a/R_\star$                         & $14.43 \pm 0.29$              & $23.37 \pm 0.47$          \\[0.1 cm]
        $i_{\rm p}$ (deg)                   & $87.35 \pm 0.14$              & $87.79 \pm 0.08$                \\[0.1 cm]
        $t_T$ (h)\tablefootmark{(b)}        & $1.15 \pm 0.02$               & $0.98 \pm 0.02$        \\[0.1 cm]
        \noalign{\smallskip}
        \multicolumn{3}{c}{\it Derived physical parameters} \\[0.1cm]
        \noalign{\smallskip}
        $M_{\rm p}$ ($M_\oplus$)            & $2.14 \pm 0.34$       & $3.09 \pm 0.48$              \\[0.1 cm]
        $R_{\rm p}$ ($R_\oplus$)            & $1.240 \pm 0.023$     & $1.533^{+0.051}_{-0.046}$            \\[0.1 cm]
        $\rho_{\rm p}$ (g cm$^{-3}$)        & $6.2 \pm 1.0$         & $4.7^{+0.9}_{-0.8}$        \\[0.1 cm]
        $g_{\rm p}$ (m s$^{-2}$)            & $13.7 \pm 2.2$        & $12.9 \pm 2.2$             \\[0.1 cm]
        $a_{\rm p}$ (au)                    & $0.02933 \pm 0.00024$ & $0.04749 \pm 0.00039$        \\[0.1 cm]
        $T_\textnormal{eq}$ (K)\tablefootmark{(c)}  & $709 \pm 4$   & $557 \pm 3$             \\[0.1 cm]
        $S$ ($S_\oplus$)                    & $42.2 \pm 0.7$        & $16.1 \pm 0.3$              \\[0.1 cm]
        \noalign{\smallskip}
        \hline
    \end{tabular}
    \tablefoot{
      \tablefoottext{a}{Error bars denote the $68\%$ posterior credibility intervals.}
      \tablefoottext{b}{Transit duration as measured from the first to the last contact.}
      \tablefoottext{c}{Equilibrium temperatures were calculated assuming zero Bond albedo and perfect energy redistribution.}
      }
\end{table}

\section{Discussion} \label{sec:discussion}

The HD~260655 system contains at least two transiting terrestrial planets: HD~260655~b, with a period of 2.77\,d, a radius of $R_{\rm b}=1.240\pm0.023\,R_\oplus$, a mass of $M_{\rm b}=2.14\pm0.34\,M_\oplus$, and a density of $\rho_{\rm b}=6.2\pm1.0\,\mathrm{g\,cm^{-3}}$; and HD~260655~c, with a period of 5.71\,d, a radius of $R_{\rm c}=1.533^{+0.051}_{-0.046}\,R_\oplus$, a mass of $M_{\rm c}=3.09\pm0.48\,M_\oplus$, and a density of $\rho_{\rm c}=4.7^{+0.9}_{-0.8}\,\mathrm{g\,cm^{-3}}$.

\subsection{Planet composition and formation history}

\begin{figure*}
    \centering
    \includegraphics[width=0.49\hsize]{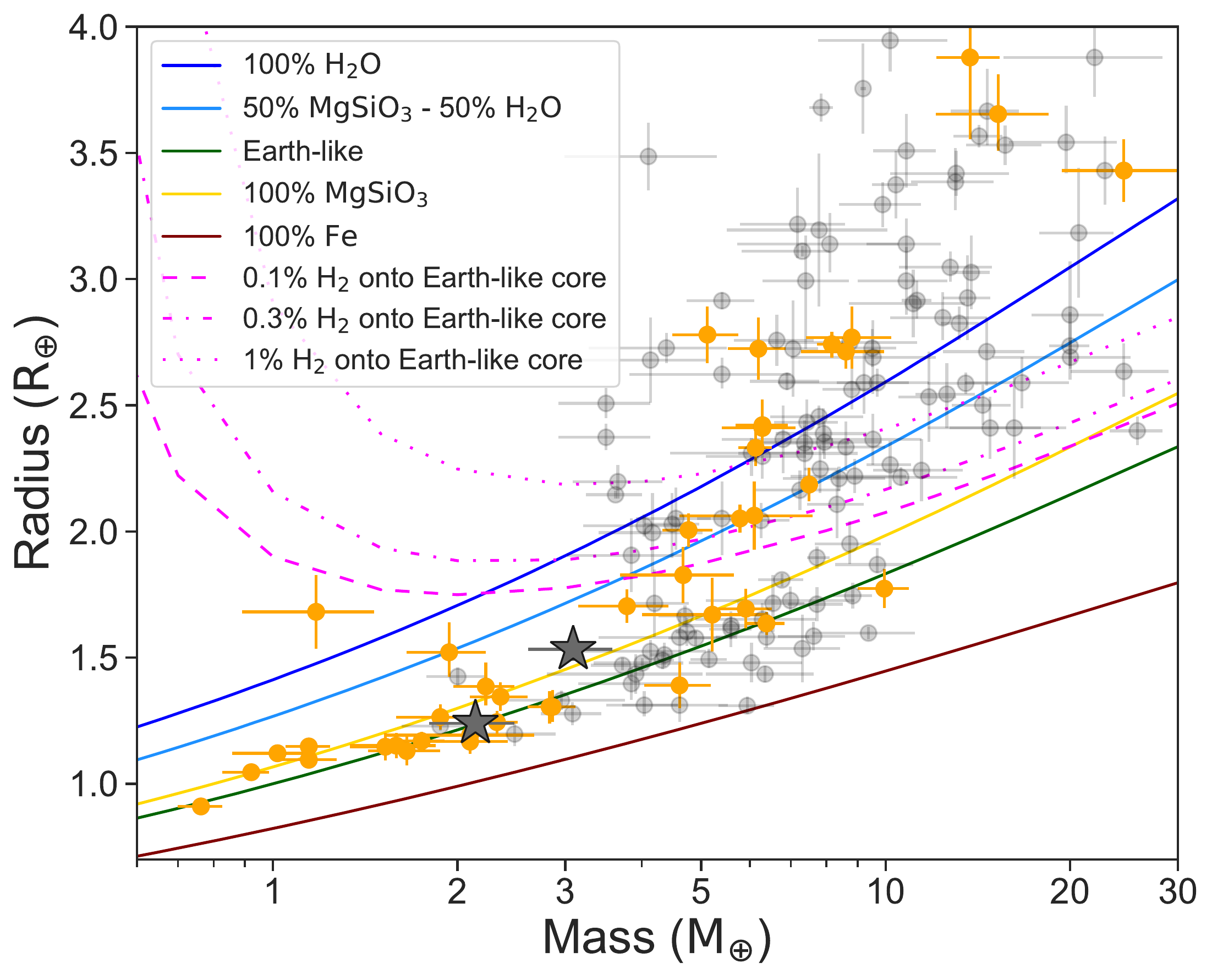}
    \includegraphics[width=0.49\hsize]{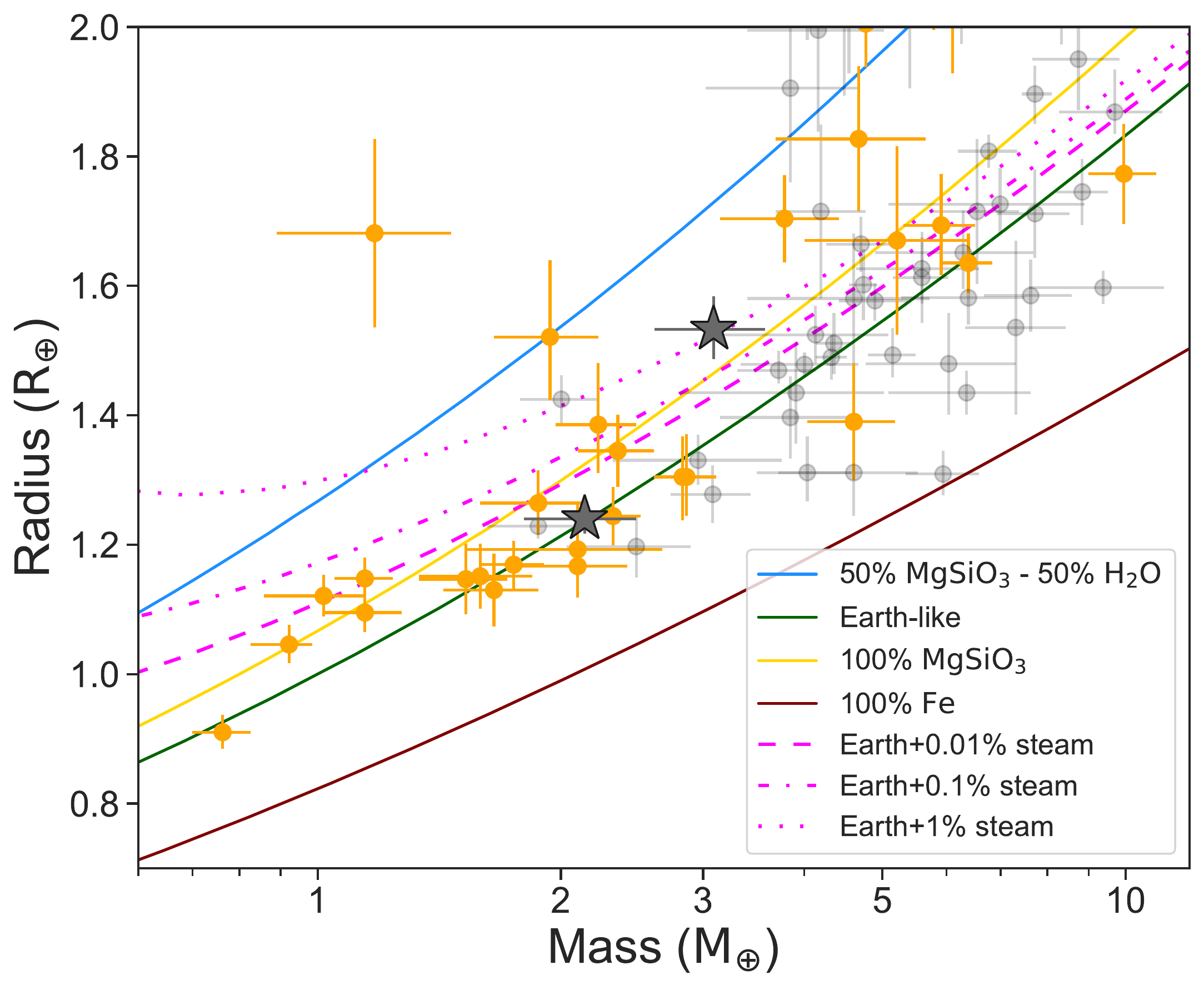}
    \caption{Mass-radius diagram for all known transiting planets with dynamical masses determined with a precision better than 30\% from the TEPCat data base of well-characterized planets \citep{Southworth2011MNRAS.417.2166S}. Planets orbiting M dwarfs ($T_{\rm eff} < 4000\,\mathrm{K}$) are shown in orange, others gray. The planets orbiting HD~260655 are shown with stars. The left panel shows theoretical models for the internal composition from \citet{Zeng2019}, while the right panel shows a zoom-in version of the left panel with theoretical models from \citet{Turbet2020}.}
    \label{fig:massradius}
\end{figure*}

Given their measured physical properties, we can estimate the internal composition of the planets in the HD~260655 system. Figure~\ref{fig:massradius} shows HD~260655~b and c in the context of all known transiting planets with measured dynamical masses via RVs or transit timing variations (TTVs) to a precision better than 30\%. We see that HD~260655~b belongs to a typical class of exoplanets orbiting M~dwarfs, with masses between 2 and 3\,$M_\oplus$ and radii smaller than $1.3\,R_\oplus$. Rocky planets such as GJ~357~b \citep{GJ357}, GJ~1252~b \citep{GJ1252}, GJ~3473~b \citep{GJ3473}, GJ~486~b \citep{GJ486}, LTT~3780~b \citep{LTT3780-1,cloutier2020}, LHS~1478~b \citep{LHS1478}, and L98-59~c \citep{L98-59} belong to this class. They are typically small, hot and, in the case of known multi-planetary systems, the closest to the host. On the other hand, HD~260655~c seems to have no counterparts\footnote{Except for TOI-1634~b, although its bulk density measurement and internal composition are under discussion in the literature \citep{TOI-1634-1,TOI-1634-2}.} among the small planet population orbiting M~dwarfs. With a density of $\rho_{\rm c}=4.7^{+0.9}_{-0.8}\,\mathrm{g\,cm^{-3}}$, it is less dense than the rocky planets of its size. 

When comparing with theoretical mass-radius curves from \citet{Zeng2016ApJ...819..127Z,Zeng2019}, both planets are consistent with rocky compositions. However, while HD~260655~b has a bulk density in perfect agreement with the Earth's, HD~260655~c is more consistent with an internal composition void of iron and fully made of silicates if it is assumed to be free of volatiles. 

The observed variation in the bulk density of the planets is very likely to be a consequence of their different volatile contents or the observational uncertainties. Using the models from \citet{Zeng2019}, we can estimate the amount of hydrogen necessary to reproduce the mass and radius of planet c assuming an Earth-like interior composition. The magenta lines in the left panel of Fig.~\ref{fig:massradius} show a set of such models with hydrogen envelopes of 0.1\%, 0.3\%, and 1\% by mass, assuming 
a surface temperature of 500\,K (a proxy of the measured equilibirum temperature $T_{\rm eq} = 557\pm3\,\mathrm{K}$). We find that a $\sim 3\,M_\oplus$ Earth-like planet with just a 0.1\% hydrogen-rich envelope by mass would have a much larger radius than the one measured for HD~260655~c. 

Another volatile molecule abundant in protoplanetary disks is water \citep[e.g.,][]{Lodders2003ApJ...591.1220L,Terada2007ApJ...667..303T}. Mass-radius relationships for water-rich rocky planets are usually calculated assuming that most water is present in condensed (either liquid or solid) form \citep[e.g.,][]{Seager2007,Swift2012,Zeng2016ApJ...819..127Z}. However, HD~260655~c receives an irradiation much larger than the runaway greenhouse irradiation limit \citep{Kasting1993Icar..101..108K,Kopparapu13}, for which water has been shown to be unstable in condensed form and would instead form a thick water-dominated atmosphere \citep{Valencia2013,Turbet2019}. The right panel of Fig.~\ref{fig:massradius} shows a new set of mass-radius curves (magenta lines) for small planets in this regime computed by \citet{Turbet2020}. These new mass-radius relationships lead to planet bulk densities much lower than calculated when water is assumed to be in condensed form. For HD~260655~c, assuming an Earth-like composition, a water fraction of just 1\% by mass is able to account for the difference in bulk density with respect to planet b. In this scenario, however, the water content of the innermost planet must be negligible. Otherwise, considering that planet b receives more irradiation than planet c, it would lead to a much larger radius than measured.

\begin{figure}
    \centering
    \includegraphics[width=\hsize]{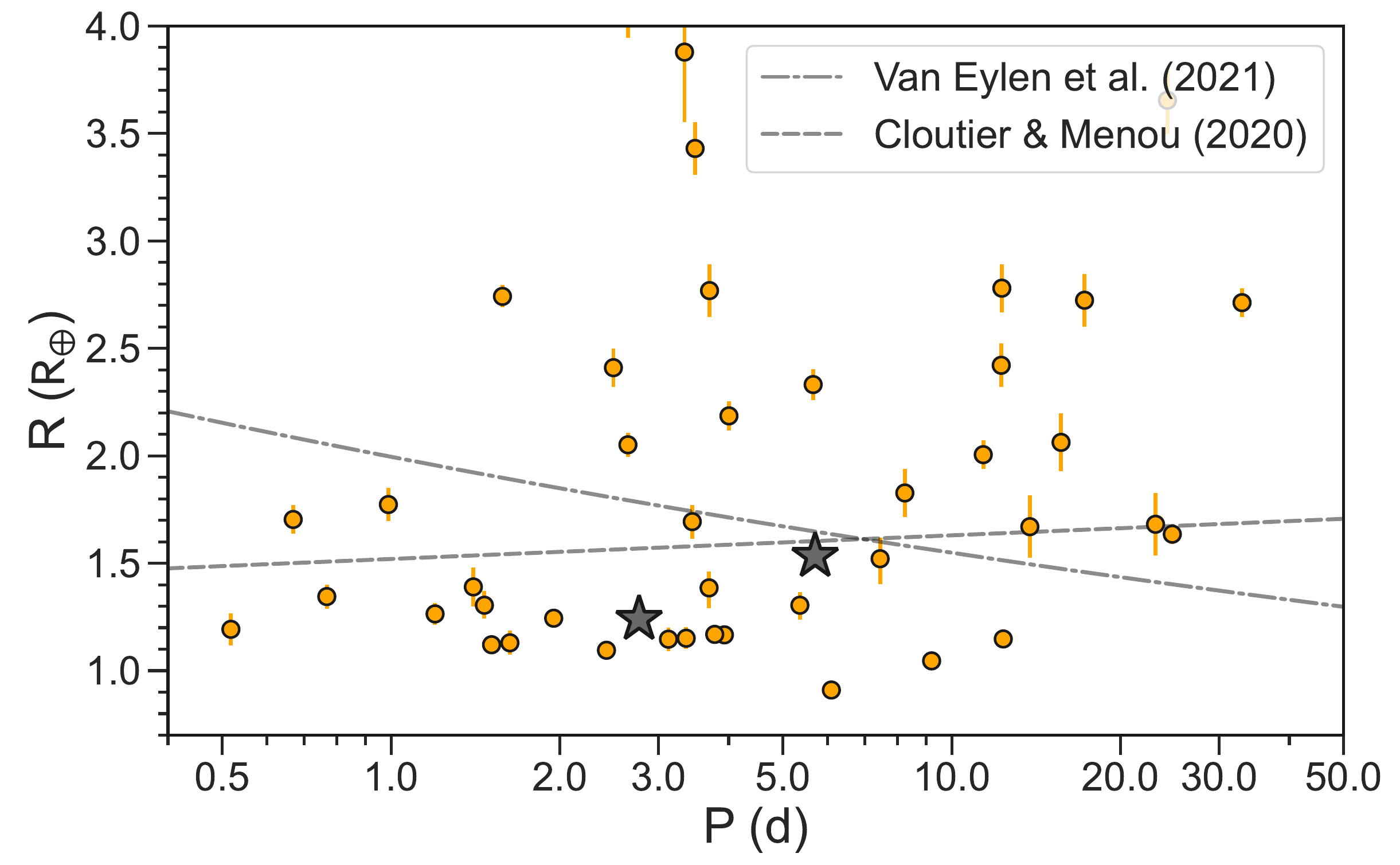}
    \caption{Period-radius diagram of the M-dwarf transiting planets from Fig.~\ref{fig:massradius}. The location of the radius valley for M~dwarfs derived in \citet{CloutierMenou2020} and \citet{VanEylen2021} is shown. The planets orbiting HD~260655 are marked with stars. }
    \label{fig:period-radius}
\end{figure}

Figure~\ref{fig:period-radius} shows that, according to their periods and radii, both planets are located below the so-called radius valley for planets orbiting M~dwarfs \citep{CloutierMenou2020,VanEylen2021}. Despite the different empirical locations of the valley separating rocky from non-rocky planets in these works, the planets in the HD~260655 system are compatible with not having retained a substantial atmosphere as discussed above. The origin of this division is typically attributed to thermally driven atmospheric mass-loss mechanisms, such as photoevaporation \citep[e.g.,][]{OwenWu2017ApJ...847...29O,JinMordasini2018ApJ...853..163J} or core-powered mass-loss models \citep{Ginzburg2018MNRAS.476..759G,GuptaSchlichting2019MNRAS.487...24G}, in which planets below the radius valley have been stripped of their atmosphere, whereas planets above the valley have held on to it. 

An estimate of the expected mass-loss rate due to photoevaporation in the planets can be made assuming that the atmosphere is composed essentially by hydrogen, and that all photoevaporation is produced by XUV (X-rays+EUV) photons as explained in \citet{san11}. Ideally, a coronal model should be employed for that purpose, but we have not enough data either in X-rays (see Sect.~\ref{subsec:rotation}), or in the UV. The available UV observations of the star taken with the International Ultraviolet Explorer observatory \citep{Elgaroy1997A&A...326..165E,Elgaroy1999A&A...343..222E} do not have enough statistics nor S/N to measure any transition region lines useful for building a coronal model. Instead, we can calculate the EUV (100--920\,\AA) flux following the X-ray/EUV relations in \citet{san11}, resulting in $L_{\rm EUV}=1.9 \times 10^{28}$\,erg\,s$^{-1}$. For planets b and c, the current mass loss rate expected is 0.61 and 0.22\,$M_\oplus\,\mathrm{Gyr}^{-1}$, respectively, and should have been larger in the past. Therefore, if the planets originally accreted any amount of hydrogen and/or helium during formation, the host star would have stripped it during the system's evolution. Thus, we conclude that both planets are primarily rocky, without extended H/He atmospheres. Their different bulk densities may be related to minimum differences in their water-mass fractions or observational uncertainties.

This is consistent with planet formation models of core accretion, which predict similar density regimes for rocky planets around stars in the mass range of HD~260655~\citep{Burn2021}. The planetesimal accretion-based Bern model of planet formation and evolution~\citep{Emsenhuber2020} shows a high occurrence density of planets in this region of mass-radius space, and HD~260655~b would correspond to their typical volatile-poor rocky planet without any atmosphere. The individual planets can thus both be well explained in a core accretion framework. However, the same model typically produces planets of similar bulk density within the same system. The absolute value of this density depends on the system architecture: if an outer giant planet companion is present, inner small planets are found to be devoid of volatiles delivered from the outer regions of the protoplanetary disk~\citep{Schlecker2021}. As a consequence, this would lead to more dense planets in systems with cold Jupiters. The present data for HD~260655 does not show any indications of a giant planet, and the small eccentricities of its planets do not suggest a close-by massive body. Planets that accreted volatile-rich material in distant regions could thus have migrated into the inner system, leading to the characteristics of HD~260655~c. The higher and significantly different density of HD~260655~b remains unexplained in this scenario. 

If the volatiles of the planets were delivered via a pebble flux to the inner system, the chronological order of their formation might explain the architecture of the system: in the case where the outer planet grows first and reaches its pebble isolation mass, it would block the interior regions from the supply of pebbles~\citep[e.g.,][]{Haghighipour2003,Pinilla2012,Morbidelli2012}. This would leave only classical planetesimal accretion for the inner planet, potentially explaining its smaller mass and higher bulk density. Regarding the specific architecture of the HD~260655 planetary system, it will be interesting to see what future pebble accretion models that also model planetary radii will predict.


\subsection{Searches for additional planets and detection limits}
\label{sec:limits}
We explored the \textit{TESS} data searching for extra transiting planets that may remain unnoticed by the automatic SPOC and the QLP pipelines due to their detection thresholds. To this aim we used the {\fontfamily{pcr}\selectfont  SHERLOCK}\footnote{{The \fontfamily{pcr}\selectfont  SHERLOCK} (\textbf{S}earching for \textbf{H}ints of \textbf{E}xoplanets f\textbf{R}om \textbf{L}ightcurves \textbf{O}f spa\textbf{C}e-based see\textbf{K}ers) code is fully available on GitHub: \url{https://github.com/franpoz/SHERLOCK}} pipeline \citep{pozuelos2020,demory2020}. {\fontfamily{pcr}\selectfont  SHERLOCK} is an end-to-end pipeline that employs six different modules to (1) acquire and prepare the light curves from their repositories, (2) search for planetary candidates, (3) vet the interesting signals, (4) perform a statistical validation, (5) model the signals to refine their ephemerides, and (6) compute the observational windows from ground-based observatories to trigger a follow-up campaign. To optimize the transit search, {\fontfamily{pcr}\selectfont  SHERLOCK} applies a multi-detrend approach to the nominal PDC-corrected \textit{TESS} light curve, employing the \texttt{w{\={o}}tan} package \citep{wotan2019}, that is, the PDC light curve is detrended several times using a bi-weight filter by varying the window size. Then the transit search is performed over the PDC light curve jointly with the new detrended light curve through the \texttt{transit least squares} (TLS) package \citep{hippke_TLS_2019}, which is optimized to detect shallow periodic transits using an analytical transit model based on the stellar parameters. The transit search is performed in a loop; once a signal is found, it is recorded and masked, and then the search keeps running until no more signals with S/N$\geqslant$5 are found in the data set. 

Following the strategy presented in \cite{nicole2022}, we recovered the planets HD~260655~b and c in the first and second runs, respectively. Then, these planets were masked, and we performed three types of transit searches by simultaneously considering all the sectors available. We focused our first search on orbital periods ranging from 0.5 to 30\,d, requiring a minimum of two transits to identify a potential signal. We focused on longer orbital periods in the second trial, ranging from 30 to 80\,d, where single events could be recovered. In the final trial, we explored all the sectors independently, focusing on short orbital periods ranging from 0.5 to 15\,d and using a two times denser period grid. This strategy allowed us to explore more accurate short orbital periods and avoid having sectors with different photometric precision affecting the search.
 
None of these three strategies yielded positive results. All the signals found were attributable to systematics, noise, or variability. As stated by \cite{wells2021}, the lack of extra signals suggests that either: (1) no other planets exist in the system; (2) they do exist, but they do not transit; or (3) they do exist and transit, but the precision of the data set is not good enough to detect them, or (4) they have orbital periods longer than those explored in this study. Any massive enough planets might be detected during our RV follow-up if scenario (2), (3) or (4) is true. However, no other prominent signal was found. To evaluate scenarios (3) and (4), we studied the detection limits in both the \textit{TESS} photometric data and the RV measurements.         

To test the planetary detectability in the \textit{TESS} data we used the {\fontfamily{pcr}\selectfont  MATRIX ToolKit}\footnote{{The \fontfamily{pcr}\selectfont  MATRIX ToolKit} (\textbf{M}ulti-ph\textbf{A}se \textbf{T}ransits \textbf{R}ecovery from \textbf{I}njected e\textbf{X}oplanets \textbf{T}ool\textbf{K}it) code is available on GitHub: \url{https://github.com/martindevora/tkmatrix}}. {\fontfamily{pcr}\selectfont MATRIX} performs an inject-and-recovery experiment of synthetic planets in the \textit{TESS} PDC-corrected light curve, combining the three sectors available, allowing the user to define the ranges in the $R_{\mathrm{planet}}$--$P_{\mathrm{planet}}$ parameter space to be examined. In our case we explored the ranges of 0.5 to 3.5\,R$_{\oplus}$ with steps of 0.16\,R$_{\oplus}$, and 1.0 to 30.0\,d with steps of 1.0\,d. Each combination of $R_{\mathrm{planet}}$--$P_{\mathrm{planet}}$ was explored using five different phases, that is, different values of $t_{0}$, which allows us to improve the statistics of our recovery rates. Hence, we explored 2887 scenarios. For simplicity, the synthetic planets are injected into the light curve assuming their impact parameters and eccentricities are equal to zero. We detrend the light curves using a bi-weight filter with a window size of 0.5\,d, which was found to be the optimal value during the {\fontfamily{pcr}\selectfont SHERLOCK} runs, and masked the transits corresponding to the known planets in the system.

A synthetic planet was recovered when its epoch was detected with 1\,h accuracy and the recovered period was within 5\,\% of the injected period. Since we used the PDC-corrected light curve, the signals were not affected by the PDC systematic corrections; therefore, the detection limits should be considered as the most optimistic scenario \citep[see, e.g.,][]{pozuelos2020,eisner2020}.

\begin{figure}
    \centering
    \includegraphics[width=0.50\textwidth]{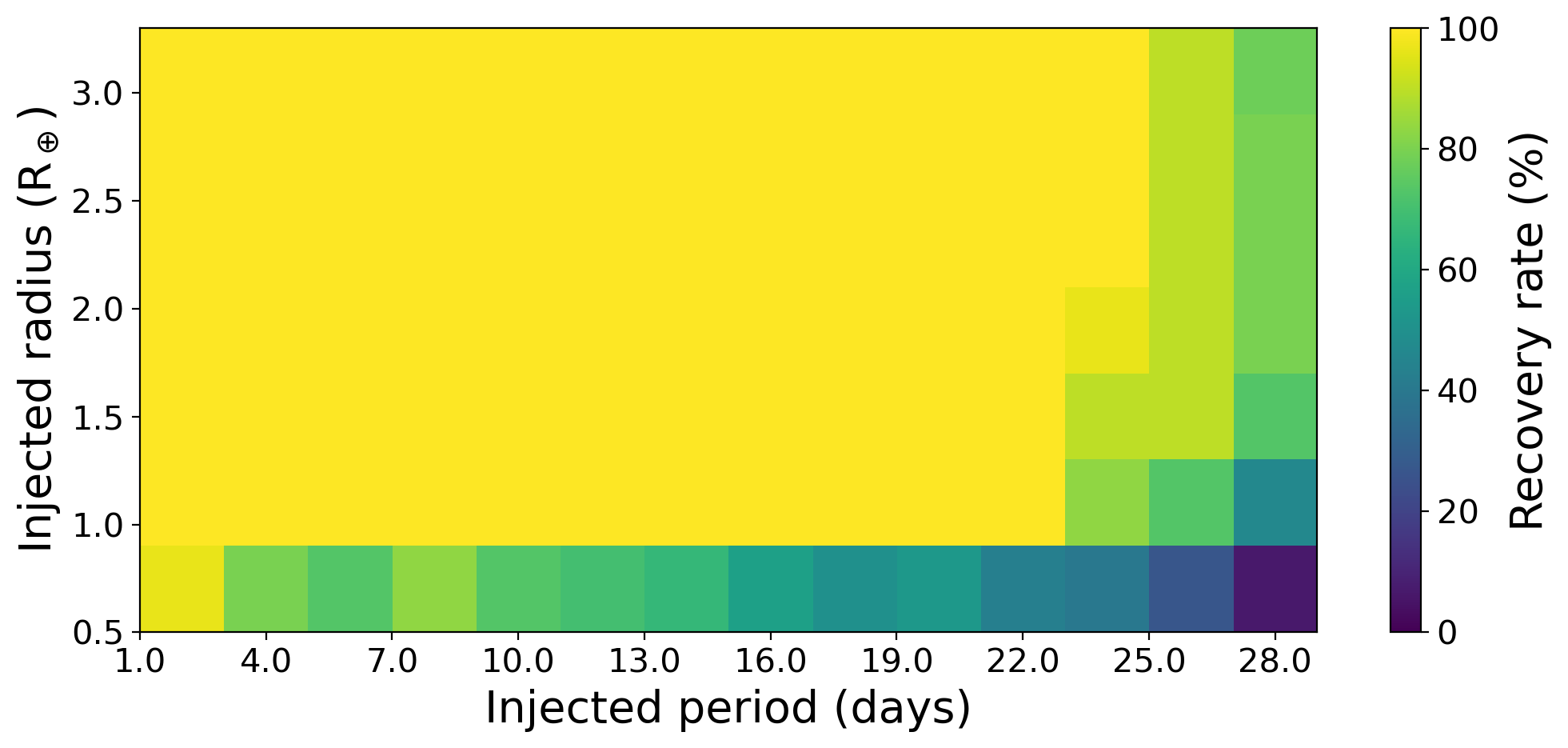}
    \caption{Injection-and-recovery tests performed on \textit{TESS} data to check the detectability of extra planets in the system. We explored a total of 2887 different scenarios. Each pixel evaluated 30 scenarios, that is, 30 light curves with injected planets having different $P_{\mathrm{planet}}$, $R_{\mathrm{planet}}$ and $t_{0}$. Larger recovery rates are presented in yellow and green, while lower recovery rates are shown in blue and darker hues. We can rule out the presence of additional transiting planets in the system with sizes $>$1.0\,R$_{\oplus}$ and periods shorter than 22\,d.}
    \label{fig:recovery}
\end{figure}

The results, shown in Figure~\ref{fig:recovery}, allowed us to rule out planets with sizes $>$1.0\,R$_{\oplus}$, with recovery rates ranging from 80 to 100\,$\%$ for the full range of periods explored. On the other hand, planets with sizes $<$1.0\,R$_{\oplus}$ would be more challenging to detect with recovery rates of about 50\,$\%$ for orbital periods shorter than 20\,d. Longer orbital periods yielded recovery rates $<$50\,$\%$.
 
On the other hand, the RV data presented in this work provide constraints on the existence of larger outer planets in this system. We estimated the maximum $M\sin{i}$ value compatible with the RV measurements as a function of prospective orbital period. First, we verified that the RVs of our star did not show long-term trends, which would have needed correction, and fitted a circular orbit to the data using a partially-linearized, least-squares fitting procedure \citep{James1975CoPhC..10..343J} as a function of varying orbital period, $P$. For each prospective $P$, we then determined the best-fit semi-amplitude $K(P$) and computed a planet mass value ($M\sin{i})_{\rm max}(P) = M_{\rm m}(P)$. This mass is then an optimistic 1$\sigma$ maximum value a planet could have to be non-detectable in our RV time series, considering its noise characteristics.

\begin{figure}
    \centering
    \includegraphics[width=0.99\hsize]{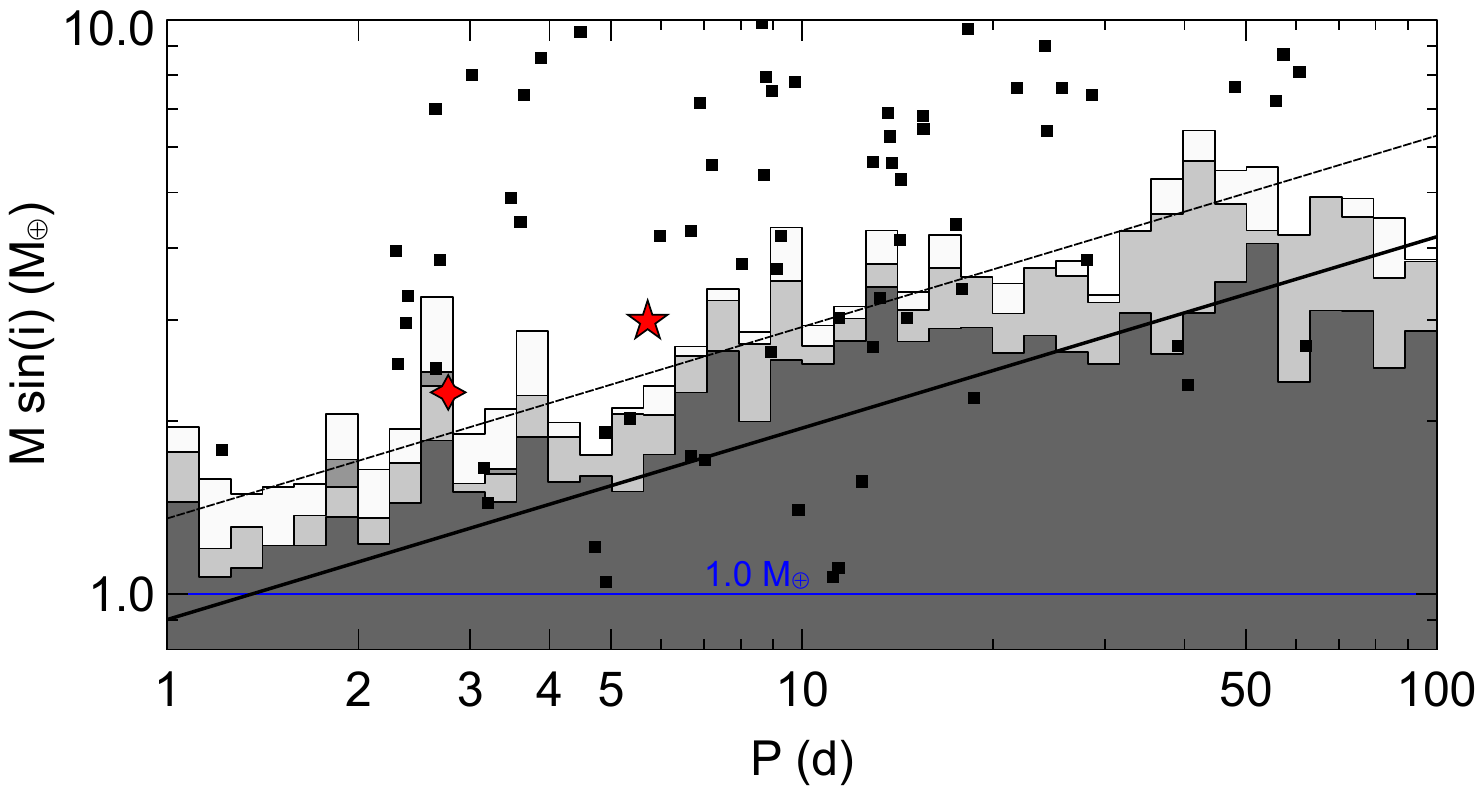}
    \caption{Maximum $M\sin{i}$ value compatible with the RV measurements as a function of prospective orbital period. From light to darker gray: detectability from original RV data (no modeling of planets or activity), data corrected for the signal of the two planets, and data corrected for activity with GP and the two planets. The horizontal blue line shows the signal expected for 1.0 M$_\oplus$ planets. The two diagonal black lines show masses corresponding to semi-amplitudes of 1.0 and 1.5 m\,s$^{-1}$ in the RVs. Small black squares represent exoplanets around M dwarfs from \url{exoplanet.eu} with mass determinations from RVs. Red symbols represent HD\,260655 b (four-point star) and c (five-point star).}
    \label{fig:sensitivity_rv}
\end{figure}

Figure~\ref{fig:sensitivity_rv} shows $M_{\rm m}(P)$ as a function of the orbital period for HD~260655. The quality of the data acquired is good, both the HIRES and CARMENES data showing similar internal precision for this star. The much longer time span covered by the HIRES data during the first years of observations, together with the better sampling of HIRES during its last eight years and the CARMENES data, generate a rather uniform detectability at both short and long periods, making the whole data set sensitive to planets above masses between $M\sin{i} \approx 1.0$ and 3.0\,$M_{\oplus}$ for periods between 1.0 and 10\,d respectively, and above 3.0\,$M_{\oplus}$ for periods larger than 10\,d. For this star, the change of behavior of the detectability for periods beyond the length of the data set is apparent at around 8500\,d. The plot also shows the differences in detectability of our raw data (no cleaning of planets or activity) compared with data cleaned from the planetary signals, and data cleaned fully from planetary and stellar activity signals.

In summary, with \textit{TESS} data, we can exclude any additional Earth-sized or larger planets with orbital periods shorter than 20\,d transiting in the system. On the other hand, with RV data, we can rule out the presence of planets with 1--3\,$M_\oplus$ and orbital periods of less than 10\,d, and rule out planets with masses above 3\,$M_\oplus$ for periods between 10 and 100\,d.

\subsection{Dynamics and TTV analysis}

The periods of HD~260655~b and c lie moderately close to the first order mean-motion resonance 2:1. Hence, due to their gravitational interaction in such a configuration, one may expect some level of mutual orbital excitation, which in turn may induce measurable TTVs \citep{agol2005,holman2005}. The amplitude of these TTVs depends on the masses and eccentricities of the existing planets, which, when combined with RV observations, are especially powerful to probe the dynamics of a given system \citep[see, e.g.,][]{holman2010,weiss2017,kaye2021}.  

In this context, we aim to predict the amplitudes of the TTVs in the HD~260655 system to evaluate the reliability of such combined analysis. To this end, we used the \texttt{TTVFast2Furious} package \citep{hadden2019} following the strategy presented by \cite{cloutier2020}. We ran 10$^{3}$ realizations with planetary masses, orbital periods, times of mid-transit, eccentricities, and arguments of periastron sampled from their joint fit posterior distributions given in Table~\ref{tab:posteriors}. Each realization corresponds to a unique set of parameters for which we compute, for each planet, the maximum deviation from a linear ephemeris over a one-year baseline, that is, about three times the super-period of the system, which is found to be 95\,d \citep{lithwick2012}. We obtained maximum TTV amplitudes for both planets lower than 2\,min. The small amplitudes of the expected TTVs combined with the unlikely existence of additional massive planets in the system as probed in Sect.~\ref{sec:limits} hint that the HD~260655 system is not optimal for intense TTV follow-up.  

Besides, we assessed the dynamical stability of the system from 100 orbital simulations with the \texttt{mercury6} N-body integrator \citep{Chambers1999}. We randomly set the initial conditions for each simulation by drawing from the joint fit posteriors. We used a time-step of 0.07\,d (approximately 1/40 the period of the inner planet, HD~260655~b), integrated the system through $10^{5}$\,yr with the hybrid symplectic and Bulirsch-Stoer integrator, and set the integration accuracy parameter to $10^{-12}$.  As expected from the TTV simulations, all 100 simulations remained stable through the entire time span, with no collisions, ejections, close encounters, or orbit crossings. We find no hints of an increase in the orbital eccentricity of the planets that could lead to dynamical instabilities during the age of the system.

\subsection{Star-planet interaction. Prospects for detecting coherent radio emission}

Auroral radio emission from stars and planets alike is due to the electron cyclotron maser (ECM) instability \citep{Melrose1982}, whereby plasma processes within the stellar (or planetary) magnetosphere generate a population of unstable electrons that amplifies the  emission. The characteristic frequency of the ECM emission is given by the electron gyrofrequency, $\nu_G = 2.8 \times B$\,MHz, where $B$ is the local magnetic field in the source region, measured in Gauss. ECM emission is a coherent mechanism that yields broadband ($\Delta\,\nu \sim \nu_G/2$), highly polarized (sometimes reaching 100\%), amplified non-thermal radiation.  

For Jupiter-like planets, which have magnetic fields $B_{\rm pl} \simeq 10\,\mathrm{G}$, the direct detection of radio emission from them is plausible, as the associated gyrosynchrotron frequency falls above the $\simeq$10\,MHz ionospheric cutoff. However, the detection of radio emission from Earth-sized exoplanets is doomed to fail, as the associated frequency falls below the cutoff. 

Fortunately, if the velocity, $v_{\rm rel}$, of the plasma relative to the planetary body is less than the Alfv\'en speed, $v_A$, i.e., $M_A = v_{\rm rel}/v_A < 1$, where $M_A$ is the Alfv\'en Mach number, then energy and momentum can be transported upstream of the flow along Alfv\'en wings. Jupiter’s interaction with its Galilean satellites is a well-known example of sub-Alfvénic interaction, producing auroral radio emission \citep{Zarka2007}.  In the case of star-planet  interaction, the radio emission arises  from the magnetosphere of the host star, induced by the exoplanet crossing the stellar magnetosphere, and the  relevant magnetic field is that of the star,  $B_\star$, not the exoplanet magnetic field.  Since M-dwarf stars have magnetic fields ranging from about 100\,G to above 2--3\,kG, their auroral emission falls in the range from a few hundred MHz up to a few GHz.  This interaction is expected to  yield detectable auroral radio emission via the cyclotron emission mechanism  (e.g., \citealt{Turnpenney2018,PerezTorres2021}).

We followed the prescriptions in Appendix~B of \citet{PerezTorres2021} to estimate the flux density expected to arise from the interaction between the planets HD~260655~b and HD~260655~c and their host star, at a frequency of $\sim$504\,MHz, which corresponds to the cyclotron frequency of the stellar magnetic field of 180\,G, from \citet{Reiners2022arXiv220400342R}. We computed the radio emission arising from star-planet interaction for a closed dipolar geometry and for two different models of star-planet interaction, the Zarka-Lanza model and the Saur-Turnpenney model (see \citealt{PerezTorres2021} for details).  The interaction between the planet and its host star happens in the sub-Alfv\'enic regime, so that energy and momentum can be transferred to the star and the ECM mechanism can be at work.  We estimated the magnetic field of the planet by using the Sano scaling law \citep{Sano1993}, which assumes the planet is a rocky one. We find values of  $B_{\rm pl,b} \simeq 0.46\,B_\oplus$ and  $B_{\rm pl,c} \simeq 0.24\,B_\oplus$ for HD~260655~b and HD~260655~c, respectively.

\begin{figure}
\centering
\includegraphics[width=\hsize]{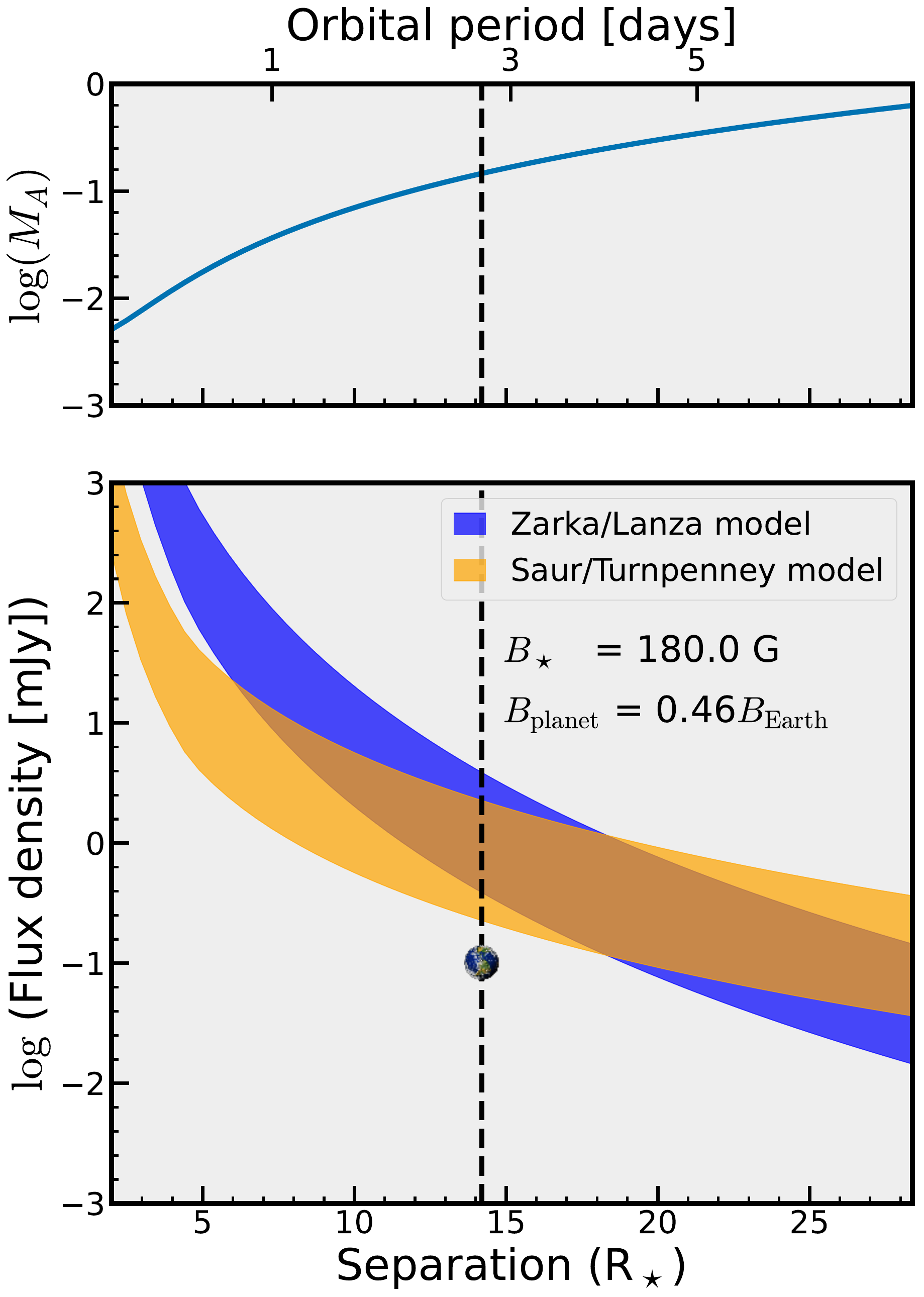}\\
\caption{Expected flux density for auroral radio emission arising from star-planet interaction in the system HD~260655 - HD~260655~b, as a function of orbital distance. The interaction is expected to be in the sub-Alfv\'enic regime (i.e. Alfven Mach number $M_A = v_{\rm rel}/v_{\rm Alfv} \leq 1$; top panel) at the location of the planet (vertical dashed line). The planet in the bottom panel is drawn at 0.1\,mJy in the y-axis. The radio emission expected from the interaction between the planet and its host star holds promises for detection, and we encourage observations to test star-planet interaction scenarios.}\label{fig:toi4559b-spi-radio} 
\end{figure}

As an example, we show in Fig.  \ref{fig:toi4559b-spi-radio} the predicted flux density as a function of orbital distance arising from the interaction of HD~260655~b for the case of a closed dipolar magnetic field geometry of the star.  The yellow and blue shaded areas encompasses the range of values from 0.01 to 0.1 for the efficiency factor, $\epsilon$, in converting Poynting flux into ECM radio emission. The flux density arising from star-planet interaction is expected to be from $\sim$0.24\,mJy up to 4.1\,mJy. The radio emission arising from the interaction between HD~260655~c and its host star is about four times weaker, ranging from $\sim$0.07\,mJy up to 1.1\,mJy, so it also holds promises for detection. We encourage radio observations of this target aimed at detecting a periodic signal that would test those model predictions.

\subsection{Atmospheric characterization}

\begin{figure*}
    \centering
    \includegraphics[width=\hsize]{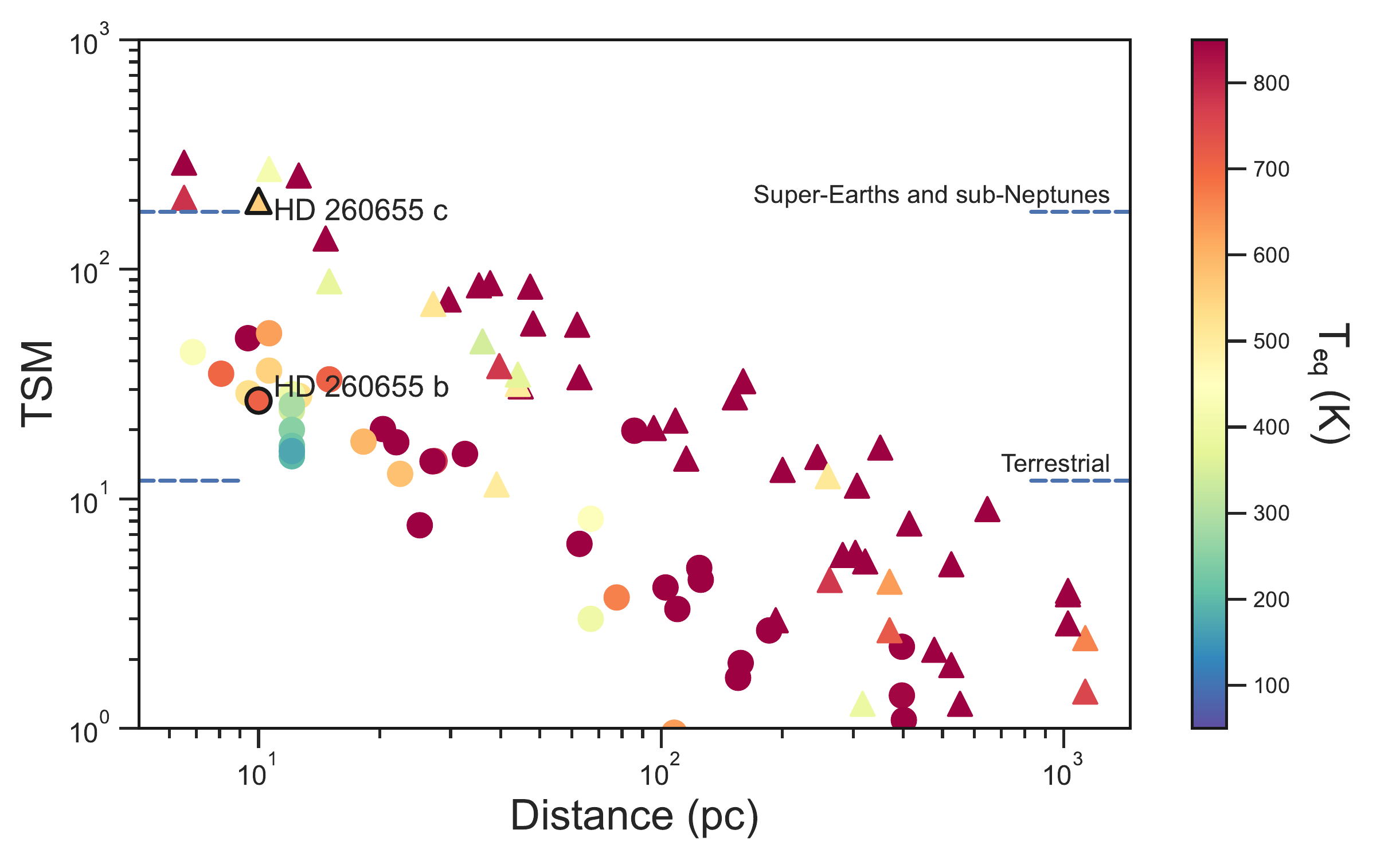}
    \caption{Transiting planets (from the NASA Exoplanet Archive as of January 2022) with radii smaller than $2\,R_\oplus$ and dynamical masses measured via TTVs or RVs as a function of distance from the Sun and the transmission spectroscopy metric (TSM) from \citet{Kempton2018PASP..130k4401K}. The color indicates the equilibrium temperature of the planet. Planets in the habitable zone have equilibrium temperatures ranging between 200 and 300 K and are depicted in turquoise. Following the division in \citet{Kempton2018PASP..130k4401K}, circles indicate terrestrial planets ($R_p < 1.5\,R_\oplus$) and triangles indicate planets in the super-Earth/sub-Neptune range ($1.5 < R_p < 2.8\,R_\oplus$). The blue dashed lines indicate the TSM top quartile threshold to select the most amenable targets for atmospheric characterization studies with the \textit{JWST} for each class.}
    \label{fig:tsm}
\end{figure*}

To evaluate the suitability of the HD~260655 planets for atmospheric characterization studies we computed the Transmission Spectroscopy Metric (TSM) and Emission Spectroscopy Metric (ESM) proposed by \citet{Kempton2018PASP..130k4401K}. Figure~\ref{fig:tsm} shows the TSM for all exoplanets in the NASA Exoplanet Archive\footnote{\url{exoplanetarchive.ipac.caltech.edu/}, (accessed on January 2022).} with a radius smaller than $2\,R_\oplus$ and measured dynamical masses as a function of their distance from the solar system. The estimated TSM values of HD~260655~b and c are 26.8 and 198.6, respectively. These numbers place both targets in the top quartile in their respective categories (terrestrial planets for HD~260655~b and super-Earths and sub-Neptunes for HD~260655~c) according to the simulations by \citet{Kempton2018PASP..130k4401K}. HD~260655~b is among the top 10 terrestrial planets for atmospheric characterization, which includes LTT~1445~A~b \citep{Winters2019AJ....158..152W}, GJ~486~b \citep{GJ486}, GJ~367~b \citep{GJ367}, GJ~357~b \citep{GJ357}, L~98-59~b and c \citep{Kostov2019AJ....158...32K}, TRAPPIST-1~b \citep{Gillon2016Natur.533..221G}, GJ~1132~b \citep{GJ1132}, and LHS~1140~c \citep{Ment2019AJ....157...32M,Lillo-Box2020A&A...642A.121L}. For HD~260655~c the list is even smaller, with only 55~Cnc~e \citep{Winn2011ApJ...737L..18W}, HD~219134~b and c \citep{Motalebi2015A&A...584A..72M,Gillon2017NatAs...1E..56G}, and L~98-59~d \citep{Kostov2019AJ....158...32K}. All of these planets (except for the HD~219134 system, which is too bright) will be observed with the \textit{JWST} as part of its Guaranteed Time Observations, Early Release Science or Guest Observing programs, highlighting the relevance of the HD~260655 system as a prime target for exoplanet atmospheres studies.
The ESM values of HD~260655~b and c are 11.8 and 8.9, respectively. Both values are also above the threshold of 7.5 indicated by \cite{Kempton2018PASP..130k4401K} to select promising targets for the \textit{JWST} terrestrial emission sample.

\begin{figure*}
    \centering
    \includegraphics[width=\hsize]{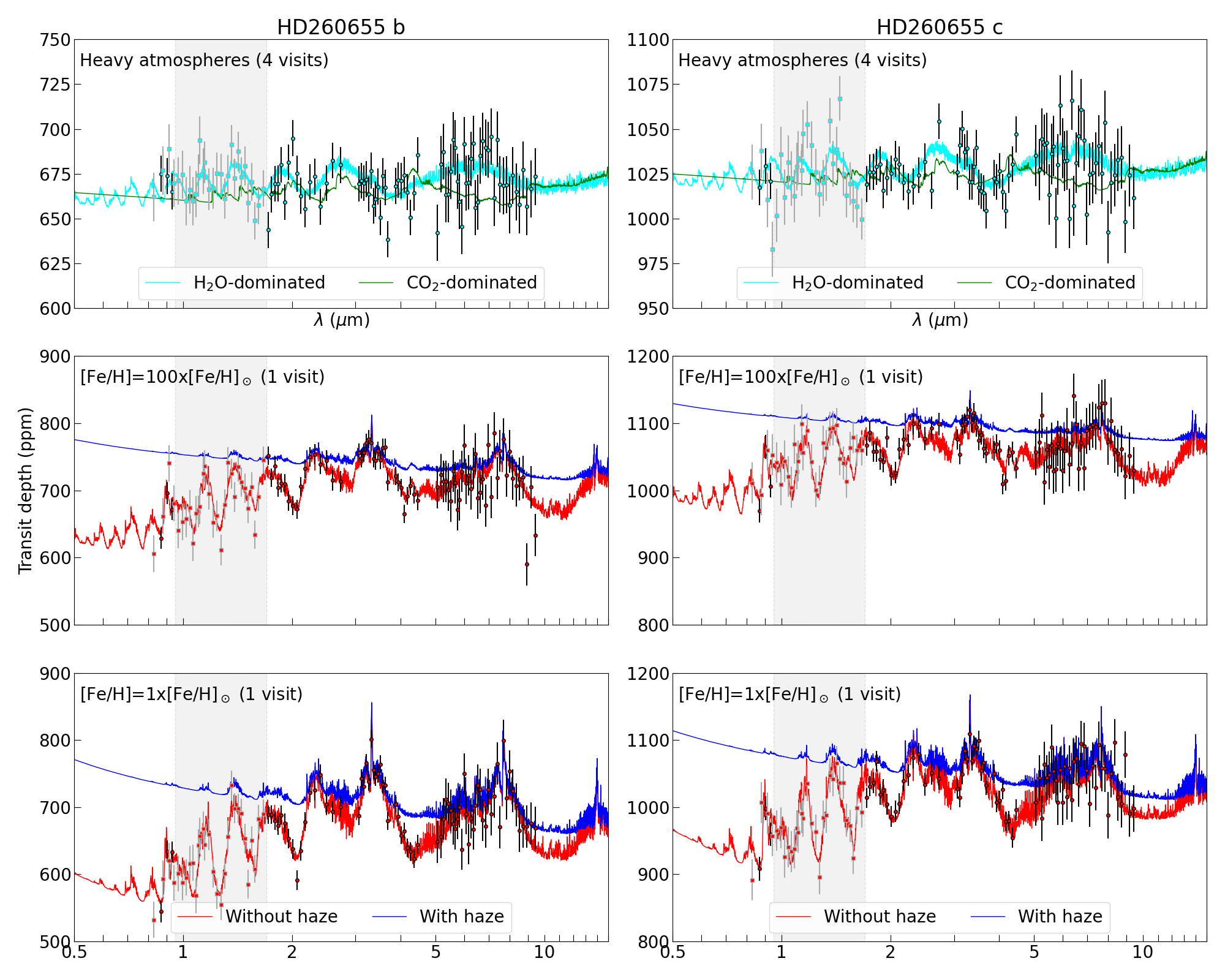}
    \caption{Synthetic atmospheric transmission spectra of HD~260655~b (left) and c (right). \textit{Top}: Models of secondary H$_2$O (cyan) and CO$_2$-dominated (green) atmospheres. \textit{Middle}: Fiducial models of H/He atmospheres with solar abundances and no haze (solid red lines) and with haze (blue). \textit{Bottom}: Models of H/He atmospheres with enhanced metallicity by a factor of 100. Estimated uncertainties are shown for the observation of four/one transits with \textit{JWST} NIRISS-SOSS, NIRSpec-G395H, MIRI-LRS configurations (cyan/red circles with black error bars), and with \textit{HST} WFC3-G102 and G141 scanning modes (cyan/red squares with gray error bars). The gray region denotes the wavelength range 0.95-1.7\,$\mu$m for which the host star brightness is beyond the saturation limits.}
    \label{fig:jwst}
\end{figure*}

We generated synthetic transmission spectra of both planets around HD~260655 for a range of atmospheric scenarios. We adopted the photo-chemical model \texttt{ChemKM} \citep{molaverdikhani2019cold, molaverdikhani2019cold2, molaverdikhani2020} and the radiative transfer code \texttt{petitRADTRANS} \citep{petitRADTRANS} to simulate the native spectra. For each planet, we considered models of H$_2$O- and CO$_2$-dominated atmospheres without H/He, and four models with H/He gaseous envelope, assuming 1$\times$ or 100$\times$ solar abundances, without or with haze. We made use of the online Exoplanet Characterization Toolkit (ExoCTK, \citealp{bourque2021_exoctk})\footnote{\url{https://exoctk.stsci.edu}} and of the \textit{JWST} Exposure Time Calculator (ETC)\footnote{\url{https://jwst.etc.stsci.edu}} to assess the observability of HD~260655 with various spectroscopic modes. The largest spectral coverage can be achieved by combining NIRISS-SOSS (0.6--2.8\,$\mu$m), NIRSpec-G395H (2.87--5.27\,$\mu$m) and MIRI-LRS (5--12\,$\mu$m) instrumental modes. However, the NIRISS-SOSS detector will saturate after the first group on the 0.95--1.7\,$\mu$m range, even using the small subarray (\texttt{SUBSTRIP96}). The NIRSpec-G140H could observe small intervals around 1.1--1.2\,$\mu$m or 1.4--1.6\,$\mu$m, depending on the subarray. We conclude that the HD~260655 system cannot be fully explored with \textit{JWST} in the near-infrared. This gap can be covered by \textit{Hubble Space Telescope} (\textit{HST}) observations using the Wide Field Camera 3 (WFC3) scanning mode with G102 (0.8--1.15\,$\mu$m) and G141 (1.075--1.7\,$\mu$m) grisms.

Finally, we used \texttt{ExoTETHyS}\footnote{\url{https://github.com/ucl-exoplanets/ExoTETHyS}} \citep{morello2021} to compute realistic transmission spectra as they could be observed with \textit{HST} and \textit{JWST} selected instrument modes. For each mode, the wavelength bins were automatically adjusted to have similar count rates. The photon noise error bars were calculated assuming observing windows of twice the transit duration for the \textit{JWST} modes, and three useful orbits for the \textit{HST} modes, including the overheads. The final error bars were inflated by a factor 1.2 to account for correlated noise. Figure~\ref{fig:jwst} shows the simulated transmission spectra.

The spectra of atmospheres depleted of H/He show very weak modulations of $\sim$20-30\,ppm, depending on the dominant molecule. These small amplitudes are comparable with the noise floor of 10\,ppm that is expected for \textit{JWST} transit spectroscopy \citep{beichman2014}. We estimated realistic error bars of 13--18\,ppm for the \textit{JWST} NIRISS-SOSS and NIRSpec-G395H modes with median spectral resolution of R$\sim$50, 30--35\,ppm for the MIRI-LRS with wavelength bin sizes of 0.1--0.2\,$\mu$m, and 20--28\,ppm for the \textit{HST} WFC3-G102 and G141 scanning modes with 12 and 18 bins (R$\sim$40), assuming just one transit observation for each mode. Our simulations indicate that the absorption features in the case of H$_2$O-dominated atmospheres are difficult to detect even combining up to four visits with any \textit{HST} and \textit{JWST} instrument mode.

For the cases of H/He-dominated secondary atmospheres, the spectroscopic modulations are on the order of 100--200\,ppm, mostly attributable to H$_2$O and CH$_4$ absorption. The spectral features are damped by a factor $<$2 in the cases with 100$\times$ solar metallicity. The presence of haze also damps the spectral features, especially at shorter wavelengths, and more severely in case of enhanced metallicity. Similar trends with enhanced metallicity or haze were also observed in simulations made for other planets (e.g., \citealp{LTT3780-1,GJ486,TOI-1759}). 
We conclude that a single transit observation with any of these \textit{JWST} and \textit{HST} modes would be sufficient for robust detection of the molecular features in the H/He-dominated scenarios, otherwise placing tight constraints on the presence of such species and/or of an H/He envelope. In the former scenario, the larger wavelength coverage provided by multiple modes can help distinguishing between the effects of metallicity and haze.

\section{Summary and conclusions} \label{sec:conclusions}

This work presents the discovery and characterization of a multi-planetary system orbiting the nearby M~dwarf HD~260655. Transit observations from \textit{TESS} detected two small planet candidates that were confirmed with independent RV data from the HIRES and CARMENES instruments taken since 1998 and 2016, respectively. The M0\,V star hosts two planets, HD~260655~b and HD~260655~c, on orbits with periods of 2.77 and 5.71\,d. HD~260655~b has a radius of $R_{\rm b} = 1.240 \pm 0.023\,R_\oplus$, a mass of $M_{\rm b} = 2.14 \pm 0.34\,M_\oplus$, and a density of $\rho_{\rm b} = 6.2 \pm 1.0\,\mathrm{g\,cm^{-3}}$, consistent with an Earth-like composition. HD~260655~c has a radius of $R_{\rm c}=1.533^{+0.051}_{-0.046}\,R_\oplus$, a mass of $M_{\rm c}=3.09\pm0.48\,M_\oplus$, and a density of $\rho_{\rm c}=4.7^{+0.9}_{-0.8}\,\mathrm{g\,cm^{-3}}$, implying that it either is nearly devoid of iron, which would be difficult to form, or contains a significant amount of volatiles. Although a very low-mass H/He atmosphere surrounding an Earth-composition would explain the data for planet c, such an atmosphere would likely be removed rapidly. In contrast, an atmosphere made of water surrounding an Earth-composition core fits the observations without invoking astrophysically-unlikely processes.  Nevertheless, the bulk densities from both planets are consistent at the 1$\sigma$ level and the apparent discrepancy may be due to observational uncertainties. 

The HD~260655 system presents a unique opportunity for comparative planetology studies of rocky worlds. At a distance of only 10\,pc, it is the third closest M-dwarf multi-planet transiting system to the Sun (fourth considering all spectral types) and the second brightest after AU\,Mic. Both planets rank among the best targets for transmission and emission spectroscopy observations with \textit{JWST}, which could detect secondary volatile-rich atmospheres or confirm the presence of water and carbon species in one or multiple visits, respectively. Moreover, the radio emission arising from the interaction between the planets and its host could be measured in radio wavelengths. These follow-up observations will improve our knowledge about the formation and evolution history of the system and open a new observational avenue to study the magnetic fields of low-mass stars and their imprint in planetary systems.

\begin{acknowledgements}

This paper includes data collected by the \textit{TESS} mission. Funding for the \textit{TESS} mission is provided by the NASA Explorer Program. We acknowledge the use of public TESS data from pipelines at the TESS Science Office and at the TESS Science Processing Operations Center. Resources supporting this work were provided by the NASA High-End Computing (HEC) Program through the NASA Advanced Supercomputing (NAS) Division at Ames Research Center for the production of the SPOC data products. This research has made use of the Exoplanet Follow-up Observation Program website, which is operated by the California Institute of Technology, under contract with the National Aeronautics and Space Administration under the Exoplanet Exploration Program. 

CARMENES is an instrument at the Centro Astron\'omico Hispano-Alem\'an (CAHA) at Calar Alto (Almer\'{\i}a, Spain), operated jointly by the Junta de Andaluc\'ia and the Instituto de Astrof\'isica de Andaluc\'ia (CSIC). CARMENES was funded by the Max-Planck-Gesellschaft (MPG), the Consejo Superior de Investigaciones Cient\'{\i}ficas (CSIC), the Ministerio de Econom\'ia y Competitividad (MINECO) and the European Regional Development Fund (ERDF) through projects FICTS-2011-02, ICTS-2017-07-CAHA-4, and CAHA16-CE-3978, and the members of the CARMENES Consortium with additional contributions.


Some of the observations in this paper made use of the NN-EXPLORE Exoplanet and Stellar Speckle Imager (NESSI). NESSI was funded by the NASA Exoplanet Exploration Program and the NASA Ames Research Center. NESSI was built at the Ames Research Center by Steve B. Howell, Nic Scott, Elliott P. Horch, and Emmett Quigley.


R.L. acknowledges funding from University of La Laguna through the Margarita Salas Fellowship from the Spanish Ministry of Universities ref. UNI/551/2021-May 26, and under the EU Next Generation funds. 

We acknowledge financial support from the Agencia Estatal de Investigaci\'on of the Ministerio de Ciencia e Innovaci\'on and the ERDF ``A way of making Europe'' through projects
PID2019-109522GB-C5[1:4]/AEI/10.13039/501100011033	
and the Centre of Excellence ``Severo Ochoa'' and ``Mar\'ia de Maeztu'' awards to the Instituto de Astrof\'isica de Canarias (CEX2019-000920-S), Instituto de Astrof\'isica de Andaluc\'ia (SEV-2017-0709), and Centro de Astrobiolog\'ia (MDM-2017-0737); 
the Generalitat de Catalunya/CERCA programme;
the Deutsche Forschungsgemeinschaft (DFG) through the Major Research Instrumentation Programme and Research Unit FOR2544 ``Blue Planets around Red Stars'' (RE 2694/8-1, KU 3625/2-1), the Excellence Cluster ORIGINS (EXC-2094 - 390783311) and the Priority Programme ``Exploring the Diversity of Extrasolar Planets'' (JE 701/5-1);
the National Aeronautics and Space Administration under grants 80NSSC21K0367 and 80NSSC22K0165 in support of Cycles 3 and 4 of the TESS Guest Investigator program; 
the National Science Foundation, Tennessee State University, and the State of Tennessee through its Centers of Excellence Program;
and the Bulgarian BNSF program ``VIHREN-2021'' project No. KP-06-DV/5.
The results reported herein benefited from collaborations and/or information exchange within the program ``Alien Earths'' (supported by the National Aeronautics and Space Administration under agreement No. 80NSSC21K0593) for NASA’s Nexus for Exoplanet System Science (NExSS) research coordination network sponsored by NASA’s Science Mission Directorate.
The authors wish to recognize and acknowledge the very significant cultural role and reverence that the summit of Maunakea has always had within the indigenous Hawaiian community. We are most fortunate to have the opportunity to conduct observations from this mountain.
We thank Vera\,M. Passegger for a helpful discussion on the photospheric parameters of HD~260655.

\end{acknowledgements}

\bibliographystyle{aa} 
\bibliography{biblio} 


\begin{appendix} 

\section{Corner plot of the ground-based T8 APT photometry fit}

\begin{figure*}
    \centering
    \includegraphics[width=\hsize]{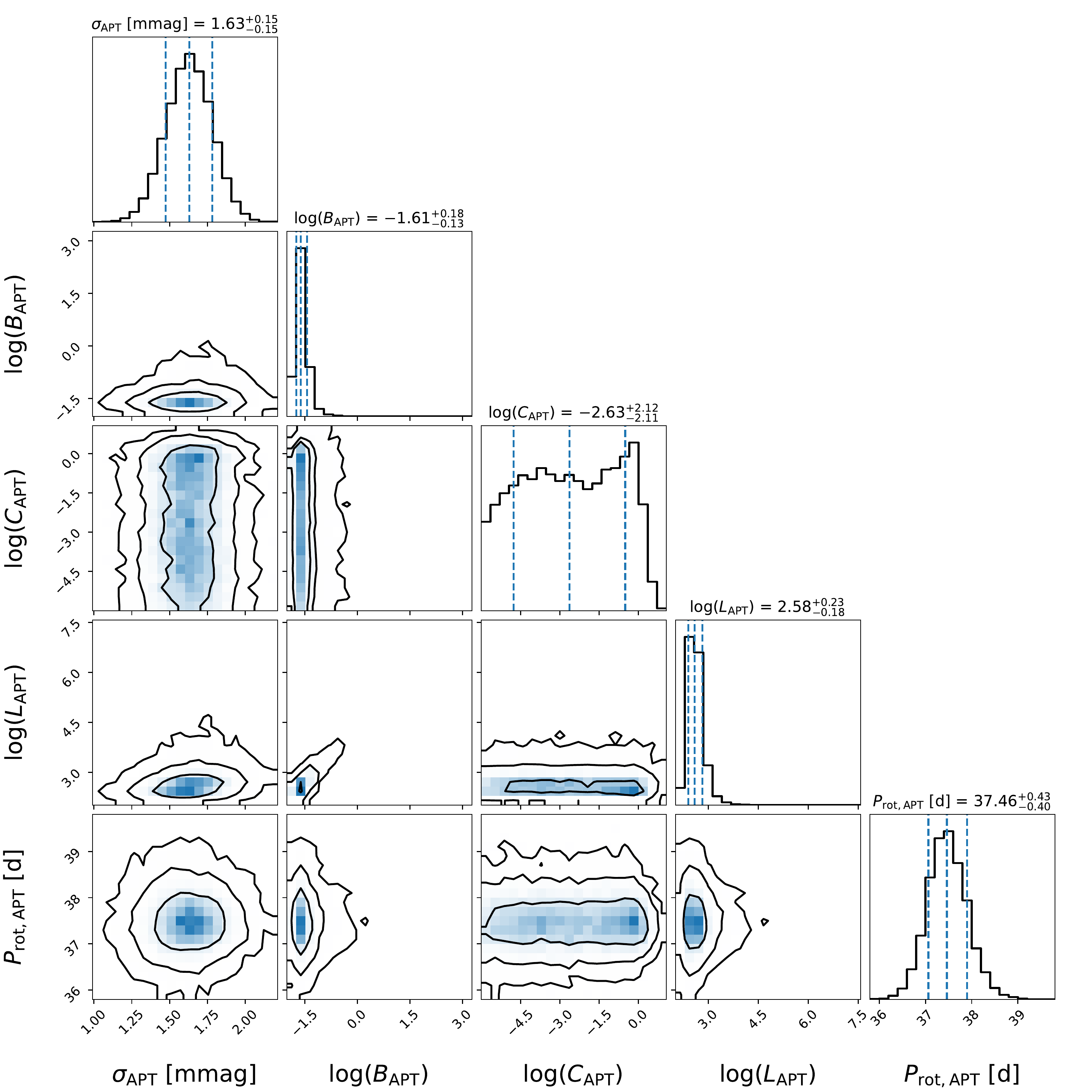}
    \caption{Posterior distributions from the GP modeling of the T8 APT ground-based photometry. The top panels of the corner plot show the probability density function of each fitted parameter. The vertical dashed lines indicate the 16th, 50th, and the 84th percentiles of the samples. Contours are drawn to improve the visualization of the 2D histograms and indicate the 68.3\%, 95.5\%, and 99.7\% confidence interval levels (i.e., 1$\sigma$, 2$\sigma$, and 3$\sigma$). }
    \label{fig:cornerAPT}
\end{figure*}

\section{Analysis of the ground-based SuperWASP photometry}

\begin{figure*}
    \centering
    \includegraphics[width=\hsize]{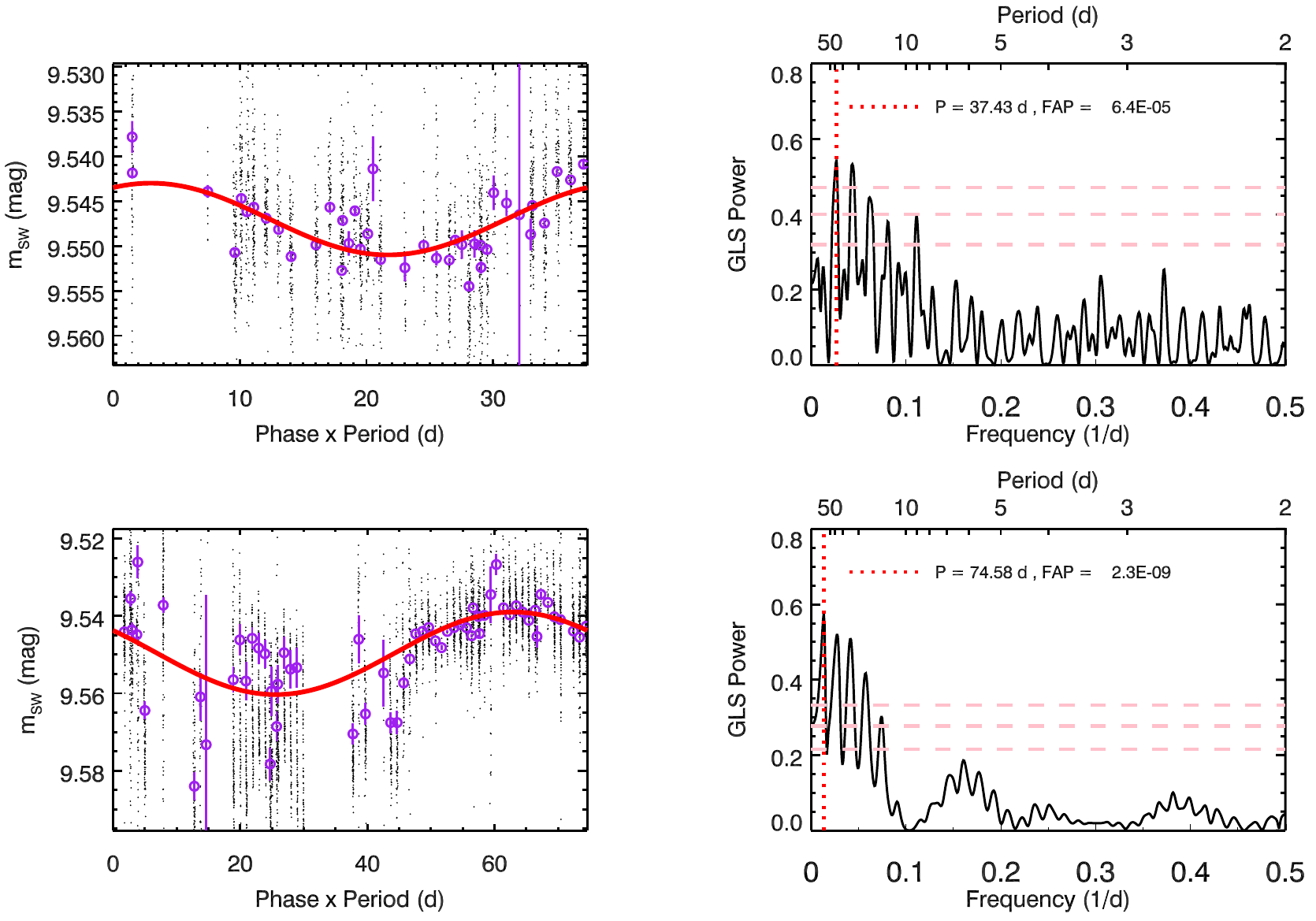}
    \caption{GLS analysis of light curves from SuperWASP. \textit{Left}: phase-folded light curve on the most significant peak period found from each season, overplotted with the best-fit sinusoidal signal in red. Daily-binned data are shown in purple circles. \textit{Right}: GLS periodogram of the data, where the highest peak is highlighted in red dotted lines. FAP levels at 10\%, 1\%, and 0.1\% are shown in pink dashed lines. \textit{Top}: 2009--10 season showing the best period at 37.4\,d. \textit{Bottom}: 2010--11 season showing the top peak period at 74.58d, double that of the first season. While the coverage of SuperWASP data for this star is less complete, it broadly supports the rotation period of 37.5d derived from T8 APT data.  }
    \label{fig:swasp}
\end{figure*}

\end{appendix}

\end{document}